\documentclass[manuscript, screen]{acmart}
\AtBeginDocument{%
  }

\setcopyright{acmlicensed}
\copyrightyear{2018}
\acmYear{2018}
\acmDOI{XXXXXXX.XXXXXXX}

\acmJournal{POMACS}
\acmVolume{37}
\acmNumber{4}
\acmArticle{111}
\acmMonth{8}

\usepackage{makecell}
\usepackage{longtable}
\usepackage{multirow}
\usepackage{bbding}

\usepackage{adjustbox}
\usepackage{graphicx} 
\usepackage{forest}
\useforestlibrary{edges}

\definecolor{diagblue}{RGB}{52, 148, 186}
\definecolor{LSCgreen}{RGB}{117, 189, 167}
\definecolor{EPpurple}{RGB}{212, 211, 221}
\definecolor{VWAcyan}{RGB}{88, 182, 192}
\definecolor{GSblue}{RGB}{38, 131, 198}
\definecolor{customlightgray}{rgb}{0.8, 0.8, 0.8}

\begin{document}

\title{Reinventing Clinical Dialogue: Agentic Paradigms for LLM‑Enabled Healthcare Communication}

\author{Xiaoquan Zhi}
\affiliation{%
  \institution{College of Management and Economics, Laboratory of Computation and Analytics of Complex Management Systems(CACMS), Tianjin University}
  \city{Tianjin}
  \country{China}}
\email{zhixiaoquan@tju.edu.cn}

\author{Hongke Zhao}
\authornote{Hongke Zhao is the corresponding author.}
\email{hongke@tju.edu.cn}
\affiliation{%
  \institution{College of Management and Economics, Laboratory of Computation and Analytics of Complex Management Systems(CACMS), Tianjin University}
  \city{Tianjin}
  \country{China}}

\author{Likang Wu}
\affiliation{%
  \institution{College of Management and Economics, Laboratory of Computation and Analytics of Complex Management Systems(CACMS), Tianjin University}
  \city{Tianjin}
  \country{China}}
\email{wulikang@tju.edu.cn}

\author{Chuang Zhao}
\affiliation{%
  \institution{College of Management and Economics, Laboratory of Computation and Analytics of Complex Management Systems(CACMS), Tianjin University}
  \city{Tianjin}
  \country{China}}
\email{zhaochuang@tju.edu.cn}

\author{Hengshu Zhu}
\affiliation{%
  \institution{Computer Network Information Center, Chinese Academy of Sciences} 
  \city{Beijing}
  \country{China}}
\email{zhuhengshu@gmail.com}

\renewcommand{\shortauthors}{Zhi et al.}

\begin{abstract} 

Clinical dialogue represents a complex duality requiring both the empathetic fluency of natural conversation and the rigorous precision of evidence-based medicine. While Large Language Models possess unprecedented linguistic capabilities, their architectural reliance on reactive and stateless processing often favors probabilistic plausibility over factual veracity. This structural limitation has catalyzed a paradigm shift in medical AI from generative text prediction to agentic autonomy, where the model functions as a central reasoning engine capable of deliberate planning and persistent memory. Moving beyond existing reviews that primarily catalog downstream applications, this survey provides a first-principles analysis of the cognitive architecture underpinning this shift. We introduce a novel taxonomy structured along the orthogonal axes of knowledge source and agency objective to delineate the provenance of clinical knowledge against the system's operational scope. This framework facilitates a systematic analysis of the intrinsic trade-offs between creativity and reliability by categorizing methods into four archetypes: \textit{Latent Space Clinicians}, \textit{Emergent Planners}, \textit{Grounded Synthesizers}, and \textit{Verifiable Workflow Automators}. For each paradigm, we deconstruct the technical realization across the entire cognitive pipeline, encompassing strategic planning, memory management, action execution, collaboration, and evolution to reveal how distinct architectural choices balance the tension between autonomy and safety. Furthermore, we bridge abstract design philosophies with the pragmatic implementation ecosystem. By mapping real-world applications to our taxonomy and systematically reviewing benchmarks and evaluation metrics specific to clinical agents, we provide a comprehensive reference for future development. Finally, we identify critical frontiers regarding trustworthiness, outlining a roadmap for future research to foster reliable and ethically aligned healthcare AI. The latest papers and related resources are maintained on our \href{https://github.com/xqz614/Reinventing-Clinical-Dialogue-Agentic-Paradigms-for-LLM-Enabled-Healthcare-Communication.git}{website}.

\end{abstract}



\maketitle

\section{Introduction} 



Clinical dialogue represents far more than a mere exchange of information, as it constitutes a goal-oriented, sequential decision-making process that is the cornerstone of healthcare delivery. It encompasses a wide spectrum of high-stakes interactions, ranging from patient education and diagnostic information gathering to intricate treatment deliberation \cite{SUPPORT, MedAlign} and behavior change counseling \cite{KGAREVIS}. The effectiveness of this dialogue acts as a primary determinant of clinical outcomes, directly influencing diagnostic accuracy, patient adherence to treatment regimens, and the long-term therapeutic alliance between provider and patient. However, the global scarcity of medical professionals often creates a bottleneck, limiting the time and attention available for high-quality interactions. Consequently, the automation of clinical dialogue has become an imperative research frontier, aiming to scale personalized care without compromising quality.

Historically, the automation of clinical dialogue has progressed through distinct paradigms, each grappling with the trade-off between control and flexibility. The earliest attempts were dominated by pipeline-based dialogue systems \cite{JOINT23}, which employed a modular architecture comprising natural language understanding, dialogue state tracking \cite{LLMasA23}, and policy execution \cite{SEMANK22}. While effective for structured, slot-filling tasks like appointment scheduling, these systems were inherently rigid, suffering from cascading errors where a failure in one module propagated downstream. Subsequently, a retrieval-based paradigm emerged \cite{RETLLM23}. These systems operate by matching user queries to predefined responses from a curated knowledge base, offering high scalability and factual safety. However, they lack the generative flexibility to address the nuanced, long-tail queries typical of authentic patient interactions, often resulting in impersonal and disjointed communication that fails to address specific patient needs \cite{EVENWORSERAN}.

The advent of Large Language Models (LLMs) marked a significant departure from these rigid approaches, introducing a probabilistic generative paradigm. By leveraging linguistic patterns and world knowledge learned from massive corpora during self-supervised pre-training \cite{EVALLLM24}, these models function as powerful sequence-to-sequence generators\cite{EVALLLM24}. Specialized models such as BioGPT \cite{BioGPT22} and Med-PaLM \cite{MEDPALM225} demonstrate unprecedented fluency in understanding complex medical semantics. Their capabilities have been rigorously validated not only on high-stakes benchmarks like the USMLE \cite{MEDPLANA} and MedMCQA \cite{MedMCQA22}, but also across a wide array of pragmatic clinical tasks, ranging from clinical text summarization \cite{HSCR25} and medical note generation \cite{MEDIQA19} to patient-friendly text simplification \cite{AGENT11}. This success across diverse scenarios \cite{MEDQA20, HeadQA} underscores the potential to democratize access to medical knowledge, offering a versatile, general-purpose interface for healthcare applications.

 Despite their linguistic prowess, the paradigm reveals significant design flaws when applied to the rigorous demands of complex clinical environments. Primarily, operating as probabilistic engines, these models often prioritize fluency over factual accuracy without an intrinsic mechanism to verify outputs against real-time clinical evidence, leading to plausible but clinically dangerous hallucinations \cite{RAGITER}. Furthermore, their inherently \textbf{reactive and stateless} nature forces them to treat each patient encounter as a discrete episode, preventing the construction of a longitudinal and coherent model of a patient's evolving condition, which is a prerequisite for chronic disease management \cite{MEDSENTRY, medla}. Compounding these internal limitations is their operational isolation. Functioning as systems disconnected from the dynamic clinical environment, standard LLMs lack the agency to \textbf{actively} query external tools or interact with essential data repositories like Electronic Health Records (EHRs) and clinical guidelines, rendering them insufficient for orchestrating the multi-step, mixed-initiative workflows that define modern clinical practice \cite{Baymax25}.

\subsection{Difference between Agentic Paradigm with Traditional Methods}


To transcend these inherent limitations, the field is rapidly converging on an Agentic Paradigm, which fundamentally reconceptualizes the LLM from a passive text generator into the central reasoning engine of an autonomous, goal-directed system \cite{MEDIATOR, ADVB25}. Unlike traditional models that merely predict the next token based on immediate context, a clinical agent actively perceives its environment, engages in deliberate reasoning, and executes a sequential trajectory of actions to achieve specific clinical objectives. This paradigm is operationalized through a set of disentangled yet synergistic technical components, each rigorously designed to address a specific cognitive deficit of the foundational model. To resolve the challenge of complex clinical reasoning, \textbf{strategic planning modules} enable the agent to break down high-level, ambiguous objectives (e.g., diagnosing a rare disease) into a structured sequence of executable sub-goals \cite{COD25, MEDPLANA}. To overcome the stateless nature of standard interactions, sophisticated \textbf{memory management systems} are employed to maintain a persistent, longitudinal context of the patient's history and the evolving dialogue state, ensuring continuity of care across sessions \cite{MEDIKNOWLED, CliCR}. Bridging the critical gap between the model's internal knowledge and the real-world clinical environment, \textbf{action execution} mechanisms empower the agent to invoke verifiable external tools \cite{ATOMRAG} and query authoritative knowledge sources \cite{DIALMED22}, thereby grounding its decisions in reality rather than in hallucination. Furthermore, recognizing that effective healthcare is rarely a solitary endeavor, \textbf{collaboration layers} orchestrate dynamic interactions among multiple specialized agents \cite{ENHANCECLITRIAL}, simulating the multi-disciplinary team (MDT) approach found in real-world medical practice. Finally, distinguishing themselves from static pre-trained models, these systems incorporate \textbf{evolutionary mechanisms} for self-reflection and learning from environmental feedback \cite{MedBullet}, allowing the agent to continuously refine its clinical strategies over time.

\begin{figure}[!htp]
    \centering
    \setlength{\abovecaptionskip}{0cm}   
    \setlength{\belowcaptionskip}{0cm}   
    \includegraphics[width=0.8\linewidth]{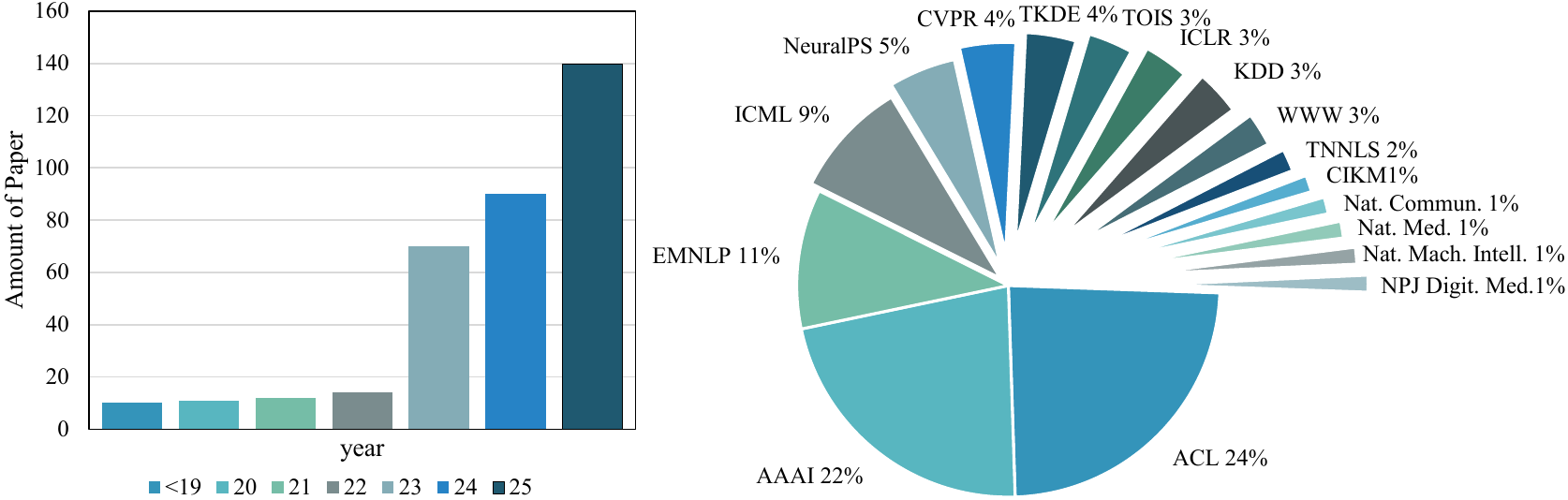}
    \caption{Statistical overview of the literature surveyed. The bar chart (left) illustrates the temporal distribution of cited works, highlighting the surge in recent research ("<19" denotes 2019 and prior). The pie chart (right) depicts the distribution of articles across top-tier venues. Venue abbreviations follow the standard of \href{https://www.acm.org/publications/authors/reference-formatting}{ACM}.}
    \label{fig:stats}
\end{figure}

To construct a comprehensive overview of this emerging field, we have conducted a thorough review of \textbf{over 300 papers} investigating the theory and application of LLM-based agents in healthcare and related scientific domains. Our search was carried out using reputable academic databases such as PubMed, Google Scholar, ACM Digital Library, and DBLP, utilizing specific keywords, including "medical agent", "clinical agent", "llm for medicine", "agentic paradigm", in conjunction with "healthcare", "clinical dialog", and "question answering". The surveyed papers were meticulously sourced from esteemed computer science and medical informatics conferences and journals such as ACL, NeurIPS, ICLR, ICML, KDD, WWW, Nature Medicine, Nature Machine Intelligence, and NPJ Digital Medicine, as depicted in Fig.~\ref{fig:stats}. To ensure the inclusion of state-of-the-art research, we also explored citation networks and incorporated relevant, high-impact preprints from \href{https://arxiv.org/}{arXiv}. 

\subsection{Existing Surveys and Our Contribution} 





While the rapid proliferation of LLMs in medicine has prompted several reviews, our work, as demonstrated in Table~\ref{tab:compsurv}, addresses a distinct void by shifting the focus from downstream applications to the intrinsic architectural philosophy of agentic systems. A critical analysis of the existing literature reveals three major gaps that this survey uniquely aims to bridge. First, surveys on LLMs in healthcare predominantly offer a task-centric taxonomy, categorizing papers by functions such as summarization, triage, or coding \cite{LLMINHEALTH1, LLMINHEALTH5}. While valuable for understanding utility, these works treat the LLM primarily as a static information processor, failing to capture the fundamental paradigm shift required to transform these passive models into proactive, state-tracking agents capable of autonomous workflow execution. Second, general surveys on Agentic AI establish foundational concepts of planning and tool use but operate largely within open domains \cite{AGENT11, AGENTH22}. Consequently, they overlook the stringent constraints of the medical field—specifically, the non-negotiable requirements for safety, interpretability, and verifiable knowledge grounding—necessitating a re-examination of agent architectures through the lens of clinical rigor. Third, and most critically, while recent reviews of medical agents have begun to emerge \cite{Baymax25, AGENTRAG44}, they largely remain at a phenomenological level, classifying systems based on their clinical application scenarios or merely listing technical modules. These reviews often lack a unified framework to explain why an agent is designed with specific memory or planning mechanisms \cite{MEDDIAG1, MEDDIAG2, ASYSTEMETICREV}. Distinct from these works, we move beyond surface-level categorization to provide a first-principles analysis of clinical agents, as shown in Table \ref{tab:compsurv}. We argue that the differentiation of agents should be based on how they derive knowledge and how they exercise agency, as these dimensions dictate the intrinsic trade-offs between generative creativity and factual reliability, and between operational autonomy and clinical safety. By bridging abstract design philosophies with concrete technological implementations, we offer a conceptual lens that explains the trade-offs inherent in building reliable digital doctors.
\begin{table}
\centering
\setlength{\abovecaptionskip}{0cm}   
\setlength{\belowcaptionskip}{0cm}   
\caption{Comparison between existing surveys and our survey. $\checkmark$ denotes the presence of an attribute.}
\label{tab:compsurv}
\resizebox{0.85\textwidth}{!}{
\begin{tabular}{ccclccccccc} 
\hline
\multirow{3}{*}{Ref.} & \multicolumn{2}{c}{Paradigm of Clinical Dialogue}     &                   & \multicolumn{7}{c}{In-depth Review of Agentic Clinical Dialgue}                                                        \\ 
\cline{2-3}\cline{5-11}
                      & \multirow{2}{*}{LLM-Based} & \multirow{2}{*}{Agentic} & \multirow{2}{*}{} & \multicolumn{5}{c}{Technical Component}                & \multirow{2}{*}{Agentic Knowledge} & \multirow{2}{*}{Agency}  \\ 
\cline{5-9}
                      &                            &                          &                   & Planning & Action & Memory & Collaboration & Evolution &                                    &                          \\ 
\hline
2019\cite{ASYSTEMETICREV} & \checkmark & & &  &  &  &  &  &  &  \\ \hline
2022\cite{MEDDIAG2} & \checkmark & \checkmark &&  & \checkmark & \checkmark &  &  & \checkmark &  \\ \hline
 2024\cite{LLMINHEALTH1}& \checkmark & & &  &  &  &  &  &  &  \\ \hline
 2024\cite{AGENT11}&  & & & \checkmark &  &  &  &  &\checkmark  &  \\ \hline
 2024\cite{AGENTH22}&  & & & \checkmark & \checkmark &  &  &  &  &  \\ \hline
2024\cite{MEDDIAG1} &  & \checkmark& & \checkmark & \checkmark & \checkmark &  &  &  & \checkmark \\ \hline

2024\cite{LLMINHEALTH5} &  &  &&  & \checkmark & \checkmark &  &  &  &  \\ \hline
2024\cite{AGENTH3} &  & & &  &  & \checkmark &  &  & \checkmark &  \\ \hline
2025\cite{ACOMPSURV55} &  & & &  &  &  &  & \checkmark &  &  \\ \hline
2025\cite{Baymax25} &  & \checkmark && \checkmark & \checkmark &  \checkmark & \checkmark &  &  &  \\ \hline
2025\cite{AGENTRAG44} &  &  &  & \checkmark & \checkmark &  &  &  &  &\checkmark  \\ \hline
\textbf{Our Survey} & \CheckmarkBold & \CheckmarkBold & & \CheckmarkBold & \CheckmarkBold & \CheckmarkBold & \CheckmarkBold & \CheckmarkBold & \CheckmarkBold & \CheckmarkBold \\ \hline
\end{tabular}}
\end{table}
In a nutshell, this survey fills this critical gap by providing a principled, in-depth analysis of the agentic paradigms shaping modern healthcare communication. Our primary contributions are \textbf{threefold}:

\begin{itemize}
    \item \textbf{A novel taxonomy based on fundamental trade-offs}. We introduce a novel taxonomy for clinical dialogue agents, organized along two critical and conceptually orthogonal perspectives: \textit{Knowledge Source} and \textit{Agency Objective}. We argue that these axes are fundamental because they dictate the two most critical trade-offs in medical AI, implying the balance between creativity and reliability, and the balance between safety and autonomy. This framework allows for a systematic analysis of the field, revealing four distinct paradigms.
    \item \textbf{Comprehensive and in-depth component analysis}. Moving beyond mere categorization, we conduct a granular analysis of the core technical components within each of the four paradigms. We dissect how different paradigms operationalize these components differently to meet their specific objectives, highlighting distinct engineering challenges and delineating potential application scenarios to provide a clear roadmap for developers.
    \item \textbf{Systematic review of resources and future directions}. We provide a systematic review of the implementation tools, benchmark datasets, and evaluation metrics that are critical for developing and assessing clinical agents. Furthermore, we synthesize open challenges, such as neuro-symbolic integration and holistic patient management, to guide future research toward more reliable and ethically aligned healthcare AI.
\end{itemize}

\subsection{Survey Organizations}
The rest of this survey is organized as follows. Section~\ref{sec:pre} establishes the preliminaries and defines the key concepts of clinical dialogue and the agentic paradigm. Section \ref{sec:agentic} forms the core of our analysis, where we systematically dissect each of the four proposed agentic paradigms. Section \ref{sec:data_eval} provides an overview of implementation tools, datasets, and evaluation metrics. Section \ref{sec:appl} maps real-world applications to our taxonomy. Finally, Section \ref{sec:futurecha} discusses open research challenges and future directions before we conclude the survey.

\section{Preliminaries and Definition}\label{sec:pre}

This section establishes the formal conceptual groundwork for our survey. We rigorously define clinical dialogue as a stochastic decision process under uncertainty. Subsequently, we delineate the pivotal paradigm shift from the foundational LLM paradigm to the emerging agentic paradigm. To facilitate a clear understanding, we summarize the primary mathematical notations in Table \ref{tab:nota}.

\begin{table}[!htp]
\centering
\setlength{\abovecaptionskip}{0.1cm} 
\setlength{\belowcaptionskip}{0cm}
\caption{Important mathematical notations used in this survey.}
\label{tab:nota}
\resizebox{0.85\textwidth}{!}{
\begin{tabular}{cl} 
\toprule
\textbf{Notation} & \textbf{Detailed Description} \\ 
\midrule
$t, T$ & Current turn index and the total number of dialogue turns \\
$s_t$ & The unobservable, latent state space and specific state at turn $t$, $s_t\in\mathcal{S}$\\
$o_t $ & The observation space and observation at turn $t$ including user utterance $u_t$, $o_t\in\mathcal{O}$ \\
$a_t$ & The action space and action at turn $t$ including system response $r_t$ and tool use, $a_t\in\mathcal{A}$ \\
$H_t$ & Observable dialogue history at turn $t$, $H_t = \{(o_1, a_1),(o_2, a_2), \dots, (o_{t-1}, a_{t-1})\}$ \\
$\tau$ & A interaction trajectory $\tau = \{(s_1, a_1),(s_2, a_2) \dots, (s_{T}, a_{T})\}$ \\
$\Phi$ & The latent pathology/medical reality of the patient ($\Phi \subset s_t$) \\
$G$ & The overarching clinical goal, including diagnosis, triage, and education \\
$\mathcal{K}$ & External knowledge space, including EHRs, guidelines, and tools\\
$\pi$ & The policy function mapping history or belief states to actions \\
$R(\cdot)$ & The reward function evaluating clinical utility and safety \\
$\theta$ & Trainable parameters of the foundation model \\
$b_t$ & The theoretical belief state distribution over $\mathcal{S}$ at turn $t$ \\
$S_t$ & The agent's internal memory state approximating $s_t$ \\
$\mathcal{P}_t$ & The generated plan or sequence of sub-goals at turn $t$ \\
$\mu$ & The memory transition function updating the internal state $S_t$ \\
$Q_\theta$ & The action-value function estimating the expected cumulative reward \\
$\mathcal{G}_{conn}$ & Topology graph defining agent connections in multi-agent systems \\
\bottomrule
\end{tabular}}
\end{table}

\subsection{What is Clinical Dialogue?}
From a formal perspective, clinical dialogue represents a goal-oriented, sequential decision-making process. Unlike open-domain chitchat, which primarily relies on surface-level semantics, clinical dialogue is characterized by a fundamental information asymmetry: a substantial gap exists between the observable dialogue surface, including patient utterances and reports, and the underlying, often unobservable, medical reality.

To model this complex interaction rigorously, we conceptualize it as a \textbf{Partially Observable Markov Decision Process} (POMDP) defined by the tuple $\mathcal{M} = \langle \mathcal{S}, \mathcal{A}, \mathcal{O}, \mathcal{T}, \mathcal{Z}, R, \gamma \rangle$. Central to this framework is the latent state $s_t \in \mathcal{S}$, which encapsulates the ground-truth patient status. While in diagnostic scenarios this corresponds to the latent pathology $\Phi$, in broader contexts, it encompasses variables such as urgency levels, patient information gaps, or psychological states. The system interacts with this environment by executing actions $a_t \in \mathcal{A}$ (e.g., inquiries, education, or reassurance) and receiving observations $o_t \in \mathcal{O}$ (e.g., user responses $u_t$), which serve as noisy probabilistic projections of the hidden state. The temporal decision horizon is regulated by a discount factor $\gamma \in [0,1]$, which mathematically balances the trade-off between immediate patient satisfaction and long-term health outcomes. The interaction follows a biological and cognitive logic governed by two probability functions: the transition function $\mathcal{T}(s_{t+1} \mid s_t, a_t)$, modeling how the patient state evolves such as symptom worsening or anxiety reduction, and the observation function $\mathcal{Z}(o_t \mid s_t, a_{t-1})$, modeling how internal states manifest externally. Since the true medical reality $\Phi \subset s_t$ is hidden, the system relies on the observable history $H_t = \{(o_1, a_1), \dots, (o_{t-1}, a_{t-1})\}$ to infer the state. Effective clinical reasoning requires maintaining a belief state $b_t$, a probability distribution over $\mathcal{S}$ derived from $H_t$. Mathematically, updating this belief upon new evidence follows Bayesian inference:
\begin{equation} \label{eq:belief_update_bayes}
    b_t(s') \propto \underbrace{\mathcal{Z}(o_t \mid s', a_{t-1})}_{\text{Observation Probability}} \sum_{s \in \mathcal{S}} \underbrace{\mathcal{T}(s' \mid s, a_{t-1})}_{\text{State Evolution}} b_{t-1}(s),
\end{equation}
where $s'\in\mathcal{S}$ denotes the candidate state. Eq. \ref{eq:belief_update_bayes} mathematically formalizes the clinical reasoning process: iteratively refining the system's understanding of the patient's needs and condition based on new evidence. The ultimate objective is to learn a policy $\pi$ that maximizes the expected cumulative clinical utility (Reward $R(\cdot)$) over the interaction trajectory $\tau$:
\begin{equation} \label{eq:pomdp_ideal}
    \pi^* = \operatorname*{arg\,max}_{\pi} \mathbb{E}_{\tau \sim \pi} \left[ \sum_{t=1}^{T} \gamma^{t-1} R(s_t, a_t, G) \right].
\end{equation}
This formulation highlights that clinical dialogue systems must solve the dual challenges of state estimation to reduce entropy regarding $\Phi$ and strategic planning to optimize the long-term utility defined in Eq. \ref{eq:pomdp_ideal}, thereby achieving the clinical goal.

\subsection{What is LLM and Agentic Paradigm?}
\begin{figure}[!htp]
    \centering
    \setlength{\abovecaptionskip}{0cm}
    \setlength{\belowcaptionskip}{0cm}
    \includegraphics[width=0.8\linewidth]{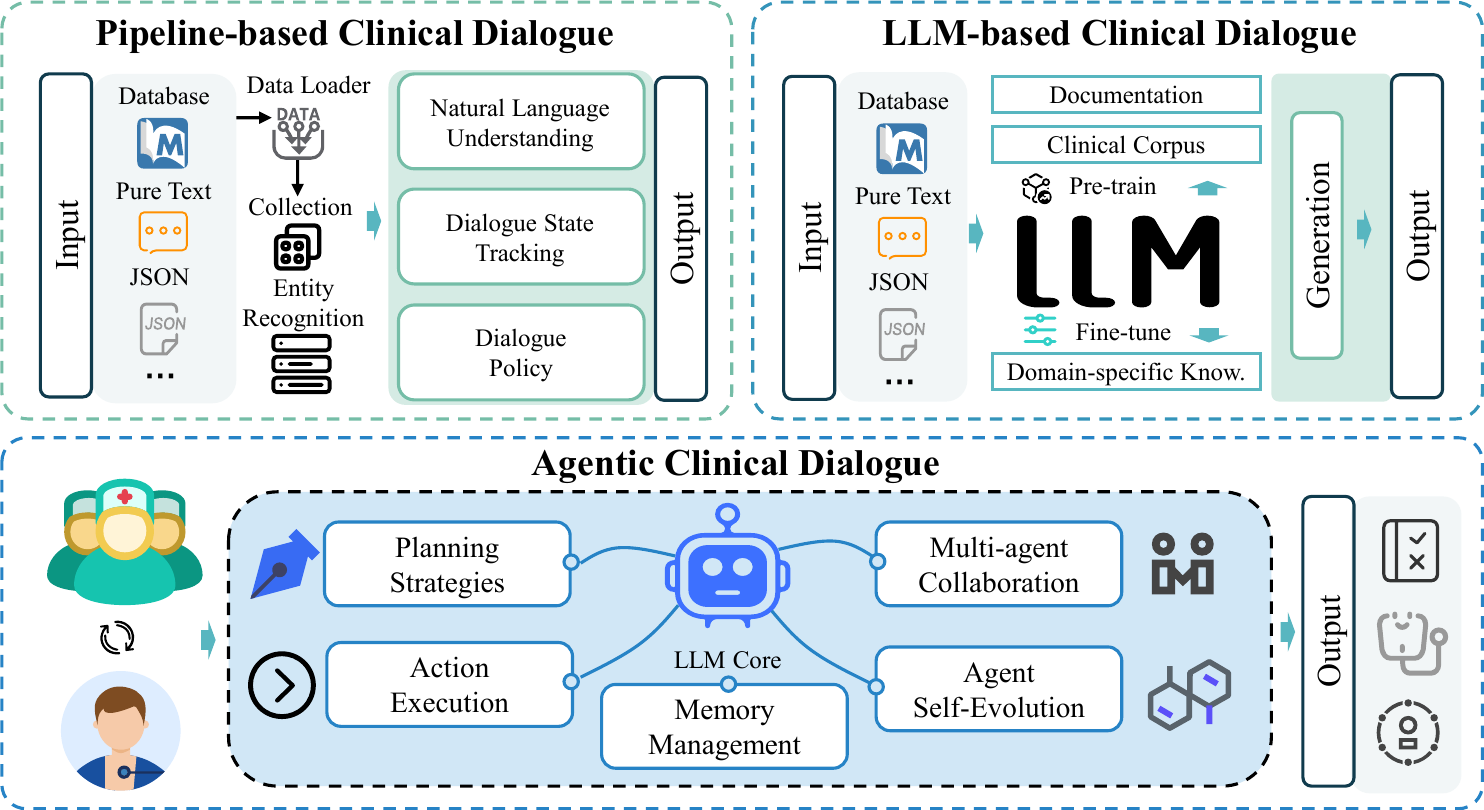}
    \caption{Comparison of three paradigms of clinical dialogue. The evolution progresses from modular pipeline-based systems (upper left) and reactive LLM-based generators (upper right) to he agentic paradigm (bottom). It represents a shift, elevating the LLM from a passive text predictor to an autonomous controller by integrating strategic planning, persistent memory, tool execution, and others.}
    \label{fig:path}
\end{figure}

LLMs represent a technological inflection point; however, their application in medicine necessitates a rigorous distinction between generative capabilities and agentic reasoning. We delineate this evolution from the foundational paradigm to the agentic paradigm, as visualized in Fig. \ref{fig:path}.

\noindent\textbf{The foundational LLM paradigm in clinical dialogue}. 
Traditional approaches utilizing LLMs simplify the POMDP by effectively assuming that the observable dialogue history is a sufficient proxy for the latent state (i.e., $s_t \approx H_t$). Under this assumption, the complex decision process collapses into a conditional sequence generation task. In this paradigm, the system response $r_t$ is sampled from a conditional probability distribution parameterized by $\theta$, implicitly encoding specific knowledge $\mathcal{K}_{\text{implicit}}$:
\begin{equation}
    r_t \sim P_\theta\left(r_t \mid u_t, H_t, G; \mathcal{K}_{\text{implicit}}\right).
\end{equation}
Consequently, the learning strategy transitions from maximizing the intractable cumulative reward (Eq. \ref{eq:pomdp_ideal}) to supervised fine-tuning based on expert demonstrations. The optimization objective shifts to the maximum likelihood estimation of a gold-standard response sequence $D = \{(u_1, r_1^*), \dots, (u_T, r_T^*)\}$, where the model is trained to minimize the divergence from human expert behavior:
\begin{equation}\label{eq:mle}
    \theta^* = \operatorname*{arg\,max}_{\theta} \sum_{(u_t, r_t^*) \in D} \log P_\theta(r_t^* \mid u_t, H_t, G; \mathcal{K}_{\text{implicit}}).
\end{equation}
Despite its efficacy in fluency and short-term context understanding \cite{CLISUM25}, this paradigm is inherently reactive. By bypassing the explicit modeling of the latent medical state $s_t$, it often struggles with long-horizon reasoning and factual consistency when the parametric knowledge conflicts with patient-specific data.

\noindent\textbf{The agentic paradigm: a shift towards autonomous systems}. The Agentic Paradigm elevates the LLM from a passive text generator to the cognitive controller of an autonomous system \cite{COD25, MMEDAG24}. Unlike the reactive foundational paradigm, the agent explicitly addresses the POMDP structure by maintaining a comprehensive internal state $S_t$ to approximate the unobservable medical reality $s_t$. The agent optimizes a dynamic policy $\pi_\theta(a_t | S_t)$ to maximize the expected return:
\begin{equation} \label{eq:agent_objective}
    \mathcal{J}(\pi_\theta) = \mathbb{E}_{\tau \sim \pi_{\theta}} \left[ \sum_{k=0}^{\infty} \gamma^k R(S_{t+k}, a_{t+k}, G) \right].
\end{equation}
This paradigm is operationalized through five orthogonal yet synergistic components, each playing a distinct mathematical role in solving the clinical POMDP:

\begin{itemize}
    \item \textbf{Strategic planning}. 
    To tackle the complexity of clinical goals $G$, this module bridges the gap between high-level objectives and executable actions. It generates a plan $\mathcal{P}_t$ (a sequence of sub-goals or reasoning steps) conditioned on the current belief state:
    \begin{equation}
        \mathcal{P}_t = \Pi_{\text{plan}}(G, S_t, \mathcal{K}) = \{g_1, g_2, \dots, g_k\}.
    \end{equation}
    In the POMDP context, $\mathcal{P}_t$ serves as a constraint on the policy space, guiding the agent to select actions that systematically reduce the entropy of the latent pathology $\Phi$ rather than engaging in greedy token generation \cite{REACT23, AGENMD24}.

    \item \textbf{Memory management}. 
    To overcome the stateless nature of raw LLMs, agents must maintain a persistent internal state $S_t$. This is formalized as a memory transition function $\mu$, which approximates the Bayesian belief update (Eq. \ref{eq:belief_update_bayes}) defined in Section \ref{sec:pre}:
    \begin{equation}
       S_t = \mu(S_{t-1}, o_t, a_{t-1}, \mathcal{K}).
    \end{equation}
    Here, $S_t$ accumulates evidence from observations $o_t$ and tool outputs, constructing an evolving patient model that tracks variables like symptom duration and medication history over time \cite{MEDRAG25}.

    \item \textbf{Action execution}. 
    The agent interacts with the physical and digital worlds through an expanded action space $\mathcal{A} = \mathcal{A}_{\text{text}} \cup \mathcal{A}_{\text{tool}} \cup \mathcal{A}_{\text{plan}}$. The policy selects specific actions not randomly but by evaluating their expected utility. We formalize this via an action-value function $Q_\theta$, which estimates the cumulative reward (Eq. \ref{eq:agent_objective}) conditioned on the current plan:
    \begin{equation}
        a_t = \operatorname*{arg\,max}_{a \in \mathcal{A}} Q_\theta(S_t, a, \mathcal{P}_t).
    \end{equation}
    By invoking search engines or clinical tools ($a \in \mathcal{A}_{\text{tool}}$) \cite{SEAR125}, the agent actively gathers new information to bridge the gap between its belief $S_t$ and the true medical reality $s_t$, grounding its reasoning in verifiable evidence.

    \item \textbf{Collaboration}. 
    In complex scenarios, the clinical outcome is an emergent property of the interaction between $N$ distinct agents. We formalize this interaction as a consensus protocol $\Psi$ defined over a topology graph $\mathcal{G}_{conn}$. The system's final action is sampled from a joint distribution derived by aggregating individual policies:
    \begin{equation}
        a_t \sim \Psi_{\text{consensus}}\left( \big\{ \pi_i(\cdot \mid S_t^{(i)}) \big\}_{i=1}^N ; \mathcal{G}_{conn} \right),
    \end{equation}
    where $\Psi$ represents the mechanism (e.g., weighted voting, iterative debate, or a meta-policy) that minimizes the divergence between conflicting agents, mimicking a medical board where diverse expert perspectives are synthesized to ensure decision robustness \cite{COLAC25}.

    \item \textbf{Agent evolution}. 
    Distinct from static deployment, agentic systems possess the capacity for self-improvement. Evolution is modeled as a parameter update mechanism based on historical trajectories $\tau_{\text{history}}$ and feedback rewards $R$:
    \begin{equation}
        \theta_{k+1} \leftarrow \theta_k + \alpha \nabla_\theta \mathbb{E}_{\tau} [R(\tau)].
    \end{equation}
    This allows the agent to internalize successful diagnostic patterns from past experiences \cite{SELFEVT25}, progressively refining its policy $\pi_\theta$ to better approximate the optimal solution without extensive human retraining.
\end{itemize}

These components collectively transform the LLM from a static knowledge repository into a dynamic problem-solving entity. They form the conceptual basis for the taxonomy of agentic paradigms for clinical dialogue, which will be systematically analyzed in the subsequent chapters.

\subsection{Why Agents for Healthcare Communication?} 


Unlike open-domain conversation, a clinical encounter is structured to achieve a specific outcome, such as a diagnosis or treatment plan. This requires proactive state management, task decomposition, and the ability to steer the conversation\cite{MEDAGENT24, MEDSYN24}. The agentic paradigm, through strategic planning, explicitly models this workflow, transforming the LLM into a purposeful conversational director rather than a passive respondent \cite{MEDAGENT24}.
Secondly, safe clinical dialogue must be radically contextual and grounded in an external, dynamic information ecosystem. The veracity of a medical statement is contingent upon real-time patient data from EHRs and the latest evidence from clinical guidelines \cite{BEYONDEHR}. Relying solely on the model's internal parameters creates an unacceptable risk of factual hallucination and clinically unsafe outputs \cite{SAFERAG25}. An agent, via action execution, fundamentally reframes the dialogue as a process of reasoning over retrieved authoritative evidence, making the communication verifiable and trustworthy.
Thirdly, effective healthcare communication is inherently longitudinal, building a cumulative and evolving understanding of a patient over time. To avoid suffering from catastrophic forgetting between encounters \cite{TOOLCLINIAGEN}, the agentic paradigm addresses this by introducing persistent memory management and agent evolution. This allows the agent to construct and continuously refine a dynamic patient model, enabling context-aware engagement and ensuring continuity of care. This is a hallmark of a robust therapeutic alliance \cite{MULTIEV25}.

\subsection{Taxonomy of Agents} 



To systematically analyze the burgeoning landscape of LLM-based agents in healthcare, we propose a novel taxonomy structured along two conceptually orthogonal axes. These axes address the two most fundamental questions of any autonomous system: "From where does it derive its knowledge?" and "What is its objective?". The intersection of these axes defines a $2\times2$ matrix, yielding four distinct agentic paradigms that frame the subsequent analysis of this survey.

\textbf{The Knowledge Source} axis delineates the primary source of clinical knowledge that an agent utilizes. It spans a spectrum from Implicit Knowledge Navigation, which emphasizes creativity by leveraging the vast, unstructured knowledge embedded within the LLM's parameters, to Explicit Knowledge Grounding, which prioritizes reliability by compelling the agent to anchor its reasoning in external, verifiable knowledge sources.\textbf{The Agency Objective} axis defines the agent's primary operational objective. This axis ranges from Event Cognition, where the goal is to understand and summarize a clinical situation, emphasizing safety and acting as an advisor, to Goal Execution, where the aim is to autonomously complete a multi-step clinical workflow, emphasizing autonomy and acting as a collaborator. The interplay between these two axes gives rise to \textbf{four archetypal agent categories}, as shown in Fig. \ref{fig:cat}:

\begin{figure}[!htp] 
    \centering
    \setlength{\abovecaptionskip}{0cm}   
    \setlength{\belowcaptionskip}{0cm}   
    \includegraphics[width=0.6\linewidth,height=0.45\linewidth]{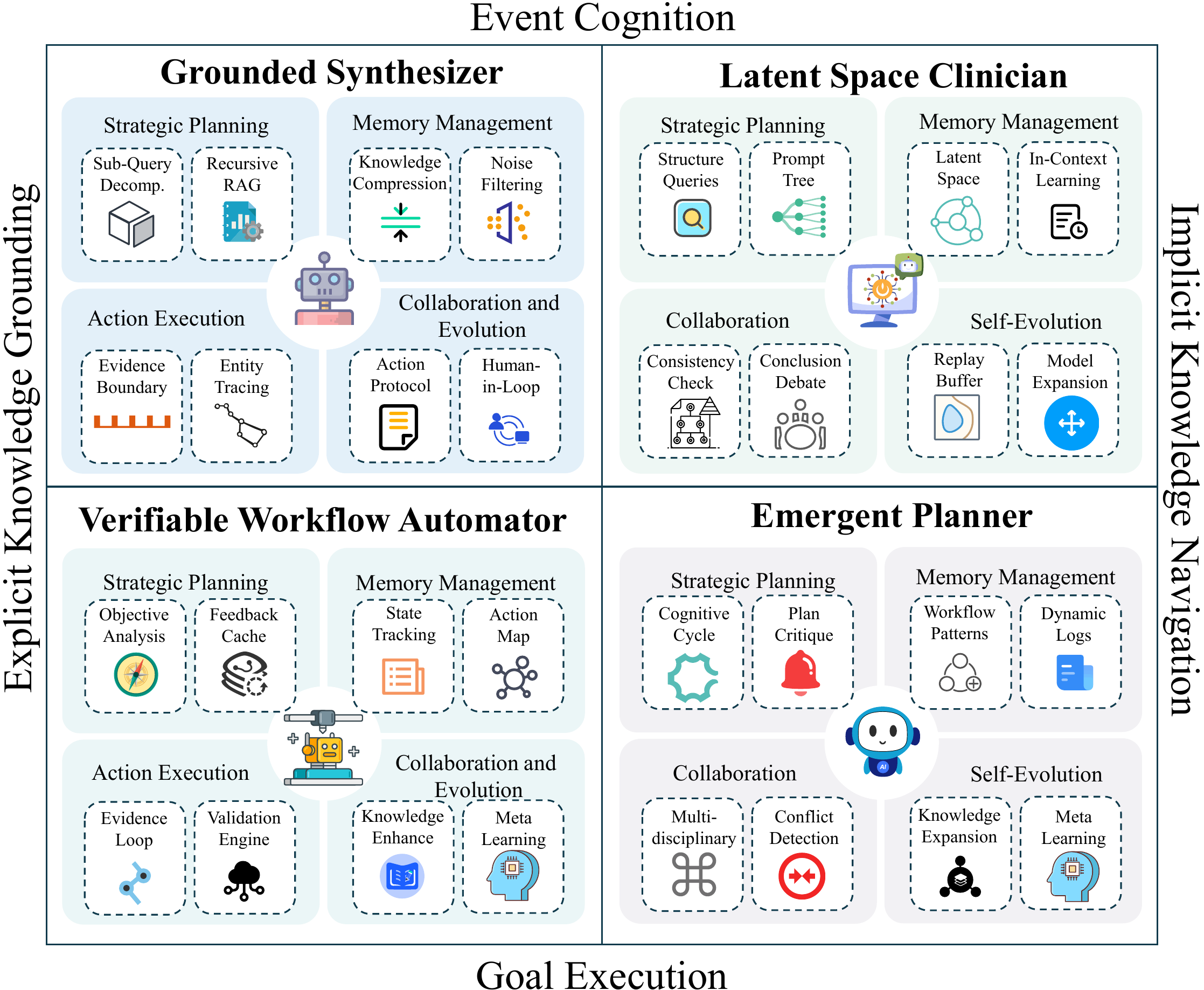}
    \caption{Taxonomy of agentic clinical dialogue in our survey. Structured along the orthogonal axes of knowledge source and agency objective, this framework reveals that the design of these agents is governed by two trade-offs: the balance between generative creativity and factual reliability (dictated by knowledge source), and the tension between operational autonomy and clinical safety (dictated by agency objective). }
    \label{fig:cat}
\end{figure}

\begin{itemize}
    \item \textbf{Latent Space Clinician (LSC)}. These agents leverage the LLM's vast internal knowledge for creative synthesis and forming a coherent understanding of a clinical situation. Their philosophy is to trust the model's emergent reasoning capabilities to function like an experienced clinical assistant, providing insights. For instance, the zero/few-shot reasoning capabilities of Med-PaLM \cite{MEDPALM225} or MedAgents \cite{MEDAGENT24} exemplify this paradigm.
   
    \item  \textbf{Grounded Synthesizer (GS)}. These agents operate under the principle that LLMs should function as powerful natural language interfaces to reliable external information rather than as knowledge creators. Their primary role is to retrieve, integrate, and accurately summarize information from verifiable sources such as medical databases or imaging data. Exemplars include the foundational frameworks for medical retrieval and indexing techniques such as Med-RAG \cite{MEDRAG25} and MA-COIR \cite{MACOIR25}.
    
    \item \textbf{Emergent Planner (EP)}. This paradigm grants the LLM a high degree of autonomy, allowing it to dynamically devise its own multi-step plan to achieve a complex clinical goal. The agent's behavior is emergent, as it independently determines the necessary steps and goals. Frameworks like AgentMD, which use ReAct-style prompting \cite{REACT23, AGENMD24}.
    
    \item \textbf{Verifiable Workflow Automator (VWA)}. In this paradigm, agent autonomy is strictly constrained within pre-defined, verifiable clinical workflows or decision trees. The LLM acts as a natural language front-end to a structured process, executing tasks rather than making open-ended decisions, which ensures maximum safety and predictability. This approach is exemplified by commercial triage bots, the structured conversational framework of systems like Google's AMIE\cite{AIME25}, and principles from classic task-oriented dialogue systems such as MeDi-TODER\cite{MEDTOD25}.
\end{itemize}

\begin{figure*}[!htp]
    \centering
    \setlength{\abovecaptionskip}{0cm}   
    \setlength{\belowcaptionskip}{0cm}   
    \resizebox{0.85\textwidth}{!}{\begin{forest}
            forked edges,
            for tree={
                grow=east,
                reversed=true,
                anchor=base west,
                parent anchor=east,
                child anchor=west,
                base=left,
                font=\small,
                rectangle,
                draw=black,
                rounded corners,
                align=center,
                minimum width=4em,
                edge+={darkgray, line width=1pt},
                s sep=3pt,
                inner xsep=2pt,
                inner ysep=3pt,
                ver/.style={rotate=90, child anchor=north, parent anchor=south, anchor=center},
            },
            where level=1{text width=10.4em,font=\scriptsize,}{},
            where level=2{text width=8.5em,font=\scriptsize,}{},
            where level=3{text width=19em,font=\scriptsize,}{},
            [
                Medical Dialog System, ver, color=black, fill=diagblue!15, text=black
                [
                    Latent Space Clinician (\S \ref{subsec:LSC}), color=black, fill=LSCgreen!60, text=black,
                    [                    
                        Strategic Planning (\S \ref{subsubsec:LSCSP}), color=black, fill=LSCgreen!70,  text=black
                        [
                        {MedFound\cite{AGEN25}, Casual-Base Alignment\cite{LearnC25}, Foresight2\cite{FORES24}
}, color=black, fill=customlightgray!70,  text=black
                        ]
                    ]    
                    [                    
                        Memory Management (\S \ref{subsubsec:LSCMM}), color=black, fill=LSCgreen!70,  text=black
                        [
                        {$A^3$Tune\cite{FOCUSO25}, FAIIR\cite{FAIIR25}, Recursive Summary\cite{RECUR25}, Patrika\cite{AIENA25}
}, color=black, fill=customlightgray!70,  text=black
                        ]
                    ]   
                    [                    
                        Collaboration (\S \ref{subsubsec:LSCCL}), color=black, fill=LSCgreen!70,  text=black
                        [
                        {MAM\cite{MAM25}, ReConcile\cite{RECONC24}, MEDCO\cite{MEDCO24}, ColaCare\cite{COLAC25}
}, color=black, fill=customlightgray!70,  text=black
                        ]
                    ]   
                    [                    
                        Self-evolution (\S \ref{subsubsec:LSCEV}), color=black, fill=LSCgreen!70,  text=black
                        [
                        {AlphaEvolve\cite{ALPHAEV25}, STLLaVA-Med\cite{EVOVQA24}, Agent Hospital\cite{AGENHOS}}, color=black, fill=customlightgray!70,  text=black
                        ]
                    ]   
                ]
                [
                    Emergent Planner (\S \ref{subsec:EP}),
                    color=black, fill=EPpurple!60, text=black
                    [                    
                        Strategic Planning (\S \ref{subsubsec:EPSP}), color=black, fill=EPpurple!70,  text=black
                        [
                        {Self-Consistency\cite{SELFCONSIS}, Expectation-max\cite{MEDINF}, Claim Verification\cite{ADVB25}}, color=black, fill=customlightgray!70,  text=black
                        ]
                    ]    
                    [                    
                        Memory Management (\S \ref{subsubsec:EPMM}), color=black, fill=EPpurple!70,  text=black
                        [
                        {MOTOR\cite{MOTOR24}, SUFO\cite{DIAGNOSPAC}, Open-ended VQA\cite{PREFIXTUN}, Staf-LLM\cite{STAFD25}
}, color=black, fill=customlightgray!70,  text=black
                        ]
                    ]   
                    [                    
                        Collaboration (\S \ref{subsubsec:EPCL}), color=black, fill=EPpurple!70,  text=black
                        [
                        {MedLA\cite{medla}, MultiMedRes\cite{PROACT}, DynamiCare\cite{DYNAMI}, ToR\cite{TREEOFREA} 
}, color=black, fill=customlightgray!70,  text=black
                        ]
                    ]   
                    [                    
                        Self-evolution (\S \ref{subsubsec:EPEV}), color=black, fill=EPpurple!70,  text=black
                        [
                    {MedAgentSim\cite{SELEVOCL}, LLM Strategist\cite{DYNAMISIS}, MedPAO\cite{MEDPAO}, PPME\cite{IMPROVINTEACT}}, color=black, fill=customlightgray!70,  text=black
                        ]
                    ]   
                ]
                [
                Grounded Sythesizer (\S \ref{subsec:GS}), color=black, fill=GSblue!60, text=black
                    [                    
                        Strategic Planning (\S \ref{subsubsec:GSSP}), color=black, fill=GSblue!70,  text=black
                        [
                        {HyKGE\cite{HYKGE}, EvidenceMap\cite{EVIDMAP}, MHMKI\cite{MULTIHOPKNO}, MedicalGLM\cite{QEURYEVAL}}, color=black, fill=customlightgray!70,  text=black
                        ]
                    ]    
                    [                    
                        Memory Management (\S \ref{subsubsec:GSMM}), color=black, fill=GSblue!70,  text=black
                        [
                        {K-Comp\cite{KCOMP}, Rationale-Guided RAG\cite{RATIONALE}, Seek Inner\cite{SEEKINNER}, CoE\cite{CHAINOFE}
}, color=black, fill=customlightgray!70,  text=black
                        ]
                    ]   
                    [                    
                        Action Execution (\S \ref{subsubsec:GSAE}), color=black, fill=GSblue!70,  text=black
                        [
                        {MedGENIE\cite{TOGENORRET}, MediSearch\cite{MEDISEARCH}, MeNTi\cite{MENTTOOL}, BriefContext\cite{LEVARLC}
}, color=black, fill=customlightgray!70,  text=black
                        ]
                    ]   
                    [                    
                        Collaboration (\S \ref{subsubsec:GSCL}), color=black, fill=GSblue!70,  text=black
                        [
                        {MedOrch\cite{MEDIATOR}, MedConMA\cite{MEDCONMA}, Debating Framework\cite{ERRDET}}, color=black, fill=customlightgray!70,  text=black
                        ]
                    ] 
                    [                    
                        Self-evolution (\S \ref{subsubsec:GSEV}), color=black, fill=GSblue!70,  text=black
                        [
                        {AMG-RAG\cite{MedicationQA}, SFPS\cite{IMPROVESELFTRAI}, MACD\cite{MACDMULTI}, SFDA\cite{IMPROVESELFTRAI}}, color=black, fill=customlightgray!70,  text=black
                        ]
                    ]
                ]
                [
                    Verifiable Workflow Automator (\S \ref{subsec:VWA}), color=black, fill=VWAcyan!60, text=black
                    [                    
                        Strategic Planning (\S \ref{subsubsec:VWASP}), color=black, fill=VWAcyan!70,  text=black
                        [
                        {VITA\cite{VITACARE}, Quicker\cite{FROMTOBASE}, LLM-AMT\cite{AUGMENTINGBB}, MedPlan\cite{MEDPLANA}, RULE\cite{RULERELIABLEMODL}}, color=black, fill=customlightgray!70,  text=black
                        ]
                    ]    
                    [                    
                        Memory Management (\S \ref{subsubsec:VWAMM}), color=black, fill=VWAcyan!70,  text=black
                        [
                        {EMRs2CSP\cite{EMRS2CSP}, Medical Graph RAG\cite{MEDGRAGEVIB}, Deep-DxSearch\cite{E2EAGENTIC}
}, color=black, fill=customlightgray!70,  text=black
                        ]
                    ]   
                    [                    
                        Action Execution (\S \ref{subsubsec:VWAAE}), color=black, fill=VWAcyan!70,  text=black
                        [
                        {Cli-RAG\cite{CLIRAGRETRIV}, Tala-med\cite{EHANCINGMEDRET}, PPME\cite{IMPROVINTEACT}, CLI-RAG\cite{CLIRAGRETRIV}, Medex\cite{MEDEX}
}, color=black, fill=customlightgray!70,  text=black
                        ]
                    ]   
                    [                    
                        Collaboration (\S \ref{subsubsec:VWACL}), color=black, fill=VWAcyan!70,  text=black
                        [
                        {TeamMedAgents\cite{TEAMMEDAGENTSE}, TAMA\cite{TAMACOLL}, MAKAR\cite{ENHANCECLITRIAL}, ClinicalLab\cite{CLINICALLAB}
}, color=black, fill=customlightgray!70,  text=black
                        ]
                    ] 
                    [                    
                        Self-evolution (\S \ref{subsubsec:VWAEV}), color=black, fill=VWAcyan!70,  text=black
                        [
                        {HealthFlow\cite{HEALTHFLOW}, MetaAgent\cite{SELFEVOLVINH}, EvoAgentX\cite{EVOAGENTEVOL}, ZERA\cite{ZERAZEROINITINS}}, color=black, fill=customlightgray!70,  text=black
                        ]
                    ]
                ]                
        ]
        \end{forest}}
    \caption{The main content flow and categorization of this survey. We identify several key works from each paradigm.}
    \label{fig:catsurvey}
\end{figure*}
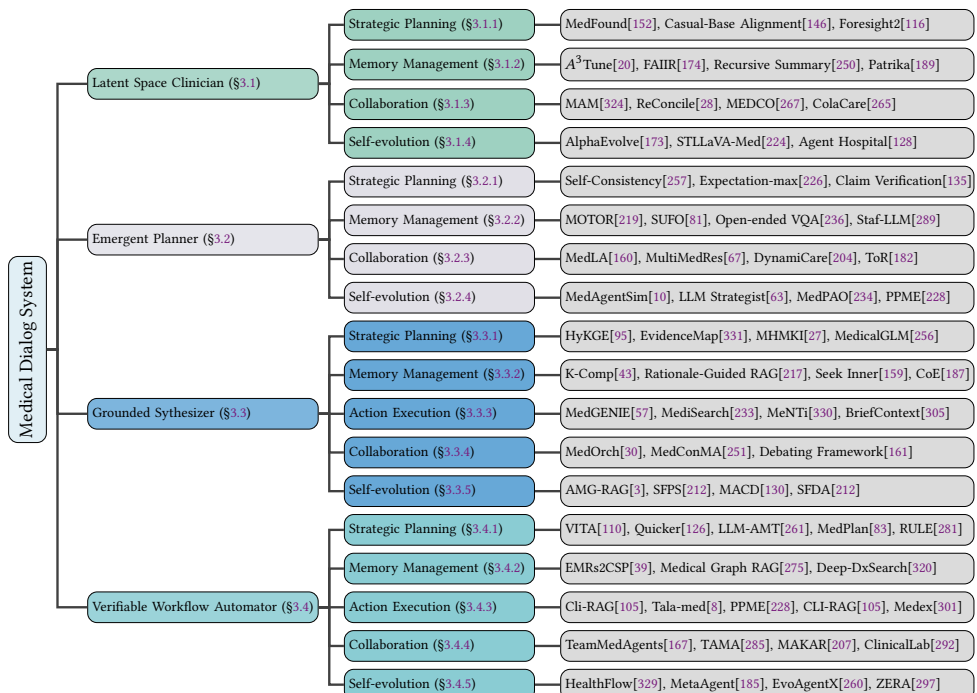

Following this taxonomy, the subsequent sections will dissect each of the four agent paradigms through the lens of five core technical components: strategic planning, memory management, action execution, collaboration, and evolution. Specific categories and exemplary papers and works are shown in Fig. \ref{fig:catsurvey}. Notably, our analysis of the Action Execution module is specific to agents on the Explicit Knowledge Grounding axis, as paradigms reliant on implicit knowledge, by definition, do not leverage external knowledge tools.

\section{Agentic Paradigm for Health Communication}\label{sec:agentic}
Following the proposed framework, this section provides a granular analysis of the four archetypal paradigms of clinical dialogue. We dissect the architectural design of each paradigm across five orthogonal dimensions, which include strategic planning, memory management, action execution, collaboration, and evolution.

\subsection{Latent Space Clinician}\label{subsec:LSC} 

The LSC paradigm represents the foundational approach in which agents primarily leverage the implicit medical knowledge encoded within the LLM's parametric memory to perform creative synthesis and clinical reasoning. Distinct from paradigms that rely on external retrieval, the LSC simulates the cognitive intuition of a physician by traversing latent representations to form a coherent understanding of clinical situations. As illustrated in Fig.~\ref{fig:LSC}, the architecture focuses on activating these internal reasoning chains through strategic prompting and memory iteration. A summary of representative works and their technical characteristics is presented in Table \ref{tab:LSCsum}.

\begin{table}[!htp]
\centering
\setlength{\abovecaptionskip}{0cm}   
\setlength{\belowcaptionskip}{0cm}   
\caption{The summarization of LSC agentic approach,  "PT" represents pre-training and fine-tuning, "PE" represents prompt engineering, "MAS" represents multi-agent system, "RL" represents reinforcement learning, and "ST" represents self-training.}
\label{tab:LSCsum}
\resizebox{0.8\textwidth}{!}{
\begin{tabular}{ccccc}
\hline
\textbf{\begin{tabular}[c]{@{}c@{}}Agentic\\ Approach\end{tabular}} & \textbf{\begin{tabular}[c]{@{}c@{}}Main Corresponding\\ Component\end{tabular}} & \textbf{\begin{tabular}[c]{@{}c@{}}Detailed\\ Method\end{tabular}} & \textbf{\begin{tabular}[c]{@{}c@{}}Downstream\\ Task\end{tabular}} & \textbf{\begin{tabular}[c]{@{}c@{}}Framework\\ Type\end{tabular}} \\ \hline
MedFound\cite{AGEN25} & Planning Strategies & Self-bootstrapping Strategy & Task-oriented & PE \\ \hline
Casual-Base Alignment\cite{LearnC25} & Planning Strategies & Casual Chain Decision Process & Medical QA & PE \\ \hline
Foresight2\cite{FORES24} & Planning Strategies & Contextual Fine-tuning & Task-oriented & PT \\ \hline
$A^3$Tune\cite{FOCUSO25} & Memory Management & Attention Alignment Tuning & Medical VQA & PT \\ \hline
FAIIR\cite{FAIIR25} & Memory Management & Domain-adapted & Supportive & PT \\ \hline
Recursive Summary\cite{RECUR25} & Memory Management & Memory Iteration & Medical QA & PE \\ \hline
MAM\cite{MAM25} & Collaboration & Role-specilization & Task-oriented & MAS \\ \hline
ReConcile\cite{RECONC24} & Collaboration & Multi-round Discussion & Medical QA & MAS \\ \hline
AlphaEvolve\cite{ALPHAEV25} & Self-evolution & Algorithm Improvement & Task-oriented & ST \\ \hline
STLLaVA-Med\cite{EVOVQA24} & Self-evolution & Instruction Generation & Medical VQA & ST \\ \hline
\end{tabular}}
\end{table}

\begin{figure}[!htp]
    \centering
    \setlength{\abovecaptionskip}{0cm}   
    \setlength{\belowcaptionskip}{0cm}   
    \includegraphics[width=0.8\linewidth]{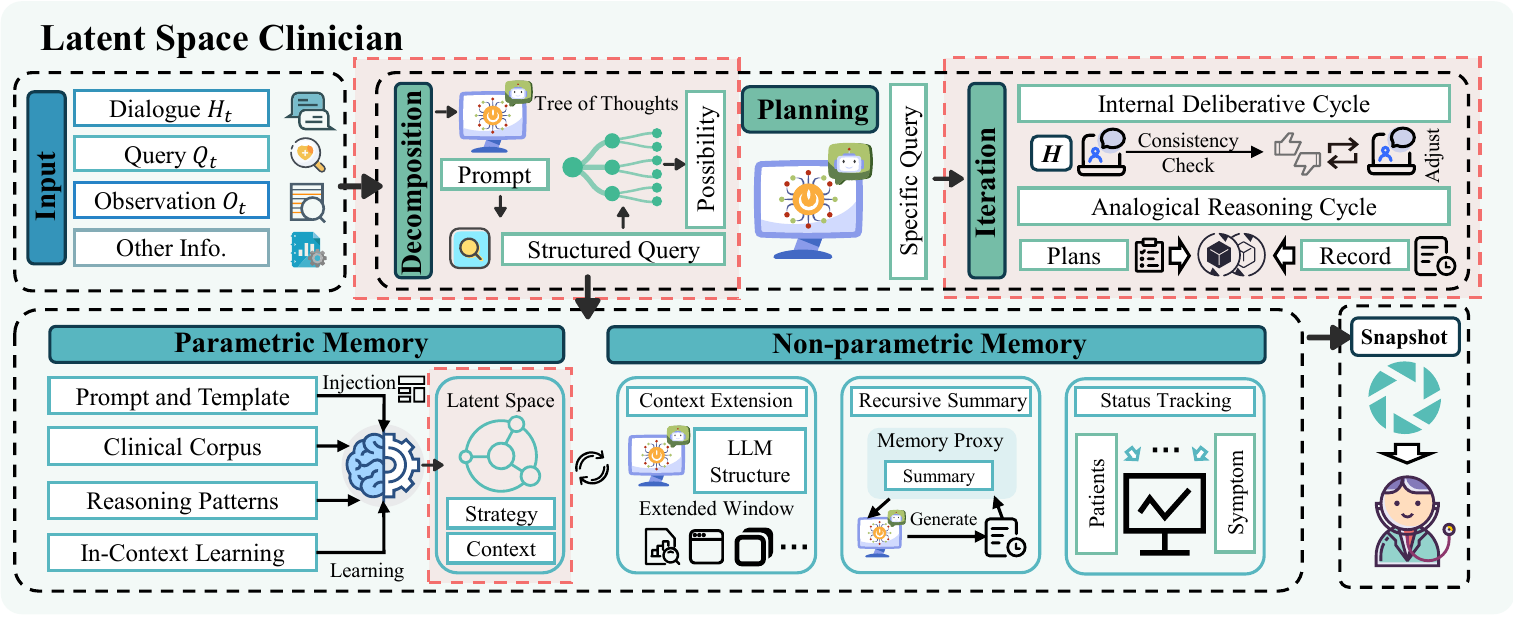}
    \caption{General structure of the single agent in the LSC paradigm. Decomposition breaks down queries into structured reasoning chains entirely within the latent space. Iteration functions as an internal deliberative cycle, employing self-correction mechanisms like consistency checks to refine hypotheses without external tools. The agent utilizes parametric memory as an internalized medical curriculum for reasoning, while non-parametric memory employs context extension techniques to maintain coherence. Core differences from other paradigms are covered with red blocks.
}
    \label{fig:LSC}
\end{figure}

\subsubsection{Strategic Planning}\label{subsubsec:LSCSP} 

Strategic planning of the LSC is the process of architecting an internal cognitive workflow to construct a comprehensive and accurate clinical snapshot through a structured path of inquiry and reasoning executed within the model's latent space.  

\noindent\textbf{Decomposition strategies}. Decomposition for the LSC functions as the architect of an internal cognitive workflow, translating a clinical query into a structured path of inquiry entirely within the model's latent space. The evolution of this strategy reveals a clear trajectory from implicit semantic association to the explicit modeling of human-like clinical reasoning chains, showcasing the development of a more deliberate and inspectable clinical logic.

An underlying premise is that effective decomposition depends on a rich, structured internal knowledge base. Early generative task reframing demonstrated that models such as BioGPT \cite{BioGPT22} and BioBART \cite{BioBART22} could perform ordered analysis based primarily on probabilistic associations. However, such implicit decomposition is clinically meaningful only when grounded in deep domain expertise, requiring specialized models like ClinicalBERT \cite{CliBERT22} and BioMegatron \cite{BioMega20} to robustly encode clinical semantics. To transcend the limitations of implicit associations, contemporary research has adapted generic chain-of-thought methodologies into explicit clinical reasoning chains. Unlike general-purpose reasoning, this approach structures the latent space traversal to mirror the review of systems utilized by physicians. Med-PaLM \cite{MEDPALM225} demonstrates that instruction fine-tuning serves as a soft decomposition mechanism, compelling the model to externalize its latent reasoning into a structured explanation before concluding. Moving beyond simple step-by-step generation, frameworks such as the chain of diagnosis \cite{COD25} and HuatuoGPT-II \cite{HUATUO23} impose a strict medical ontology on the decomposition process. By constraining the thought process to follow a specific clinical sequence, ranging from symptoms and history to etiology and diagnosis, this transformation ensures that decomposition is not merely a linguistic artifact but a deliberate simulation of diagnostic logic, rendering the LSC's intuition transparent and inspectable.

Representing the frontier of this paradigm, advanced decomposition strategies capture the distributed and causal nature of expert cognition, reflecting a shift toward simulated multi-disciplinary collaboration. In systems like MedAgents \cite{MEDAGENT24}, a central coordinator breaks down a complex clinical case into distinct medical perspectives, assigning sub-questions to specialized "expert" roles (e.g., a pathologist or cardiologist) instantiated within the latent space. This trend culminates in Causal Alignment, where problems are decomposed based on domain-specific causal mechanisms rather than heuristics. When guided to follow human-like causal reasoning, the LSC's internal plan mirrors the nuances of clinical judgment \cite{LearnC25}. This granular cognitive modeling is architecturally supported by findings in knowledge editing, which indicate that factual medical knowledge is localized in modular LLM components, thereby enabling step-wise, targeted analysis without the need for external tools \cite{KE23}.

\noindent\textbf{Iteration strategies}. Reasoning within the LSC operates as an internal deliberative cycle, distinct from the physical trial-and-error observed in agentic planning \cite{REASONLLM24, EVALLLM24}. This mechanism simulates the cognitive process of a clinician who pauses to refine a hypothesis before speaking. Rather than seeking new external data, the cycle re-processes existing latent information to produce a reliable clinical snapshot, effectively turning the generation process into a rigorous differential diagnosis loop \cite{EMPOWER24}. A prerequisite for this internal iteration is the capability to map high-dimensional patient data into a structured cognitive space. Leveraging deep learning architectures such as FNAE \cite{AFAST23}, REGLE \cite{MULTIV24}, and Autosurv \cite{AUTOSURV24}, the LSC compresses complex, multi-modal clinical signals into a low-dimensional latent space. In a medical context, this represents the construction of a clinical manifold, where patient states are dynamically adjusted through iterative clustering and prototyping. Within this manifold, the agent can simulate potential disease progressions or treatment outcomes by analyzing vector trajectories, projecting the patient's future state to validate current decisions without requiring real-world intervention.

To enhance diagnostic accuracy, the LSC employs multi-round reasoning that mirrors clinical self-correction. Instead of generating a conclusion in a single forward pass, the agent performs internal cycles of hypothesis generation and validation. Frameworks like Qilin-Med \cite{QILIN24} and counterfactual reasoning \cite{Count25} operationalize this by generating intermediate clinical questions or alternative scenarios, such as assessing whether a diagnosis changes in the absence of a specific symptom, to challenge and refine the initial hypothesis. Furthermore, predictive models like Foresight \cite{FORES24} enforce logical consistency by updating the internal representation of the patient's condition after each dialogue turn. This ensures that the agent's evolving understanding remains coherent with the longitudinal patient history, significantly reducing contradictions in both open-ended and task-oriented clinical conversations.

At a higher cognitive level, iteration strategies empower the LSC to emulate the experience-based intuition of human physicians through emergent analogical reasoning. Research indicates that by iteratively referencing known cases encoded within its parameters, the LSC can interpret novel clinical presentations by analogy \cite{ONTO25}. This process transcends monolithic prediction, executing instead as a series of incremental, stepwise inferences \cite{AGEN25, TAIYI23}. By iteratively mapping the current patient's symptoms against implicit case prototypes learned during training, the agent constructs a diagnostic pathway that is closer to expert clinical judgment than to simple pattern matching.

\subsubsection{Memory Management}\label{subsubsec:LSCMM}

The LSC's memory can be conceptualized along a spectrum defined by two primary modalities: the static, foundational knowledge encoded within the model's parameters and the dynamic, context-specific information managed during a single interactive session.

\noindent\textbf{Parametric memory}. Parametric memory constitutes the cognitive foundation of the LSC, functioning effectively as an internalized medical curriculum. Unlike GSs (please refer to Section~\ref{subsec:GS}) that strictly rely on external databases for verification \cite{ATTRIP25}, LSC agents encode vast repositories of biomedical literature, clinical guidelines, and reasoning patterns directly into the model's weights during large-scale pre-training and fine-tuning \cite{FOCUSO25}. This implicit encoding enables agents to perform complex inference and creative diagnostic reasoning solely through internal computation, establishing the model's parameters as the primary source of clinical truth within this paradigm.

The primary mechanism for leveraging this internalized knowledge is dynamic activation during inference, most effectively achieved via sophisticated prompt engineering. Methodologies such as in-context learning \cite{ICL22} function as cognitive scaffolds that guide the model to activate specific parametric pathways relevant to a clinical query. Foundational research demonstrates that instruction fine-tuning \cite{HUATUO225} and contextual prompting \cite{DIAS24} do not merely retrieve facts but restructure the model's latent space to align with clinical tasks, thereby enhancing zero-shot summarization and reasoning capabilities \cite{ATTRIP25}. Further elucidating this mechanism, studies analyzing the activation of parametric memory during in-context learning \cite{CONTEXTB23, MEDVH24} reveal that the transformer architecture performs an implicit model fitting, adapting its pre-trained medical representations to the specific context of the current interaction.

Beyond static fact recall, parametric memory underpins the agent's capacity for dynamic semantic modeling. This capability is critical for constructing an accurate clinical snapshot, as it allows the LSC to maintain a coherent understanding of an evolving patient narrative and the intricate relationships between clinical entities. Furthermore, this memory facilitates a remarkable degree of generalization and creative inference. By drawing on deep, abstract patterns learned during pre-training \cite{FAIIR25}, the LSC can reason about unseen clinical cases or rare symptom presentations \cite{HUATUO225}, a process that simulates the intuition of a seasoned physician confronting a novel diagnostic challenge. This ability to generalize beyond explicit training data represents a distinct advantage of leveraging high-dimensional representations within parametric memory; however, it necessitates careful management to align with the rigorous factuality required in healthcare.

\noindent\textbf{Non-parametric memory}. While parametric memory serves as the medical curriculum, non-parametric memory in the LSC paradigm functions as the patient context window. Its primary objective is not to verify facts against an external database but to maintain informational continuity across the patient's narrative. This creates a robust working memory that prevents the agent from suffering from catastrophic forgetting during long, multi-turn clinical encounters \cite{AGENTH3}, ensuring that the generated advice remains consistent with the specific patient's history.

A primary challenge in processing clinical narratives is the sheer length of longitudinal patient records, which often exceeds standard model constraints. To address this, architectural innovations have been specifically adapted for the medical domain to expand contextual capacity. Long-context frameworks, such as Clinical ModernBERT \cite{CLIMODBERT25} and DK-BEHRT \cite{DKBERT25}, represent a shift from processing isolated sentences to modeling entire chronological care trajectories. By incorporating sparse or block-wise attention mechanisms tailored for sequential data \cite{TRANSHEALTH24}, these models extend their effective receptive field to tens of thousands of tokens. This enhancement is crucial for capturing long-range dependencies in EHRs, ensuring that critical diagnostic clues mentioned early in a patient's timeline are retained and integrated into current decision-making \cite{CONCLU24}.

To manage the cognitive load within a finite context window, LSC agents employ recursive summarization \cite{RECUR25} \cite{ADAP24} and memory compression techniques \cite{BIOLORD24}. Unlike generic text compression, these methods, in a clinical setting, act as filters for clinical salience. The model periodically distills the dialogue history into compact vectors or structured summaries, effectively creating a memory proxy analogous to a physician's handover note \cite{TOW25}. This process frees up contextual space for new symptoms while preserving the essential trajectory of the interaction. By maintaining this cumulative understanding, the agent ensures that its reasoning is based on a comprehensive view of the patient's status rather than on disjointed fragments.

At a more granular level, non-parametric memory facilitates the dynamic tracking of clinical entities and their evolving states \cite{FORES24}. During a dialogue, the LSC continuously updates its internal representation of critical variables, including symptom severity, medication adherence, and test results, based on new user inputs \cite{AIENA25}. This mechanism transforms static text into a dynamic clinical snapshot, allowing the agent to recognize when a patient's condition has changed and to adjust its internal reasoning accordingly. This real-time state tracking is the operational bridge that connects the agent's vast medical knowledge to the specific, evolving needs of the individual patient.

\subsubsection{Collaboration}\label{subsubsec:LSCCL}

The collaborative architecture of the LSC paradigm is designed to transcend the cognitive limitations of individual models by emulating the consultation dynamics found in clinical practice. While early implementations focused on operational simplicity, the field is progressively shifting towards architectures that simulate professional medical interaction \cite{COLAC25}, aiming to reduce the opacity of latent reasoning through structured communication.

\noindent\textbf{Single-agent system}. A single-agent system functions analogously to a general practitioner, utilizing a solitary LLM to independently process and generate the entirety of a clinical dialogue \cite{HUATUO225}. The primary advantage of this approach lies in its computational efficiency and architectural coherence, making it highly effective for routine, well-defined tasks where the reasoning path is linear and contained within general medical knowledge. However, the reliance on a single latent space reveals critical vulnerabilities in complex clinical scenarios: a solitary model often exhibits cognitive tunneling, adhering to an initially incorrect diagnosis and lacking the specialized depth required for interdisciplinary cases. This limitation necessitates a structural evolution from individual processing to collective intelligence \cite{MEDAGENT24}, mirroring the referral system in healthcare, where complex cases are escalated to a team of specialists.

\begin{figure}[!htp]
    \centering
    \setlength{\abovecaptionskip}{0cm}   
    \setlength{\belowcaptionskip}{0cm}   
    \includegraphics[width=0.8\linewidth]{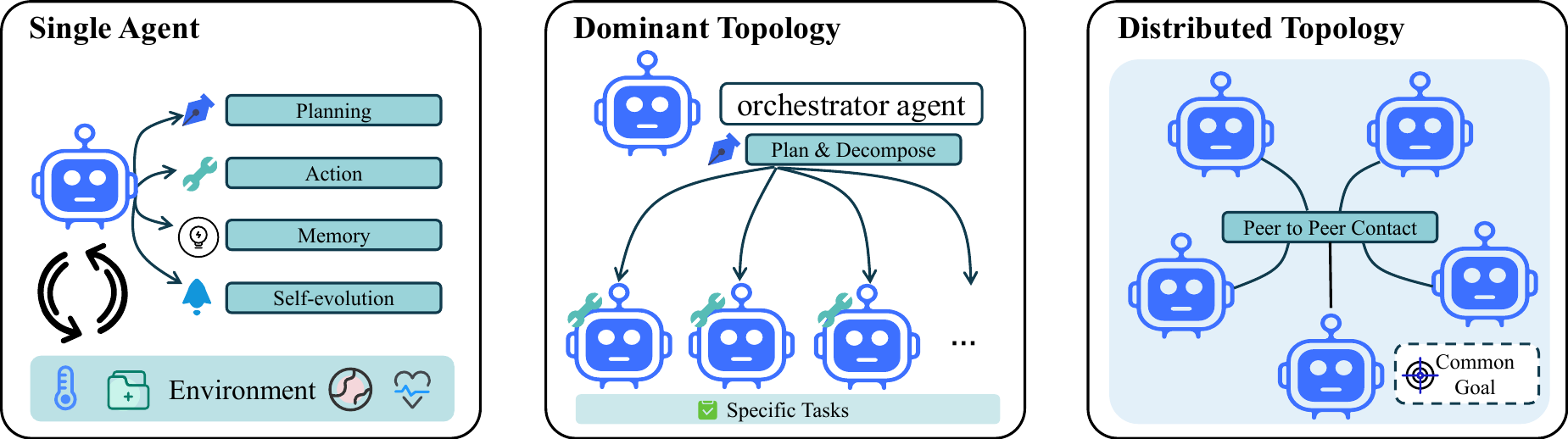}
    \caption{Multiple topologies of agent collaboration. Single-agent system (left) is a solitary LLM that handles the entire cognitive loop. Dominant topology (middle) mimics a clinical ward led by a central orchestrator agent. Distributed topology (right) simulates a peer-to-peer medical board to achieve a consensus on the common clinical goal.}
    \label{fig:topo}
\end{figure}

\noindent\textbf{Multi-agent system}. Multi-agent systems (MAS) introduce a collaborative paradigm that simulates the MDT approach inherent in modern medicine \cite{MEDCO24}. In this model, individual agents are instantiated with specialized roles—such as a clinician, pathologist, or patient advocate—each accessing different subsets of implicit knowledge or employing distinct reasoning strategies to solve a clinical task. By introducing diverse perspectives, MAS effectively operationalizes clinical peer review within the latent space, mitigating hallucinations common in single-model reasoning. As illustrated in Fig. \ref{fig:topo}, the effectiveness of this simulation is largely defined by the underlying collaboration topology.

In a dominant topology, the system mimics a hierarchical clinical ward led by a chief physician. A dominant orchestrator agent is responsible for decomposing the clinical problem, assigning sub-tasks to subordinate agents, and synthesizing their outputs into a final decision. Frameworks such as ColaCare \cite{COLAC25} operate within this structure, where a primary reasoning agent moderates findings from specialized modules to ensure high consistency and streamlined decision-making. This structure ensures that all information flows through a single adjudicating entity, maintaining strict control over the diagnostic narrative.

Conversely, distributed multi-agent systems emulate a peer-to-peer clinical case conference, where agents collaborate without a central controller to achieve consensus. In this topology, agents share information, debate hypotheses, and negotiate conclusions directly with one another. Systems like ReConcile \cite{RECONC24} and MAM \cite{MAM25} utilize this structure to foster greater interpretive depth, allowing for the organic aggregation of diverse perspectives. This paradigm is particularly well-suited for ambiguous clinical scenarios requiring multi-faceted analysis. Frameworks such as MDagents \cite{MDAGENT24} and MedAgent \cite{MEDAGENT24} have demonstrated that leveraging such peer-to-peer dynamics significantly enhances accuracy in medical exam tasks. Furthermore, the information flow in these systems can be structured through mechanisms like directed acyclic graphs \cite{MDAG25}, ensuring that the collaborative debate remains focused and converges efficiently on a robust clinical conclusion.

\subsubsection{Evolution}\label{subsubsec:LSCEV}
In the dynamic landscape of healthcare, where treatment protocols and patient demographics are in constant flux, self-evolution serves as the mechanism for transforming the LSC from a static repository of pre-trained facts into an adaptive learning entity. This process mirrors the professional development of a clinician who must continuously integrate new medical evidence without discarding foundational medical training \cite{DARWIN25}. The primary objective within this paradigm is to update the agent's parametric memory to maintain relevance and accuracy over time, effectively turning static model weights into a growing clinical experience base \cite{ALPHAEV25}.

The core engine driving this evolution is continual learning \cite{CONTINU26}, which addresses the critical challenge of catastrophic forgetting, a phenomenon where adapting to new tasks, such as a revised diabetes guideline, inadvertently erases previously learned knowledge. To mitigate this, recent research has adapted regularization techniques to the medical domain, introducing penalty terms that constrain the model to preserve established medical facts while accommodating new data \cite{ALPHAEV25}. Furthermore, systems like Agent Hospital \cite{AGENHOS} utilize experience replay buffers, which function analogously to a physician reviewing past case logs. By periodically re-training on a subset of historically successful diagnoses, the agent reinforces its long-term clinical memory. More advanced approaches employ model expansion, dynamically allocating new neural capacity to learn emerging diseases. This effectively isolates novel medical knowledge from the established parameter space, ensuring that the acquisition of new specialty skills does not interfere with the agent's generalist capabilities \cite{EVOVQA24}.

Beyond the passive absorption of new data, intrinsic motivation frameworks endow the LSC with a form of active clinical inquiry. Inspired by the learning trajectory of medical residents, these mechanisms enable agents to autonomously explore their environment and refine latent representations by pursuing tasks that maximize information gain \cite{AGENHOS}. Rather than random exploration, agents engage in self-directed curriculum learning, progressively advancing from routine common cold cases to complex multi-system pathologies. By selecting tasks and exploration strategies independently, the LSC shapes its own learning path, leading to robust and generalizable clinical reasoning that is less dependent on supervised fine-tuning and more reflective of the autonomous growth seen in human medical experts.


\subsubsection{Summary}
\textbf{The LSC represents the foundational agentic paradigm situated at the intersection of implicit knowledge navigation and event cognition.} By treating the LLM's parametric memory as an internalized medical curriculum, these agents excel at synthesizing fragmented symptoms into coherent diagnostic hypotheses \cite{LearnC25, MEDPALM225}, effectively emulating the zero-shot reasoning and pattern recognition capabilities of experienced physicians in novel or ambiguous cases \cite{CLISUM25}. However, this reliance on latent intuition introduces inherent risks regarding factual hallucinations, knowledge staleness, and the opacity of the reasoning process, which challenge the rigorous auditability required in healthcare. Consequently, the primary research frontier lies in bridging the gap between the model's powerful semantic intuition and clinical safety demands, specifically focusing on improving the interpretability of latent states and mitigating catastrophic forgetting during the continuous acquisition of new medical knowledge. Furthermore, as clinical care demands moving beyond passive case interpretation to active intervention, we next explore the EP paradigm, which harnesses this implicit intuition to autonomously orchestrate dynamic clinical workflows.

\subsection{Emergent Planner}\label{subsec:EP}

The EP paradigm grants the LLM a higher degree of autonomy, enabling it to dynamically devise and execute multi-step plans to achieve complex clinical goals based on internal procedural intuition. Unlike the LSC, which focuses on observation, EP agents operate as active digital collaborators that autonomously shape the workflow trajectory through self-directed planning and execution \cite{TOWAR25, DiaF22}. As illustrated in Fig.~\ref{fig:EP}, this paradigm relies on the model's emergent abilities to navigate open-ended clinical tasks. Key contributions to this autonomous framework are summarized in Table~\ref{tab:EPsum}.

\begin{table}[!htp]
\centering
\setlength{\abovecaptionskip}{0cm}   
\setlength{\belowcaptionskip}{-0.1cm}   
\caption{The summarization of EP agentic approach,  "PT" represents pre-training and fine-tuning, "PE" represents prompt engineering, "MAS" represents multi-agent system, "RL" represents reinforcement learning, and "ST" represents self-training.}
\label{tab:EPsum}
\resizebox{0.8\textwidth}{!}{
\begin{tabular}{ccccc}
\hline
\textbf{\begin{tabular}[c]{@{}c@{}}Agentic\\ Approach\end{tabular}} & \textbf{\begin{tabular}[c]{@{}c@{}}Main Corresponding\\ Component\end{tabular}} & \textbf{\begin{tabular}[c]{@{}c@{}}Detailed\\ Method\end{tabular}} & \textbf{\begin{tabular}[c]{@{}c@{}}Downstream\\ Task\end{tabular}} & \textbf{\begin{tabular}[c]{@{}c@{}}Framework\\ Type\end{tabular}} \\ \hline
Self-Consistency\cite{SELFCONSIS} & Planning Strategies & Reasoning Path Selection & Medical QA & PE \\ \hline
Expectation-maximization\cite{MEDINF} & Planning Strategies & Knowledge Extraction & Medical QA & PE \\ \hline
Claim Verification\cite{ADVB25} & Planning Strategies & Evidence Analysis & Task-oriented & PE \\ \hline
MOTOR\cite{MOTOR24} & Memory Management & Transfer Learning & Task-oriented & PT \\ \hline
SUFO\cite{DIAGNOSPAC} & Memory Management & Interpreting Feature Space & Task-oriented & PT \\ \hline
Open-ended VQA\cite{PREFIXTUN} & Memory Management & Prefix-tuning & Medical VQA & PT \\ \hline
MedLA\cite{medla} & Collaboration & Reasoning Organization & Medical QA & MAS \\ \hline
MultiMedRes\cite{PROACT} & Collaboration & Learner Agent & Medical QA & MAS \\ \hline
MedAgentSim\cite{SELEVOCL} & Self-evolution & Knowledge Expansion & Task-oriented & PE \\ \hline
LLM Strategist\cite{DYNAMISIS} & Self-evolution & Meta Learning & Task-oriented & ST \\ \hline
\end{tabular}}
\end{table}

\begin{figure}[!htp]
    \centering
    \setlength{\abovecaptionskip}{0cm}   
    \setlength{\belowcaptionskip}{0cm}   
    \includegraphics[width=0.8\linewidth]{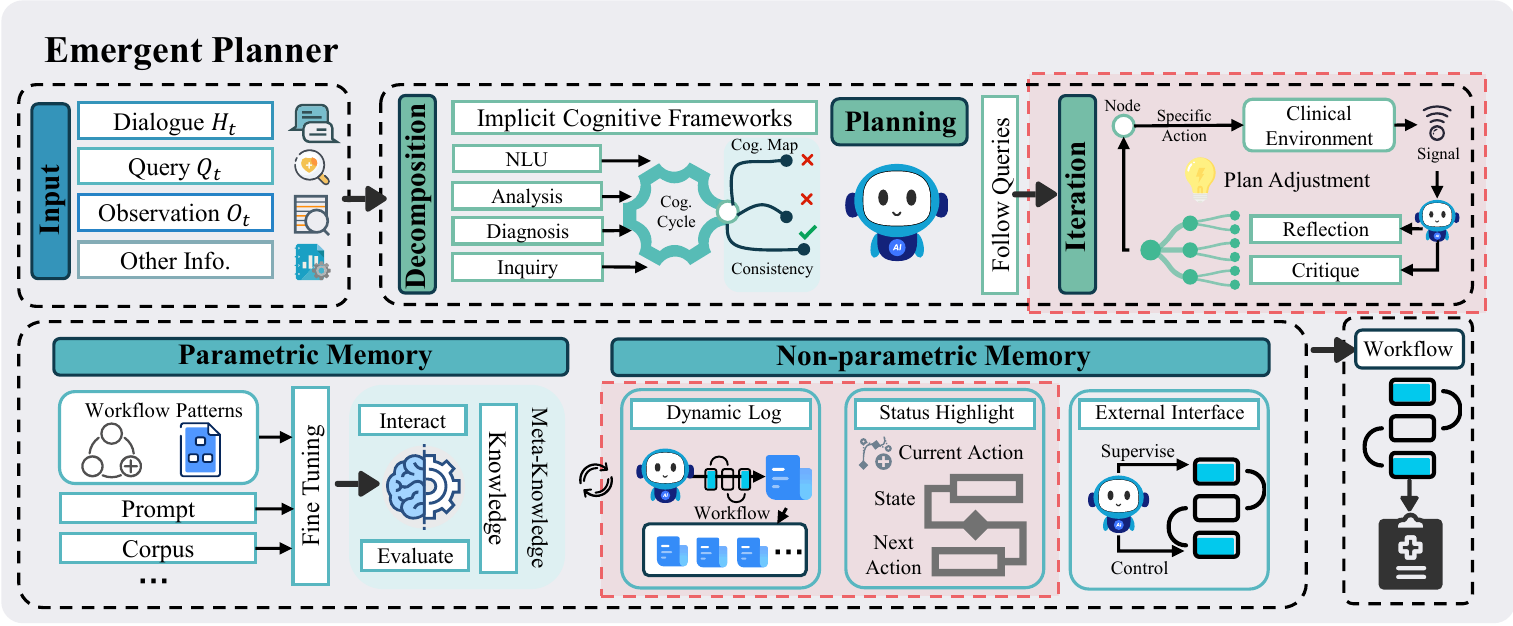}
    \caption{General structure of single agent in the EP paradigm. Decomposition translates clinical goals into sub-tasks and involves generating cognitive maps that anticipate future states. Iteration is driven by self-reflection and real-time environmental feedback to dynamically adjust plans. Parametric memory provides the procedural intuition and strategic knowledge, while non-parametric memory serves as a log or highlight to track the trajectory of the autonomous workflow execution and ensure continuity. Core differences from other paradigms are covered with red blocks.}
    \label{fig:EP}
\end{figure}

\subsubsection{Strategic Planning}\label{subsubsec:EPSP}

For EPs, strategic planning constitutes the core of their intelligence. Unlike LSC agents that primarily interpret existing information, EPs must navigate complex clinical narratives through internal deliberation and autonomous planning \cite{ZHONGJ, ADVB25}. When confronted with a high-level clinical goal, the agent is required to break it down into a sequence of logical sub-goals. This process serves as cognitive self-guidance, systematically shaping the trajectory and efficiency of the agent's actions \cite{EQUITY25}.

\noindent\textbf{Decomposition strategies}. Clinical reasoning involves navigating a landscape of uncertainty where a single symptom may point to multiple etiologies, necessitating a process akin to differential diagnosis \cite{MEDINF}. To emulate this rigorous medical logic, EPs employ the self-consistency strategy \cite{SELFCONSIS, SELFCONS}. Instead of relying on a single linear inference, the agent generates a diversity of potential reasoning paths—parallel to a physician considering various diagnostic hypotheses—and aggregates the outcomes via a voting mechanism \cite{CONSVOTE}. This approach effectively mitigates the risk of idiosyncratic errors in generation, ensuring that the final clinical conclusion represents the most robust consensus derived from the model's internal deliberation.

Furthermore, real-world clinical decision-making is rarely a straight line but rather a branching tree of possibilities, where each action influences subsequent choices. To align with this reality, EPs integrate divergent lines of thought into a structured decision blueprint \cite{MEDPROMPT}. Advanced frameworks like TREE-PLANNER \cite{TREEP24} enable the agent to construct a cognitive map prior to execution. By elevating decomposition from a deterministic sequence to the comprehensive exploration of a probabilistic action space \cite{DIAS24}, EPs can mentally simulate future clinical states—such as predicting a patient's response to different treatments—before committing to a plan. This mechanism draws inspiration from hierarchical planning methods like Least-to-Most Prompting \cite{LTMP23}, yet it is specifically adapted to navigate the prunable, tree-like structure of clinical pathways.

In addition to structural navigation, the decomposition process of EPs must adhere to established cognitive schemas used by medical experts. For instance, the verification of a medical claim typically follows a specific cognitive cycle: understanding the clinical puzzle, analyzing diagnostic goals, forming an interim verdict, and critically questioning that diagnosis. Research suggests that EPs can internalize such expert-level frameworks \cite{PREVI}, using them as an implicit scaffold to decompose complex tasks. While some studies propose explicit multi-step prompting to guide this process for claim verification \cite{ADVB25}, a sufficiently advanced EP implicitly models this workflow, ensuring that its autonomous decomposition remains clinically logical and methodologically sound.

Finally, effective planning relies on the agent's ability to model the intrinsic structure of medical data. EPs derive their planning capabilities from pre-training on vast biomedical corpora, where they learn the deep sequential patterns of disease progression and care delivery. Studies utilizing graph transformers demonstrate that capturing the relational structure among medical concepts in EHRs yields superior patient representations \cite{GRATR24}, which is a prerequisite for generating plausible plans. Similarly, breaking down lengthy medical dialogues into section-specific segments is essential for producing well-structured clinical notes \cite{CELM24}. These findings highlight that the decomposition strategy of an EP is effectively an outward expression of its deep, implicit understanding of underlying data logic.

\noindent\textbf{Iteration strategies}. While decomposition provides the initial roadmap, clinical management is inherently dynamic, requiring continuous adjustment based on evolving patient states. To address this, EPs employ iterative strategies that afford flexible planning capabilities \cite{TOWAR25}. This process is principally driven by self-reflection and critique. In a high-stakes clinical setting, generating a response without verification is hazardous; thus, the agent must not only execute actions but also scrutinize its own reasoning and conclusions \cite{BRAININS25}. This mechanism is structurally analogous to the clinical practice of differential diagnosis, where a physician proposes a hypothesis and immediately critiques it to rule out contraindications or rare pathologies \cite{SELFCRITIQUE22}. Such an internal adversarial process enables the agent to identify logical flaws or safety risks early, iteratively refining the diagnostic path before any advice is delivered to the patient. Empirical studies like Reflexion \cite{REFLEX23} demonstrate that equipping LLMs with this self-corrective capability significantly boosts performance in complex medical reasoning tasks.

Complementing this internal reflection is the feedback loop based on interaction history \cite{FRAME25}. EPs treat every conversational exchange not merely as text generation, but as an opportunity to acquire external feedback, update their internal world model, and dynamically adjust future plans \cite{FEEDB025}. Since these agents rely on implicit knowledge, extracting cues directly from the dialogue becomes the primary means of strategy refinement. For instance, if a patient expresses confusion or hesitation, the agent does not mechanically repeat prior information; instead, it iteratively modifies its communication policy to align with the patient's health literacy level. This continuous, dialogue-driven feedback loop ensures that the clinical communication process is highly adaptive and personalized, allowing the planning trajectory to emerge organically from the interaction rather than adhering to a rigid, preset script.

\subsubsection{Memory Management}\label{subsubsec:EPMM}
Memory management in an EP is fundamentally oriented toward goal execution, built upon a dual-system architecture in which both parametric and non-parametric memory are subordinated to executive function \cite{MOTOR24, ACONTP25}. Parametric memory acts as the deep, implicit foundation for strategic intuition and procedural knowledge \cite{SOULCHAT23, ZHONGJ}, while non-parametric memory serves as a transient workspace for tactical planning and real-time state tracking. Together, these systems enable the coherent generation of goal-directed actions within complex clinical workflows.

\noindent\textbf{Parametric memory}. The parametric memory of EP functions not merely serves as a static knowledge repository, but it also functions as the central engine enabling autonomous planning and multi-step action execution \cite{MOTOR24}. Unlike the LSC, which utilizes parametric memory primarily to construct a cognitive snapshot of a clinical scenario, the EP is intrinsically tasked with leveraging this internal storage to navigate and execute end-to-end clinical workflows \cite{WORKFLOWLLM}. Consequently, its parametric memory encodes not only declarative knowledge but, more crucially, embodies the procedural knowledge and strategic intuition that underlie effective clinical practice.

This procedural knowledge, embedded in the model's weights, is acquired during pre-training through exposure to vast clinical text corpora that implicitly encode workflow patterns \cite{WOKCLI}. In a clinical context, this process parallels the accumulation of experience by a resident physician. For example, the model learns not only the diagnostic criteria for myocardial infarction but also the sequential actions and underlying decision logic—spanning from evaluating a patient presenting with chest pain to ordering ECG and troponin tests, to initiating emergency interventions \cite{CLININFLOW}. This deep, implicit understanding of clinical processes enables the agent to autonomously generate coherent action sequences when provided with a high-level goal, functioning as a collaborator actively advancing clinical workflows \cite{WOKFLOWmed}.

During task execution, the management of parametric memory is realized through strategic activation and adaptive behavioral modification \cite{STAFD25}. When presented with a high-level clinical goal, the agent dynamically activates relevant procedural schemas within its parameter space rather than merely retrieving static factual knowledge \cite{CLININFLOW}. This mechanism enables a structured sequence of actions, such as assessing the patient's current level of understanding, explaining pathophysiological mechanisms, demonstrating the use of a glucose monitoring device, and collaboratively establishing initial dietary and physical activity goals \cite{CLININFLOW}. Each step is generated based on an implicit cognitive model of effective health education, ensuring a coherent progression toward the overarching clinical objective.

Correspondingly, updating this memory is less concerned with correcting factual inaccuracies and more focused on refining behavioral patterns. In this context, medical fine-tuning techniques \cite{TUNDIS} and prefix-tuning \cite{PREFIXTUN} serve as essential instruments for aligning the agent with evolving standards of care. For instance, fine-tuning the agent on a revised hypertension guideline that changes first-line medication recommendations directly alters its behavior within the initiate hypertension treatment workflow \cite{TUNDIS}. This process is analogous to reshaping a physician's clinical practice habits rather than merely updating an isolated fact in memory, thereby reflecting the goal-oriented nature of the EP \cite{ACONTP25}.

\noindent\textbf{Non-parametric memory}. If parametric memory constitutes the strategic intuition of an EP, then non-parametric memory functions as its essential tactical workspace \cite{DIAGNOSPAC}. Its primary role shifts from acting as an external source of static knowledge to serving as a dynamic and ephemeral information hub for tracking, planning, and reflecting upon the trajectory within a multi-step clinical workflow \cite{LIFECRAFT}. This specific utilization distinguishes the EP from the LSC, which employs non-parametric memory primarily to enrich context for a precise cognitive snapshot. For the EP, the value of memory is measured principally by its utility in informing and optimizing the subsequent clinical action.

Functionally, non-parametric memory operates as a dynamic log of the workflow state. For an agent designed to drive a clinical process forward, the complete dialogue history is more than a collection of information; it acts as an executive log that records completed steps, pending actions, and critical junctures \cite{HALO241}. The agent must rely on this log to orient itself within the overall care pathway, ensuring its actions are coherent and purposeful. Technically, the most direct implementation of this dynamic workspace is through in-context learning \cite{CONTEXTUAL1}, where the deliberative process and dialogue history are continuously integrated into the prompt for subsequent calls to the LLM \cite{CLICONTEXT25}. 

To support longitudinal workflows that span multiple sessions, external storage mechanisms are utilized to extend the agent's episodic reach, preventing the catastrophic forgetting of patient history \cite{LIFECRAFT}. Crucially, unlike the GS or VWA paradigms that query external databases for standardized medical guidelines or fixed protocols, the EP utilizes this memory strictly as a dynamic archive of patient-specific context. By encoding interaction logs and successful communication patterns into vector databases, the agent retrieves the specific narrative of the patient's care journey \cite{HALO241}. This ensures that subsequent autonomous planning is not restart-based but is continuously informed by the patient's evolving preferences and past reactions. Thus, external memory in the EP serves to maintain the continuity of the emergent persona, rather than imposing a rigid external workflow.

\subsubsection{Collaboration}\label{subsubsec:EPCL}
The essence of collaboration for EPs is defined by the capacity for dynamic coordination of multi-step, goal-oriented clinical workflows \cite{medla}. Distinct from the cognition-centric collaboration of LSCs, which prioritizes effective information transfer, collaboration in the EP paradigm centers on action synchronization, task allocation, and workflow progression, thereby embodying its core identity as an active clinical partner \cite{DYNAMI}.

\noindent\textbf{Single-agent system}. Within this framework, the agent operates as an autonomous practitioner, leveraging its implicit procedural knowledge to proactively steer the clinical dialogue. In a clinical setting, this goes beyond simple Q\&A. The agent functions as a facilitator of shared decision-making. It actively initiates health education, interprets the significance of test results, and proposes treatment options, inviting the human user to participate in the care plan \cite{CELM24}. The collaborative dynamic here is framed by the agent's role in structuring the ambiguity of patient inputs into a coherent medical narrative, using its internal planning capabilities to propel the consultation toward a defined therapeutic objective \cite{TREEOFREA}.

\noindent\textbf{Multi-agent system}. MAS expands the capability envelope of the EP by distributing complex workflows among a team of specialized agents \cite{medla}. This architecture emulates real-world MDT, enabling parallel task execution and the integration of diverse medical expertise to enhance system robustness \cite{COSTEFF}. Based on their organizational structure, these systems typically adopt one of two topologies.

In a dominant topology, the system mimics a hierarchical clinical ward led by a chief physician. The central agent acts as a higher-order EP responsible for meta-planning: decomposing a complex clinical objective into a series of well-defined sub-tasks \cite{PROACT, MEDIATOR}. It then delegates these sub-tasks to downstream worker agents, which may be domain-specific EPs or grounded tools. Information flows top-down: the central coordinator issues clinical directives, worker agents execute tasks such as history taking or differential diagnosis, and the coordinator synthesizes this feedback to determine the next global action \cite{COSTEFF}. This structure offers clear hierarchical control and predictable behavior, making it ideal for standardizable clinical pathways.

Conversely, distributed MAS resembles a peer-to-peer clinical case conference or tumor board assembled to address complex comorbidities \cite{ADVMUL}. In such systems, collaboration is driven not by delegation but by negotiation. Each agent, representing a specialist with distinct implicit knowledge, generates a local action plan informed by its unique perspective \cite{medla}. These agents broadcast their intentions through a shared channel \cite{DYNAMI}, engaging in conflict detection and plan integration. Through a multi-round process of counter-proposals and mutual concessions, they iteratively refine their individual plans to converge on a globally coherent and clinically optimal course of action \cite{MASPAT}. While computationally complex, this paradigm provides the flexibility required to solve integrative clinical problems where guidelines may conflict.

\subsubsection{Evolution}\label{subsubsec:EPEV}
In the EP paradigm, evolution is an intrinsic capability for autonomous self-improvement, distinct from external model updates or knowledge base expansions \cite{SELEVOCL}. Since EPs rely on implicit procedural knowledge to navigate complex workflows, they require a mechanism to dynamically refine their internal strategies based on action outcomes \cite{DYNAMISIS}. This capacity transforms them from static executors of pre-trained patterns into adaptive learners who grow expertise through clinical practice, mirroring the transition of a physician from residency to fellowship.

At the foundational level, the core of this evolutionary capability is a feedback loop driven by real-world clinical outcomes. When an EP completes a workflow, the resulting clinical impact—measured by patient recovery rates, adherence metrics, or physician oversight—is encoded as a reward or penalty signal \cite{CLITRI}. Leveraging frameworks inspired by human-AI collaboration \cite{ADAGENT}, this signal fine-tunes the agent's parametric memory. Crucially, this process does not merely correct isolated facts but reinforces or suppresses the entire neural policy associated with the action sequence. Consequently, decision paths leading to positive patient outcomes are strengthened, while those yielding suboptimal results are attenuated. This mechanism enables the agent's implicit clinical intuition to evolve toward better standards of care through thousands of such micro-adjustments \cite{MEDPAO}.

Progressing beyond case-by-case corrections, a mature evolutionary mechanism must extract generalizable clinical heuristics from successful experiences. This necessitates a meta-learning layer capable of identifying recurring success patterns across diverse tasks \cite{IMPROVINTEACT}. For instance, if an agent discovers that a specific sequence of empathetic validation followed by technical explanation consistently improves compliance in anxious patients, it abstracts this pattern into a reusable communication protocol \cite{SELEVOCL}. This protocol is then stored as a strong prior in its high-level parametric memory, allowing for rapid deployment in future encounters with similar patient profiles \cite{DYNAMISIS}. This abstraction from concrete experience to generalized strategy marks the agent's cognitive shift from a novice practitioner to an experienced expert.

Ultimately, the most advanced form of evolution is proactive self-repair. A truly evolving agent should not passively await external feedback but actively identify its capability limits and knowledge gaps to initiate targeted learning. This requires an internal auditing module to analyze the agent's past performance stability \cite{SEEKINNER}. For example, the system may detect consistently low confidence or high variance when handling cases involving rare drug-drug interactions. Upon identifying such a competency weakness, the mechanism can autonomously trigger a self-improvement task \cite{PROACT1} to fine-tune specific parameters. This self-driven learning, guided by metacognitive awareness, parallels the continuing medical education of human doctors, ensuring efficient, directed evolution into a reliable clinical collaborator.

\subsubsection{Summary}
\textbf{The EP paradigm defines a highly autonomous agent that utilizes deep, implicit procedural knowledge to generate and execute multi-step clinical workflows \cite{PROACT}, with action trajectories dynamically shaped to achieve high-level goals.} Relying on parametric memory as a strategic engine and non-parametric memory as a tactical workspace for action-oriented reasoning, EPs demonstrate exceptional flexibility and creative problem-solving in open-ended clinical tasks \cite{HALO241}. This execution-driven orientation critically distinguishes them from the more passive, cognition-focused LSCs; while both paradigms leverage internal knowledge, EPs utilize it to propel clinical processes forward rather than merely describing patient states. However, such significant autonomy introduces inherent risks regarding clinically dangerous hallucinations and the potential application of outdated procedures due to knowledge staleness. Consequently, the primary research challenge lies in balancing this generative freedom with the rigorous demands of clinical safety. This tension highlights the critical necessity for explicit verification mechanisms, effectively paving the way for the GS paradigm, which prioritizes adherence to external evidence over creative autonomy to ensure reliability.

\subsection{Grounded Synthesizer}\label{subsec:GS}
Addressing the critical challenge of hallucination in medical AI, the GS operates on the principle of strict explicit knowledge grounding. Unlike creative generators, these agents function as intelligent interfaces that retrieve, integrate, and cite verifiable external evidence to construct an accurate clinical snapshot, as illustrated in Fig.~\ref{fig:GS}. By anchoring reasoning in a traceable chain of evidence \cite{CHAINOFE}, the GS paradigm maximizes the safety and reliability necessary for clinical decision support. Table~\ref{tab:GSsum} provides a systematic overview of methods within this evidence-driven category.

\begin{table}[!htp]
\setlength{\abovecaptionskip}{0cm}   
\setlength{\belowcaptionskip}{-0.1cm}   
\caption{The summarization of grounded sythesizer agentic Approach,  "PT" represents pre-training and fine-tuning, "PE" represents prompt engineering, "MAS" represents multi-agent system, and "ST" represents self-training.}
\label{tab:GSsum}
\resizebox{0.8\textwidth}{!}{
\begin{tabular}{ccccc}
\hline
\textbf{\begin{tabular}[c]{@{}c@{}}Agentic\\ Approach\end{tabular}} & \textbf{\begin{tabular}[c]{@{}c@{}}Main Corresponding\\ Component\end{tabular}} & \textbf{\begin{tabular}[c]{@{}c@{}}Detailed\\ Method\end{tabular}} & \textbf{\begin{tabular}[c]{@{}c@{}}Downstream\\ Task\end{tabular}} & \textbf{\begin{tabular}[c]{@{}c@{}}Framework\\ Type\end{tabular}} \\ \hline
HyKGE\cite{HYKGE} & Planning Strategies & Hypothesis Enhancement & Medical QA & PE \\ \hline
EvidenceMap\cite{EVIDMAP} & Planning Strategies & Content Summarization & Biomedical QA & PT \\ \hline
MHMKI\cite{MULTIHOPKNO} & Planning Strategies & Knowledge Infusion & Biomedical QA & PT \\ \hline
MedicalGLM\cite{QEURYEVAL} & Planning Strategies & Quality Evaluation Mechanism & Medical QA & PT \\ \hline
K-Comp\cite{KCOMP} & Memory Management & Knowledge Infusion & Medical QA & PT \\ \hline
Rationale-Guided RAG\cite{RATIONALE} & Memory Management & Knowledge Retrieval & Medical QA & PT \\ \hline
Seek Inner\cite{SEEKINNER} & Memory Management & Information Mining & Medical VQA & PE \\ \hline
MedGENIE\cite{TOGENORRET} & Action Execution & Information Generation & Medical QA & PE \\ \hline
MediSearch\cite{MEDISEARCH} & Action Execution & Medical Search Engine & Task-oriented & PE \\ \hline
MeNTi\cite{MENTTOOL} & Action Execution & Calculator Tool-use & Task-oriented & PT \\ \hline
MedOrch\cite{MEDIATOR} & Collaboration & Mediator-guided Collaboration & Medical VQA& MAS \\ \hline
MedConMA\cite{MEDCONMA} & Collaboration & Voting Mechanism & Medical QA & MAS \\ \hline
AMG-RAG\cite{MedicationQA} & Self-evolution & Knowledge Updating & Medical QA & PE \\ \hline
SFPS\cite{IMPROVESELFTRAI} & Self-evolution & Domain Adaptation & Task-oriented & ST \\ \hline
\end{tabular}}
\end{table}

\begin{figure}[!htp]
    \centering
    \setlength{\abovecaptionskip}{0cm}   
    \setlength{\belowcaptionskip}{0cm}   
    \includegraphics[width=0.8\linewidth]{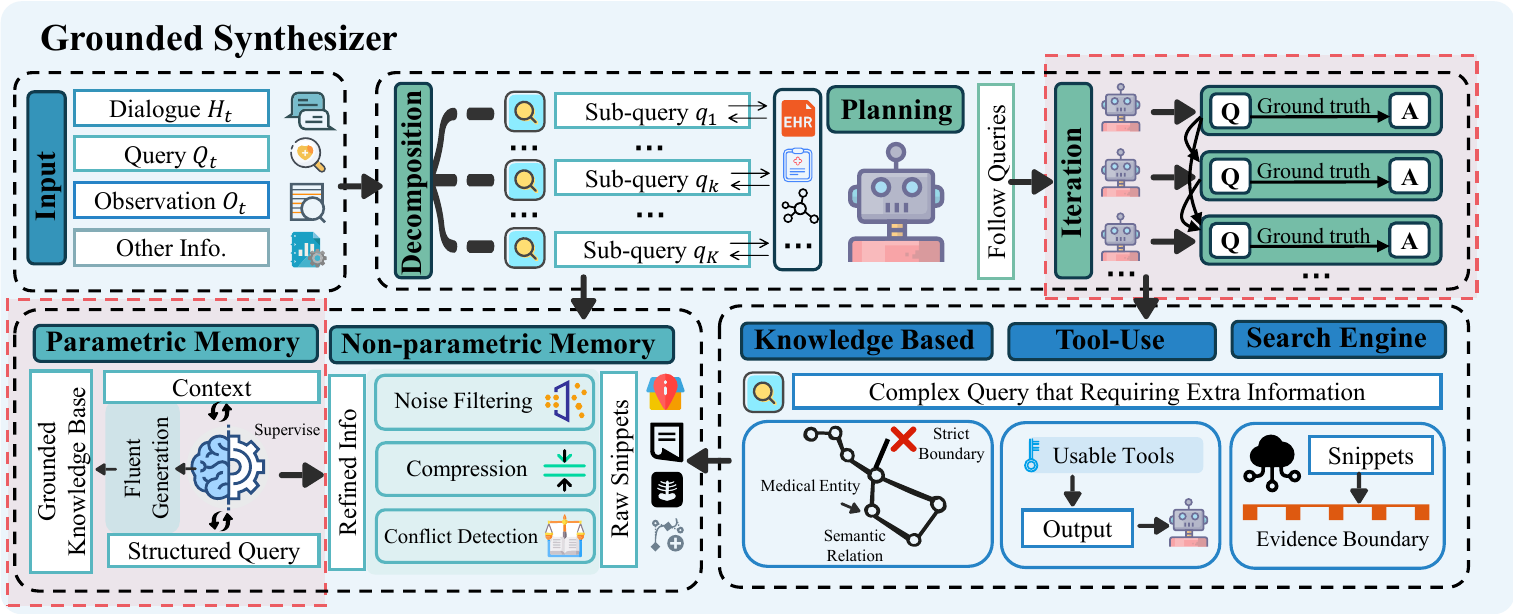}
    \caption{General structure of single agent in the GS paradigm. Strategic planning decomposes broad clinical inquiries and maps these queries to specific external sources. Action execution serves as the core mechanism for evidence acquisition, performing three distinct epistemic operations. Non-parametric memory acts as an evidence boundary, strictly buffering the retrieved snippets to prevent contamination, while parametric memory synthesizes this grounded evidence into a fluent response with traceable citations. Core differences from other paradigms are covered with red blocks.}
    \label{fig:GS}
\end{figure}

\subsubsection{Strategic Planning}\label{subsubsec:GSSP}

In the context of the GS, strategic planning is not aimed at devising a sequence of actions to alter the environment, but rather at planning how to better cognize. It breaks down a broad cognitive goal into specific, answerable sub-questions, each designed to be resolved by querying external sources such as EHRs, medical literature, or clinical guideline knowledge graphs(KGs) \cite{CHECKFACT, FRESHLLM}. This approach ensures that all information in the synthesized clinical snapshot is source-traceable, significantly reducing the risk of informational hallucinations \cite{RAGITER} and reflecting the conservative, reliable standards essential in medical research.

\noindent\textbf{Decomposition strategies}. To achieve this, the primary decomposition strategy functions as the construction of a logically rigorous inquiry pathway that mirrors professional diagnostic workflows \cite{HYKGE}. Unlike the implicit reasoning chains used in other paradigms, the GS maps complex clinical assessment tasks directly to standardized decision trees or guideline protocols. The agent generates an ordered sequence of sub-questions targeting atomic, verifiable clinical facts, such as specific biomarkers or symptom durations, that can be validated through external database queries \cite{CHECKFACT}. This logic extends to handling the complexity of heterogeneous and multi-hop medical information, where decision-making necessitates the synthesis of diverse data sources, such as laboratory information systems and medical imaging archives. The strategy employs a structured query dependency graph \cite{KGAREVIS} to orchestrate the systematic retrieval of evidence fragments. Within this graph, the output of a specific medical inquiry serves as the requisite input for subsequent diagnostic steps \cite{EVIDMAP}, thereby transforming high-complexity clinical reasoning into a traceable, multi-hop information-gathering workflow that supports both concurrent and sequential execution \cite{MULTIHOPKNO}.

Beyond static planning, advanced strategies incorporate iterative cognitive refinement driven by real-time clinical data feedback. The agent does not rely on intrinsic hallucinations for exploration; instead, it leverages specific retrieval results as conditional triggers to guide further investigation \cite{RAGITER}. These data-driven triggers enable the system to dynamically instantiate fine-grained sub-query templates adapted to emerging clinical findings \cite{QEURYEVAL}. For instance, the detection of a critical value, such as significantly elevated serum potassium, acts as a signal to initiate a new, targeted inquiry branch regarding potential causes, such as renal failure or medication side effects \cite{RAGITER}. This adaptive mechanism shifts the cognitive workflow from a rigid checklist to a dynamic inquiry tree, selectively allocating cognitive resources to the most salient aspects of the patient's condition to construct an increasingly deep and high-fidelity clinical snapshot.

\noindent\textbf{Iteration strategies}. Iteration strategies within the GS framework introduce critical dimensions of circulation and refinement, serving not as model-driven self-correction but as a rigorous mechanism to enhance the information density of the clinical snapshot through multiple structured interactions with external sources \cite{MULTIDOC, KCOMP}. Central to this process is an information-query loop \cite{MULTIHOPKNO} designed to emulate the systematic information gathering of a clinician. A primary objective is cyclical gap-filling to achieve information completeness. Unlike general open-domain inquiries, clinical data retrieval often necessitates adherence to a strict task-relevant information schema or cognitive scaffold \cite{RAGITER}, analogous to populating a diagnostic checklist or a structured admission note. Following an initial retrieval, the agent systematically compares the obtained data against this predefined schema to automatically detect missing clinical parameters. It then generates and executes a new set of highly focused sub-queries to address these specific deficits, continuing this goal-driven cycle until the patient profile is fully populated \cite{MEDINF}. This ensures that the final clinical representation maintains structural integrity and covers all medically necessary dimensions.

Parallel to gap-filling, the agent employs an evidence hierarchy-guided ambiguity resolution mechanism to handle the contradictions inherent in real-world medical records \cite{MEPNET}. When retrieved information from multiple sources conflicts—such as discrepant pathology reports or inconsistent medication lists—the agent adheres to a strict query escalation protocol rather than attempting to hallucinate a reconciliation. By referencing a predefined evidence hierarchy, the system systematically executes a cyclical inquiry process to identify a reliable anchor, such as prioritizing the clinical document with the most recent timestamp, seeking a report signed by a higher-level authority \cite{MULTIDOC}, or requesting a final consensus version \cite{INSTRUCTOFREFLEX, BEYONDEHR}. If these automated pathways fail to resolve the discrepancy, the strategy mandates explicitly flagging the conflict to the human user, thereby prioritizing safety and accountability over forced coherence. Complementing these mechanisms is granularity-driven progressive deepening, which allows the agent to construct a multi-layered clinical representation that integrates both breadth and depth. In this mode, the agent recursively executes a sequence of increasingly specific sub-queries focused on a core clinical entity \cite{BEYONDEHR}. This mimics the clinical drill-down process, whereby a physician investigates a general symptom by progressively querying for its specific attributes, severity, and associated context until the information granularity satisfies the requirements for a precise diagnosis \cite{ATOMRAG}.

\subsubsection{Memory Management}\label{subsubsec:GSMM}

For a paradigm centered on the core principle of external knowledge grounding, the design of GS's memory system presents a unique tension. On one hand, it must strictly rely on external, verifiable knowledge sources as the sole benchmark for facts \cite{KCOMP, KOSEL, CHAINOFE}. On the other hand, it cannot dispense with the parametric memory inherent to the LLM itself to execute complex cognitive tasks \cite{RATIONALE}. Therefore, the essence of memory management in the GS is not simple storage and retrieval, but rather the construction of a meticulously designed and highly controlled bridge between external truth and internal cognition \cite{UNCERTAIN}. 

\noindent\textbf{Parametric memory}. The role of parametric memory in the GS framework is to serve as an indispensable universal cognitive processor, providing the foundational cognitive capabilities necessary for the reliable integration and presentation of external knowledge. Its core value is manifested across three interconnected levels.

Clinical dialogue is inherently complex, characterized by a mix of colloquial patient descriptions and precise, domain-specific terminology. The agent must rely on the linguistic patterns encoded within its parametric memory to parse these interactions, effectively mapping lay descriptions of symptoms to standardized medical concepts and identifying latent clinical intents \cite{BIASRAGQA}. This deep natural language understanding capability enables the agent to navigate the semantic intricacies of medical dialogue, serving as a prerequisite for any subsequent evidence retrieval. Once the user's intent is identified, the agent employs its parametric knowledge to function as a universal translator, converting abstract clinical questions into concrete, executable actions. This process involves transforming natural language into structured queries tailored for specific external targets, such as SQL for EHRs or SPARQL for biomedical KGs \cite{RATIONALE}. Such capability relies heavily on the procedural knowledge stored in the model's parameters—specifically, the rules regarding database schemas, search optimization strategies, and API protocols learned during pre-training \cite{MULTIHOPKNO}.

Eventually, information integration and fluent generation represent the culmination of the GS workflow, where parametric memory plays a pivotal role in ensuring narrative quality. Data retrieved from heterogeneous sources—ranging from disjointed laboratory values to unstructured literature snippets—is often fragmented and lacks narrative flow \cite{SEEKINNER}. The agent leverages its advanced generation capabilities to synthesize these disparate evidence fragments into a coherent, fluent, and clinically appropriate output. By applying learned linguistic structures and medical reporting standards, parametric memory transforms raw, retrieved facts into a synthesized clinical summary that ensures readability without compromising the factual integrity derived from external sources \cite{RAGITER}. Thus, while the content is grounded externally, the coherence and accessibility of the final report depend entirely on the processing power of the model's parametric memory.

\noindent\textbf{Non-parametric memory}. If parametric memory functions as the cognitive processor, non-parametric memory acts as the verifiable epistemic scaffold of the GS paradigm. Unlike other agentic approaches where external memory serves merely as a conversational buffer, here it is architecturally designed as a structured log that enforces the rigorous standards of traceability and verifiability required in healthcare \cite{MMEDAG24}. Functioning primarily as an evidential ledger, this memory system ensures that every interaction with an external knowledge source is not recorded as an isolated data point but archived as a complete clinical transaction \cite{KOSEL}. Each entry captures the retrieved medical information, its precise source, and the query used to obtain it \cite{TOOLCLINIAGEN}. This design transforms memory from a passive repository into a dynamic, auditable chain of evidence \cite{CHAINOFE}. In high-stakes clinical practice, this capability is essential as it allows human experts to retrospectively verify the informational basis of any AI-generated recommendation, ensuring that synthesized outputs can be systematically traced back to their foundational evidence units for accountability.

Beyond auditability, non-parametric memory serves as the central mechanism for clinical state management. The strategic plans of the GS are explicitly instantiated and dynamically tracked within this external memory rather than being left to the opaque latent space of the LLM \cite{SEEKINNER}. This system maintains a comprehensive representation of the current diagnostic workflow, tracking which clinical sub-questions have been resolved, which information gaps remain, and how different pieces of evidence correlate \cite{RGARMEDRAG}. During the iterative reasoning process, it is the evolving state of this external memory that triggers subsequent inquiry cycles, as in the case of a patient chart \cite{ADAPTIVEKNOW}. This externalization of the cognitive state prevents procedural drift by ensuring that the agent's behavior is strictly governed by accumulated evidence rather than unpredictable internal associations \cite{MEDEX}. Consequently, when the final synthesis stage begins, the model is supplied with a structured, contextually enriched body of evidence, enabling a generation process that is faithful to the patient's actual status.

A critical architectural imperative for non-parametric memory in this paradigm is information compartmentalization \cite{EXPLAINKNOWBASE}. A significant risk in medical AI is knowledge contamination, where verified external facts blend with the LLM's vast but potentially hallucinated internal knowledge. To mitigate this, non-parametric memory is designed as a firewalled context \cite{KNOWLEDGEARGU}. The information flow is strictly unidirectional: from external clinical tools into this memory, and then to the LLM's context window for processing \cite{LEVARLC}. The system explicitly prevents the LLM from treating this memory as training data to update its internal beliefs, maintaining an absolute boundary between verifiable patient data and implicit model parameters \cite{EVALSEARCH}. This strict separation ensures that the agent functions purely as a synthesizer of external truth, embodying the conservative, safety-first philosophy essential for reliable clinical decision support.

\subsubsection{Action Execution}\label{subsubsec:GSAE}

In the GS paradigm, the concept of action execution refers to the execution of epistemic operations whose sole purpose is to acquire information from external knowledge sources to construct an internal cognition \cite{TOGENORRET}. Therefore, the function of the action execution module here is to serve as the physical interface connecting the logical inquiry pathway with the real-world data and situations \cite{MEDIKNOWLED}. It is responsible for translating abstract inquiry directives into syntactically precise API calls or query statements for specific data systems and faithfully transmitting the returned data to the memory module. 

\noindent\textbf{Knowledge-based}. A primary mode of action execution within the GS paradigm involves knowledge-based actions, which specifically entail interactions with external structured, symbolic knowledge bases \cite{KOSEL}. These interactions are designed to obtain facts characterized by explicit semantic relationships and a high degree of logical certainty \cite{MEDIKNOWLED}. In the context of the GS paradigm, which prioritizes rigorous clinical reliability, such actions are not merely optional channels for information acquisition but serve as the cornerstone providing dual logical-semantic grounding for the entire cognitive process \cite{KNOWLEDGEARGU}.

When the GS faces a complex clinical scenario that requires multi-step, cross-concept deduction, such as cross-referencing symptoms with rare disease etiologies or analyzing potential adverse drug events, it generates a formal query sequence through its strategic planning layer. This sequence traverses a specific logical path on an external medical knowledge graph \cite{MULTIHOPKNO, MEDCOTRAG}. Since the structure of these graphs is composed of nodes representing standardized medical entities and edges representing expert-defined patho-physiological relationships \cite{MEDIKNOWLED}, it inherently constitutes an axiomatic logical skeleton. Each action performed by the agent becomes an independent, formally verifiable logical operation on this skeleton \cite{MEDITRIR}. Consequently, a complex diagnostic reasoning chain is decomposed into a series of atomic, auditable transactions \cite{MEDITRIR}. This mechanism not only renders every intermediate step of the final conclusion transparent and verifiable but, more importantly, establishes a complete computational provenance trail for the entire cognitive process, which is a critical requirement for the validation and regulatory approval of clinical decision support systems.

Furthermore, the schema of the GS's knowledge base functions as a robust cognitive guardrail \cite{MEDEX}. By providing strict semantic and type constraints for the agent's inquiries, the schema imposes a strong domain-specific inductive bias. This mechanism effectively mitigates the risk of conceptual confusion and relational hallucinations often caused by natural language ambiguity \cite{KNOWMULTIHOP}. For instance, while a standard language model might statistically associate a symptom with a biologically implausible drug side effect based on text co-occurrence, a strict medical ontology prevents such invalid logical leaps by enforcing predefined inheritance hierarchies and domain constraints \cite{EXPLAINKNOWBASE}. This ensures that the agent's cognitive outputs remain rigorously aligned with established medical truths at the ontological level.

\noindent\textbf{Search engine based}. Search engine-based actions are primarily employed by the agent to explore the vast, unstructured knowledge frontier, encompassing extensive medical literature, clinical trial reports, conference abstracts, and real-time public health guidelines \cite{MEDISEARCH, EVALSEARCH}. Within the GS architecture, the strategic value of these actions lies in enabling dynamic adaptation to evolving knowledge, thereby addressing the inherent limitations of structured knowledge bases regarding coverage breadth and, critically, the timeliness of recent research advances \cite{SEARCHRAG}. However, integrating web-scale data introduces a significant cognitive challenge: preserving reliability when navigating noisy, ambiguous, and potentially biased free text. To address this, the GS paradigm redefines search engine usage not as simple information retrieval, but as a highly proceduralized process of dynamic evidential boundary construction. Rather than treating the search engine as an oracle that delivers direct answers, the agent positions it as a provider of contextual evidence \cite{SEARCHR1}. The core of this action execution consists of a rigorous three-stage workflow: forage, constrain, and attribute.

In the forage stage, the agent avoids executing a one-shot query aimed at directly hitting an answer. Instead, it emulates the methodology of a professional researcher conducting a systematic review, performing a strategic, multi-round query sequence to collect a comprehensive candidate evidence corpus \cite{MEDISEARCH}. This process is designed to ensure that the collected text snippets cover diverse aspects of the clinical question, including potential treatment controversies, varying clinical guidelines across regions, and the latest findings from ongoing trials. By systematically aggregating multi-faceted information, the agent constructs a robust evidence base that mitigates the risk of bias inherent in single-source retrieval.

The constraint stage represents the critical safety juncture of the process. All text snippets retrieved, along with their complete metadata, are injected into the non-parametric memory to construct a temporary, disposable knowledge base that is strictly firewalled from the LLM's internal parameters \cite{LEVARLC}. Subsequently, the model is instructed to perform cognitive operations exclusively within this isolated context composed entirely of external evidence. This mechanism achieves strict contextual grounding, effectively transforming the nature of the LLM's task from open-domain question answering—which relies on vast but potentially hallucinatory internal weights—to a highly constrained task of grounded text synthesis. This shift significantly reduces the risk of generating clinically unsafe advice, as the model is structurally prevented from citing any knowledge beyond the boundaries of the provided evidence \cite{KCOMP}.

Finally, in the attribute stage, when the LLM generates a summary or recommendation based on this isolated knowledge, it is mandated to provide a fine-grained citation pointer for every factual assertion. These pointers link directly back to the specific source text snippet in the non-parametric memory \cite{SEARCHHEALTH}. This achieves programmatic attributability, ensuring that the final output is not a monolithic block of text but a composite structure of assertions and their corresponding proofs. In a clinical setting, this design enables healthcare professionals to instantly verify the source and reliability of every piece of information, establishing a transparent and auditable decision support process that balances the flexibility of unstructured data with the rigor required for medical safety.

\noindent\textbf{Tool-use}. Tool-use represents the action modality with the highest cognitive certainty within the GS paradigm. It specifically entails the programmatic invocation of external, validated computational modules to execute precise clinical functions that lie beyond the stochastic capabilities of language models. These include deterministic applications such as drug dosage calculators \cite{TOOLCALAGEN}, clinical risk stratification scores \cite{MENTTOOL}, and pharmacokinetic simulation engines. By strictly delegating these tasks, the agent achieves indispensable procedural grounding \cite{MMEDAG24}, ensuring that the system's operational behavior remains predictable and compliant with rigorous clinical standards.

From an architectural perspective, these tools function as algorithmic oracles—axioms within the cognitive process that encapsulate validated clinical logic. This approach embodies the principle of cognitive liability offloading \cite{TOOLCLSSI}. The core design philosophy posits that any clinical task definable by a deterministic algorithm or a fixed mathematical formula must be strictly offloaded from the LLM's probabilistic cognitive space to a specialized external tool \cite{TOOLCALAGEN}. This distinction stems from a critical recognition of the LLM's capability boundaries: while powerful in semantic understanding and fluent generation, they lack the symbolic reasoning required for precise calculations or strict adherence to procedural guidelines.

Consequently, by executing a tool call, the agent facilitates a programmatic inheritance of authority. The credibility of the output is not derived from the model's internal intelligence but is directly inherited from the clinical validation embodied by the tool itself \cite{TOOLCLINIAGEN}. This replaces a vulnerable link in the decision chain—an opaque, probabilistic text generator—with a deterministic module that yields consistent, verifiable results. Under this framework, the LLM is redefined as a universal orchestrator of semantic interfaces. Its primary responsibility shifts to accurately parsing the clinician's natural language intent, mapping it to the appropriate function, and integrating the structured output back into the dialogue, without needing to process the internal logic of the tool \cite{ADEPTTOOL}. This design pattern significantly enhances system robustness, constructing a modular AI architecture with clear liability boundaries that align with the safety-first objectives of the GS paradigm.

\subsubsection{Collaboration}\label{subsubsec:GSCL}

Within the GS paradigm, the concept of collaboration is reframed conceptually. Unlike goal-oriented agents engaging in peer-to-peer negotiation, the GS agent functions within a strictly asymmetric human-agent partnership aimed at cognitive augmentation \cite{MEDCONMA}. Since the agent lacks the agency to alter patient states directly—such as prescribing medication without approval—its role is confined to that of an advanced cognitive tool. Collaborative behaviors focus on enhancing the situational awareness of clinicians by presenting structured, verifiable evidence rather than acting as an autonomous participant in the decision-making process. Consequently, the core value of this paradigm lies in the reliability with which the agent supplies grounded information to support the human medical cognitive process \cite{MEDIATOR}.

\noindent\textbf{Single-agent system}. The single-agent architecture represents the native implementation of the GS paradigm, designed to construct a singular, internally consistent clinical snapshot. In this model, the solitary agent acts as a centralized information orchestrator, executing optimal inquiry pathways across heterogeneous sources, such as PubMed abstracts or patient histories. Since the objective is to complete a highly focused cognitive synthesis task—such as summarizing a discharge note or answering a specific drug interaction query \cite{KCOMP,ADAPTIVEKNOW}—a single control core offers the most direct method to maintain logical consistency. This alignment with a conservative, safety-first philosophy is crucial in medicine. However, as the volume and technical heterogeneity of external sources increase (e.g., combining genomic databases with unstructured nursing notes), the maintainability of this monolithic architecture faces significant engineering challenges \cite{ERRDET}, necessitating a shift toward more modular organizational forms.

\noindent\textbf{Multi-agent system}. Introducing a MAS under the stringent constraints of the GS paradigm differs greatly from the emergent reasoning sought in LSC or the autonomous negotiation in EP. Instead, it represents a systems engineering strategy designed to handle the complexity of heterogeneous medical data through a functional cognitive division of labor. By assigning specialized agents for specific data modalities, such as one for radiology report extraction and another for clinical guideline retrieval, GS enhances the scalability and auditability of the evidence-gathering process.

The dominant topology in this context functions strictly as a verifiable evidence aggregation hub rather than a consultation team. A unique orchestrator agent acts as the absolute center, responsible not for synthesizing novel ideas, but for managing the logistics of multi-source retrieval \cite{ERRDET}. For instance, it decomposes a complex clinical query and dispatches specific fetch-tasks to worker agents—one accessing drug interaction databases and another retrieving patient history—before rigorously compiling their structured outputs \cite{LARSMACOL}. The primary advantage of this topology in GS is absolute auditability. By forcing all external information flows to pass through a single control node, the system maintains a global view of the evidence chain, ensuring that every piece of integrated data is traceable and preventing the hallucinated consensus that can occur in less constrained interactions \cite{MEDSENTRY}.

Conversely, the distributed topology adapts the concept of clinical care pathways into a deterministic data flow pipeline. Unlike the peer-to-peer negotiation characteristic of goal-oriented agents, collaboration here is governed by a fixed cognitive workflow graph, typically a directed acyclic graph \cite{MEDCONMA}. Each node represents a dedicated processor rather than a decision-maker. For example, a "symptom extractor" passes raw entities strictly to a "differential diagnosis filter" for validation \cite{MEDICONSENS}. The collaboration is driven by the rigid flow of clinical data itself rather than dynamic mediation \cite{MDTEAMGPT}. This structure is particularly effective for standardizable, multi-stage tasks like automated chart review, where the output of one specialist module serves as the immutable input for the next, ensuring execution adheres to medical protocols without deviation \cite{MEDSENTRY}.

\subsubsection{Evolution}\label{subsubsec:GSEV}

Distinct from paradigms that aim to enhance autonomous decision-making capabilities or update an internal world model through experiential learning, evolution in the GS framework focuses on optimizing the agent's efficacy as a cognitive tool \cite{ADAPTIVEKNOW}. This evolution is meta-cognitive and procedural, with its core objective being the refinement of "how to cognize better" rather than altering the factual basis of "what to cognize" \cite{TOOLCALAGEN}. Consequently, evolution is confined to a safe and controllable trajectory, primarily realized through three interconnected mechanisms: the optimization of inquiry strategies, the refinement of action policies, and the human-driven expansion of knowledge boundaries.

Inquiry pathway optimization represents the macro-evolutionary adaptation of the agent, targeting its strategic planning module. The core objective is for the GS to learn how to construct more efficient and clinically relevant inquiry pathways through experiential learning derived from its interactions \cite{MDTEAMGPT}. Critically, this learning process is decoupled from clinical outcome feedback to prevent the agent from assuming decision-making responsibilities beyond its scope. Instead, the feedback signal for policy refinement is predicated on metrics of cognitive efficiency, such as the diagnostic information gain per query \cite{ADAPTIVEKNOW}. For instance, through the offline analysis of interaction logs, the system identifies inquiry patterns that most frequently yield a comprehensive clinical snapshot with the minimum number of query steps—a principle mirroring the self-learned knowledge generation in multi-agent diagnostic frameworks \cite{MACDMULTI}. Alternatively, this can be framed as a reinforcement learning problem where the policy is refined using reward signals such as maximizing the information adoption rate in human decision-making. This evolution enhances the agent's utility as a sophisticated information-provision tool without compromising safety.

Operating at a micro-technical level, action policy refinement focuses on the efficiency of action execution when interacting with external medical sources \cite{EVIDENBASEAGEN}. Every external tool or database—whether a PubMed search API or a hospital EHR system—possesses unique operational characteristics, including specific query syntax, rate limits, and data format nuances. Action policy refinement enables the agent to adapt to these interfaces. For example, by analyzing cases of query failure, the agent acts as a procedural learner, discovering that mapping user queries to MeSH (Medical Subject Headings) terms yields a significantly higher retrieval success rate than free-text queries in biomedical databases \cite{IMPROVESELFTRAI}. Similarly, it can learn to adopt resilient calling strategies, such as automated retries or timeouts, when invoking slow-responding legacy EHR interfaces. By building an internal policy model on how to effectively converse with specific tools \cite{TOOLCALAGEN}, the agent evolves from a novice interface user \cite{TOOLCLINIAGEN} into a proficient technical expert. This results in a robust data acquisition layer that minimizes cognitive interruptions caused by technical friction.

Human-in-the-Loop knowledge boundary expansion represents the most unique form of evolution within the GS paradigm \cite{ITERACTHUMAN}, embodying the agent's role as a collaborative partner in a human-machine cognitive community \cite{HUMANCENTERED}. The agent is prohibited from autonomously updating the external knowledge bases on which it relies. However, when the agent repeatedly fails to retrieve an answer to a specific clinical question from authoritative sources during an inquiry pathway, it records this evidence gap as a structured event. These aggregated gaps form a valuable, demand-driven knowledge list that is submitted to human experts and clinical knowledge engineers \cite{HUMANCENTVER}. This feedback loop points human maintainers toward areas where clinical guidelines or databases require prioritization for expansion and updating \cite{TAMACOLL}. Thus, a symbiotic relationship is established: the agent exposes the boundaries of current medical knowledge through efficient inquiry, and human experts expand these boundaries, which in turn enhances the agent's future capabilities. This indirect, human-centered evolutionary path ensures that the system remains a safe and reliable cognitive-assistive tool.

\subsubsection{Summary}

\textbf{The GS paradigm conceptualizes a conservative agent functioning strictly as a verifiable clinical decision support tool, which anchors its entire epistemic process in external sources to produce a transparent synthesis of the clinical situation.} Characterized by an unwavering commitment to explicit grounding, this paradigm effectively mitigates factual hallucinations by constructing cognitive conclusions upon an auditable chain of evidence \cite{CHAINOFE}, prioritizing patient safety and source traceability over generative creativity. However, this strict reliance on retrieved knowledge introduces a critical trade-off: while ensuring high trustworthiness, it significantly constrains the agent's flexibility in addressing novel or undocumented clinical scenarios compared to paradigms leveraging implicit latent knowledge. Consequently, the primary research frontier lies in overcoming retrieval bottlenecks that limit synthesis quality \cite{EVALSEARCH}, managing the execution latency of complex multi-hop reasoning in time-sensitive acute care settings \cite{MULTIHOPKNO}, and solving the logistical imperative of maintaining the continuous currency of the external medical knowledge bases upon which the agent depends \cite{KNOWLEDGEARGU}.

\subsection{Verifiable Workflow Automator}\label{subsec:VWA}

The VWA represents a paradigm in which the LLM serves as an intelligent natural language interface that drives deterministic, protocol-based process engines. Distinct from the open-ended planning of EPs, VWA agents operate by strictly adhering to explicit clinical guidelines and verifiable workflows to ensure maximum safety and predictability, as depicted in Fig.~\ref{fig:VWA}. This approach balances the need for active clinical intervention with the requirement for rigorous algorithmic control. Table~\ref{tab:VWAsum} provides a comprehensive summary of methodologies categorized under this high-stakes execution framework.

\begin{table}[!htp]
\setlength{\abovecaptionskip}{0cm}   
\setlength{\belowcaptionskip}{-0.1cm}   
\caption{The summarization of VWA agentic approach,  "PT" represents pre-training and fine-tuning, "PE" represents prompt engineering, "MAS" represents multi-agent system, "RL" represents reinforcement learning, and "ST" represents self-training.}
\label{tab:VWAsum}
\resizebox{0.8\textwidth}{!}{
\begin{tabular}{ccccc}
\hline
\textbf{\begin{tabular}[c]{@{}c@{}}Agentic\\ Approach\end{tabular}} & \textbf{\begin{tabular}[c]{@{}c@{}}Main Corresponding\\ Component\end{tabular}} & \textbf{\begin{tabular}[c]{@{}c@{}}Detailed\\ Method\end{tabular}} & \textbf{\begin{tabular}[c]{@{}c@{}}Downstream\\ Task\end{tabular}} & \textbf{\begin{tabular}[c]{@{}c@{}}Framework\\ Type\end{tabular}} \\ \hline
VITA\cite{VITACARE} & Planning Strategies & Relevant-visit selectIon & Recommendation & PE \\ \hline
Quicker\cite{FROMTOBASE} & Planning Strategies & Evidence Synthesis & Recommendation & PE \\ \hline
LLM-AMT\cite{AUGMENTINGBB} & Planning Strategies & Query Augmenter & Biomedical QA & PE \\ \hline
MedPlan\cite{MEDPLANA} & Planning Strategies & Reasoning Structurizing & Task-oriented & PE \\ \hline
EMRs2CSP\cite{EMRS2CSP} & Memory Management & Clinical Representation & Medical QA & PT \\ \hline
Medical Graph RAG\cite{MEDGRAGEVIB} & Memory Management & Knowledge Retrieval & Medical QA & PE \\ \hline
Deep-DxSearch\cite{E2EAGENTIC} & Memory Management & Tracebale Knowledge Reasoning & Task-oriented & RL \\ \hline
Cli-RAG\cite{CLIRAGRETRIV} & Action Execution & Task-specific Retrival & Task-oriented & PE \\ \hline
Tala-med\cite{EHANCINGMEDRET} & Action Execution & Synonym System & Task-oriented & PE \\ \hline
PPME\cite{IMPROVINTEACT} & Action Execution & Calculator Tool-use & Task-oriented & RL \\ \hline
TeamMedAgents\cite{TEAMMEDAGENTSE} & Collaboration & Shared Mental Model & Medical QA & MAS \\ \hline
TAMA\cite{TAMACOLL} & Collaboration & Thematic Analysis & Task-oriented & MAS \\ \hline
HealthFlow\cite{HEALTHFLOW} & Self-evolution & Tool-use Strategy & Task-oriented & ST \\ \hline
MetaAgent\cite{SELFEVOLVINH} & Self-evolution & Meta Learning & Task-oriented & ST \\ \hline
\end{tabular}}
\end{table}

\begin{figure}[!htp]
    \centering
    \setlength{\abovecaptionskip}{0cm}   
    \setlength{\belowcaptionskip}{0cm}   
    \includegraphics[width=0.8\linewidth]{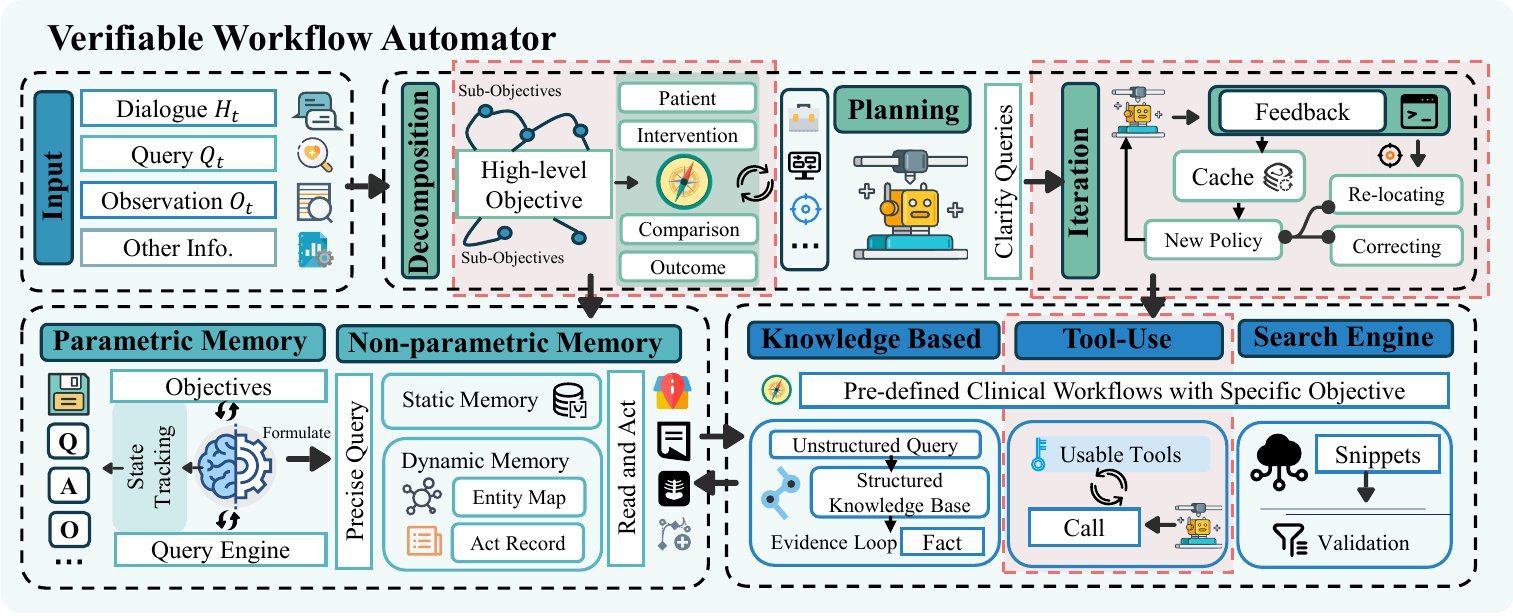}
    \caption{General structure of single agent in the VWA paradigm. Strategic planning maps the current patient state to nodes within a pre-defined clinical workflow. Action execution employs knowledge-based actions to preprocess clinical data, search engines for programmatic validation, and strictly offloads high-stakes tasks to usable tools. Non-parametric memory maintains a verifiable log of these executed actions for state tracking, while parametric memory acts as a query engine to translate intents into precise tool calls. Core differences from other paradigms are covered with red blocks.}
    \label{fig:VWA}
\end{figure}

\subsubsection{Strategic Planning}\label{subsubsec:VWASP}

Strategic planning within the VWA paradigm operates as a dual-process mechanism, synergizing decomposition and iteration to systematically navigate predefined clinical workflows \cite{VITACARE}. Unlike the EP paradigm, which leverages the implicit intuition of LLMs for open-ended reasoning, the VWA focuses on strict adherence to verified clinical protocols \cite{REALMEDQA}. In this framework, the decomposition strategy functions as a structural anchor, translating high-level and often ambiguous clinical objectives into a concrete sequence of verifiable sub-tasks \cite{HEALTHBRANCHE}. Subsequently, the iteration strategy acts as a dynamic execution engine, determining the optimal policy to traverse this established path through mechanisms such as closed-loop feedback \cite{MEDIQQA}. This synergy ensures that the agent's behavior remains autonomous in execution while being strictly verifiable in its logic, balancing the need for efficiency with the rigorous safety standards required in healthcare \cite{ASTRID}.

\noindent\textbf{Decomposition strategies}. Within the VWA framework, the decomposition strategy is principally a structured mapping task anchored in external clinical knowledge. Rather than generating novel plans from scratch, the core objective is to align the current patient state with a verified clinical workflow \cite{VITACARE}. This relies on high-quality, structured knowledge bases derived from real-world medical data, such as clinical status pathways mined from electronic medical records \cite{EMRS2CSP} or hierarchical decision trees parsed from authoritative guidelines \cite{HEALTHBRANCHE}. By standardizing the workflow into key nodes that range from problem decomposition to evidence evaluation, this approach ensures that the agent identifies the precise, protocol-compliant step within a complex environment, thereby satisfying the critical medical need for standardization \cite{FROMTOBASE}.

Operationally, this strategy mirrors the logic of clinical fact-based inferential rules, which decouple reasoning into fact verification and inferential reasoning \cite{CMQCICB}. Initially, the agent acts as an information processor, employing techniques akin to RAG to transform unstructured patient dialogue into structured clinical facts. To handle clinical ambiguity, complex inquiries are systematically broken down into standardized elements using models like PICO (population, intervention, comparison, outcome) \cite{FROMTOBASE, AUGMENTINGBB}. Leveraging the foundational reading comprehension of LLMs \cite{MQALMBM}, the agent then utilizes these verified facts to perform inference against the external workflow, decomposing high-level goals into specific, executable sub-tasks. This decoupling renders each step of the clinical decision-making process transparent and evidence-based.

However, addressing the inherent uncertainty of clinical interactions requires dynamic adaptation. The agent must effectively handle incomplete information by proactively initiating information-seeking behaviors rather than making probabilistic guesses and asking clarifying questions to bridge data gaps \cite{MEDIQQA}. Furthermore, to mitigate risks associated with patient misreporting \cite{LISTENTOP}, the strategy incorporates a pre-verification module that assesses input consistency before mapping. Crucially, strict safety boundaries are enforced. If a patient's condition falls outside the predefined workflow distribution, the agent halts execution to signal human intervention \cite{ASTRID}. This cautious handling ensures that autonomy does not exceed the verifiable knowledge base \cite{REALMEDQA}.

Advanced implementations elevate this strategy to expert-level simulation, evolving into a graph of thought processes where the agent navigates a graph of logical sub-tasks \cite{VADK}. By emulating the sequential cognitive process of physicians—assessing, diagnosing, and then planning \cite{MEDPLANA}—the strategy breaks down grand clinical tasks into concrete steps compliant with medical logic. The distinguishing characteristic of this approach is its path-following nature: whereas EP agents focus on discovering a path through open-ended reasoning, VWA agents focus on selecting the correct path from a pre-verified map, prioritizing safety and auditability over creative planning.

\noindent\textbf{Iteration strategies}. Within the VWA framework, the iteration strategy serves principally to perform efficient adaptive navigation along a predefined, verifiable clinical pathway. Unlike the EP paradigm, which relies on divergent planning, VWA iteration addresses the operational challenge of determining how the known next step can be best executed, given patient variability \cite{PIPAUNI}. This process first functions as a closed-loop locate-and-correct mechanism. In clinical practice, patient descriptions are often unstructured and ambiguous, making direct mapping to workflow nodes difficult. To resolve this, agents utilize feedback from tool retrievals to pinpoint their precise location within the protocol. The RGAR framework embodies this by iteratively querying between factual knowledge and conceptual data to accurately align ambiguous clinical objectives with specific workflow paths \cite{RGARRECRR}, effectively turning the process into precision pathfinding on a known map.

To enhance diagnostic efficiency, advanced strategies incorporate reinforcement learning for path policy optimization. Here, the clinical workflow is treated as the environment in which the agent learns to minimize diagnostic costs—such as reducing unnecessary interaction turns—while maximizing accuracy. Systems like Deep-DxSearch employ end-to-end RL to train agents in making optimal sequential decisions within retrieval-reasoning interleaved workflows \cite{E2EAGENTIC}. Similarly, optimization techniques have been applied to refine retrieval strategies in unsupervised environments \cite{LABELFREERAG} and to determine optimal regimens for personalized treatments, such as H. pylori eradication \cite{HLICOBAPYLORI}. Furthermore, this strategy supports cross-task continuous learning, enabling agents to evolve their navigation policies over time from incremental data streams, thereby ensuring that the system's handling of specific cases improves with historical experience.

In scenarios involving complex comorbidities, the iteration strategy evolves into a hierarchical and collaborative cyclic execution. A complex clinical process is modeled as a multi-stage collaborative graph where, if the initial information is insufficient, the system initiates subsequent iterations to delegate specific examination sub-tasks while strictly adhering to preset information flow paths \cite{INTEMULTISOUR}. Concurrently, microscopic iteration ensures robustness at the reasoning level, as seen in the HAR network, which dynamically adjusts attention weights on KGs based on downstream feedback \cite{STAGEAWAREHIER}. By employing risk control frameworks like RULE to filter retrieved information \cite{RULERELIABLEMODL}, these cycles ensure rigorous adherence to protocols. The key difference distinguishing this module is its focus on executional optimization. Whereas EP agents iterate to generate plausible plans, VWA agents iterate to optimize the accuracy and efficiency of traversing a fixed, verifiable clinical map.

\subsubsection{Memory Management}\label{subsubsec:VWAMM}
In the VWA paradigm, memory management transcends the simple recall of conversational history. It is the critical mechanism for maintaining the state, consistency, and traceability of a multi-step clinical workflow. The agent's memory system is not merely a passive repository but an active component that underpins its ability to execute complex procedures in a grounded and reliable manner. It ensures that each action is contextually appropriate, informed by prior steps, and anchored to verifiable information. 

\noindent\textbf{Parametric memory}. In the VWA paradigm, parametric memory functions principally as a sophisticated semantic processing unit rather than as the ultimate source of clinical truth. Its primary role is to provide the foundational linguistic capabilities required to parse complex medical terminology and translate user instructions into precise actions for external execution. While generalist models possess substantial inherent clinical knowledge, benchmarks such as M-QALM reveal that this internal memory is often prone to gaps and inconsistencies compared to human experts, particularly when lacking domain-specific fine-tuning \cite{MQALMBM}. Consequently, the VWA framework strategically repurposes this memory. It is necessary for understanding the medical context, but insufficient for asserting facts.

This operational philosophy is exemplified in frameworks like EMRS2CSP, where the pipeline relies on parametric memory not to hallucinate patient status, but to leverage its internal grasp of sequential clinical logic to decompose unstructured electronic medical records into verifiable clinical status pathways \cite{EMRS2CSP}. Here, the model's latent knowledge serves strictly to navigate the syntax of medical documents, demonstrating how parametric memory enables the execution of structured sub-tasks without becoming the factual authority. Similarly, unified evaluation protocols like PIPA highlight that an agent's ability to maintain state consistency throughout an interaction depends on using parametric memory to track the dialogue flow and evolving clinical requests \cite{PIPAUNI}. In these roles, parametric memory acts as a short-term operational buffer that bridges the gap between natural language and long-term, verifiable non-parametric knowledge stores. Unlike the LSC paradigm, which treats parametric memory as an open clinical knowledge base, the VWA restricts it to the role of a semantic controller for external tools, prioritizing procedural correctness over generative creativity.

\noindent\textbf{Non-parametric memory}. Non-parametric memory constitutes the external, explicit, and auditable information sources that form the bedrock of the VWA's reliability. Unlike the associative memory of the human brain, this component serves as the verifiable aspect of the paradigm, strictly categorized into static long-term repositories and dynamic short-term state logs.

First, static long-term memory functions as the repository of clinical axioms, providing the immutable ground truth for workflow execution. To ensure evidentiary traceability as a non-negotiable requirement in clinical audits, frameworks like Medical Graph RAG construct KGs that explicitly link retrieved clinical information to source documents and formal definitions \cite{MEDGRAGEVIB}. Similarly, MedRAG utilizes hierarchical diagnostic KGs to retrieve distinguishing features between diseases, anchoring the reasoning process in structured ontology rather than in probabilistic generation \cite{MEDRAG25}. Beyond symbolic graphs, handling high-dimensional clinical data requires vectorized representations. Systems like CardioTRAP \cite{CARDIOTRAP} and CLI-RAG \cite{CLIRAGRETRIV} utilize vector stores of EHRs as their primary memory, where every retrieval action constitutes an explicit, loggable query to this verifiable repository. More specialized implementations, such as HI-DR, enhance this concept by structuring the knowledge base as a weighted, directed EHR graph+ to accurately model the asymmetric nature of drug co-prescription relationships \cite{HIDREXPLOITING}. This ensures that the agent's medication recommendations are governed by strict pharmacological rules rather than statistical likelihood.

Conversely, dynamic short-term memory serves as the active clinical state register, explicitly tracking the trajectory of the current session. This capability is critical for multi-turn diagnostic workflows where the patient's status evolves with every interaction. The Deep-DxSearch framework formalizes this by defining the system's state at any step as the complete history of prior actions and environmental feedback, serving as the explicit input for reinforcement learning policies \cite{E2EAGENTIC}. This sequence acts as a verifiable log of the reasoning trajectory, allowing for retrospective error analysis. Addressing the specific clinical challenge of unreliable narrators, the Listening to Patients framework constructs a real-time dialogue entity graph \cite{LISTENTOP}. This graph functions as a structured working memory that the agent consults to detect inconsistencies or mitigate patient misreports during history taking. Furthermore, methods like KPL extend this verifiability to multimodal data by externalizing implicit visual cues from VLMs into a retrieval-based text store \cite{KPLTRAININGFREE}. 

\subsubsection{Action Execution}\label{subsubsec:VWAAE}

In the VWA paradigm, action execution transforms abstract workflow steps into concrete, auditable operations grounded in clinical evidence. This is achieved through three interconnected modalities, each addressing a specific dimension of medical reliability. Its foundation is knowledge-based execution, wherein agent actions are defined as precise queries over structured databases, ensuring every inference is anchored in pre-vetted sources \cite{MEDRAG25}. To integrate real-time information, the search engine functions as a programmatic evidence acquisition node where the LLM generates optimized queries and validates retrieved data, creating a mandatory evidence chain with clear provenance \cite{SEARCHRAG, EVALSEARCH}. Finally, tool-use endows the VWA with the capability to delegate specific clinical tasks—such as risk calculation or image analysis—to deterministic external tools, ensuring operational precision beyond the stochastic nature of LLMs \cite{TOWAIDOCTOR, SILENCEISNOCONS}. Regardless of the modality, the defining characteristic of VWA action execution is that every operation is explicit, traceable, and evidence-backed, guaranteeing end-to-end safety in automated workflows.

\noindent\textbf{Knowledge-based}. The core principle of this modality involves preprocessing unstructured clinical data into structured, machine-queryable knowledge bases\cite{ENHACINGDRUGPRE}. During workflow execution, the agent's actions are strictly defined as retrievals and inferences over these bases, ensuring that operational steps are grounded in verifiable evidence rather than generated from latent parameters.

Early approaches leveraged RAG techniques to transform EHR data into vector databases, supporting specific clinical tasks. The CardioTRAP system exemplifies this by constructing a specialized vector database for cardiology EHRs to manage complex, multidimensional biomedical data \cite{CARDIOTRAP}. Within this framework, workflow actions (e.g., patient risk stratification) are anchored to specific evidence retrieved from vectorized records, demonstrating outcome verifiability. To enhance retrieval precision in generating structured clinical documentation like SOAP notes, the CLI-RAG framework introduces a hierarchical chunking mechanism \cite{CLIRAGRETRIV}. This design allows the agent to first globally identify relevant note types and then perform fine-grained content extraction, respecting the intrinsic structure of clinical documents and rendering the automated generation process interpretable.

However, simple entity co-occurrence often fails to capture complex clinical relationships, necessitating structurally refined knowledge bases. Addressing this, the HI-DR framework constructs an EHR graph+ where directed edges are weighted by the degree of medication co-prescription, capturing asymmetric dependencies between drugs to improve the safety of recommendation actions \cite{HIDREXPLOITING}. Furthermore, to bridge the gap between information retrieval and diagnostic reasoning, advanced agents interact with formalized medical KGs \cite{BEYSEQ}. The MedRAG framework employs a four-tier hierarchical diagnostic KG to capture differential features between diseases \cite{MEDRAG25}. Here, the agent's action shifts from simple retrieval to querying the KG for distinguishing features, actively guiding the diagnosis. Similarly, Medical Graph RAG links user data with authoritative literature to form traceable evidence triplets, enhancing workflow trustworthiness \cite{MEDGRAGEVIB}.

Beyond text, sources for these knowledge bases have diversified to include multimodal data. The KPL framework utilizes knowledge proxy learning to distill implicit visual cues from large vision-language models into explicit textual descriptions \cite{KPLTRAININGFREE}. The agent's classification action then queries this structured base, effectively converting implicit foundation model knowledge into a verifiable resource. Ultimately, the frontier of this domain advances toward end-to-end policy learning. Frameworks like Deep-DxSearch formalize diagnosis as a reinforcement learning problem where the agent's action space consists of well-defined steps—reason, lookup, match, diagnosis—executed within a large-scale retrieval corpus \cite{E2EAGENTIC}. 

\noindent\textbf{Search engine based}. Within the VWA paradigm, the utilization of search engines presents a distinct methodological shift, transcending the passive information retrieval typical of the GS framework to address the specific deficits of implicit knowledge in clinical settings. As clinical practice indicates, directly utilizing original patient queries often yields suboptimal results due to the semantic gap between non-standardized lay descriptions and precise medical terminology \cite{EVALSEARCH}. To bridge this gap, the LLM within the VWA workflow functions as a critical context-aware query generator. Inspired by advanced methods such as SearchRAG \cite{SEARCHRAG}, the agent acts as a clinical translator, decomposing complex medical scenarios into syntactically varied synthetic queries. These optimized instructions are designed to capture latent clinical intents with greater precision, effectively mapping patient narratives to standardized medical indices and significantly enhancing the recall of high-quality guidelines.

Subsequently, the execution phase transforms the search engine from a generic tool into a programmatic evidence acquisition node. These optimized queries are routed to distinct targets, ranging from public indices to professionally engineered, domain-specific information retrieval systems like Tala-med \cite{EHANCINGMEDRET}. Such specialized engines, utilizing modular architectures and synonym expansion, offer information sources of significantly higher relevance than general-purpose counterparts. Furthermore, this conceptualization lays the groundwork for distributed agent networks, where the search engine serves as a core nexus for discovering specialized services \cite{DESADISTRIBUT}. Upon retrieval, the LLM activates its second critical role as a structured validator. It is responsible for parsing unstructured text returns and executing a programmatic validation step, often employing mechanisms based on information entropy to assess the contribution of each snippet to decision confidence. This ensures that only high-value evidence—verified against reliability metrics—is incorporated into the workflow, effectively filtering out the noise inherent in web-scale health data.

The ultimate value of this paradigm lies in its establishment of a mandatory transparency mechanism. By mandating that every decision step be linked to specific literature or data points returned by the search engine, the VWA creates a verifiable and auditable trail. This structured methodology addresses the critical issue of end-user over-trust in opaque search rankings, which often exacerbates health anxiety in online information-seeking scenarios \cite{HOWALGORAMETIC}. Through this approach, the search engine is transformed from a static repository into a dynamic component of clinical decision support, ensuring that automated workflows are founded upon traceable medical evidence. 

\noindent\textbf{Tool-use}. Whereas search engines provide the informational context, tool-use endows the VWA with the operative capability to execute concrete, non-linguistic clinical tasks with mathematical precision. Within this framework, tool-use is strictly defined as the programmatic invocation of deterministic external modules to offload high-risk computations from the probabilistic LLM, including pharmacokinetic calculators and expert systems. Primarily, the LLM functions as an intelligent clinical orchestrator, translating high-level medical intent into specific machine instructions \cite{IMPROVINTEACT}. For instance, in data-intensive scenarios, the agent converts natural language inquiries into executable structured query language for EHR databases. This mechanism is critical for patient safety, as it enables automated, error-free access to complex medical histories without the risk of hallucinating data points inherent in direct text generation \cite{TRANSMEDDATA}. By entrusting specialized tasks to reliable, rule-based tools while reserving the LLM for comprehension and planning, recent studies have demonstrated that such autonomous systems can achieve diagnostic accuracy comparable to board-certified clinicians with minimal safety violations, validating the VWA paradigm in real-world settings \cite{TOWAIDOCTOR}.

Beyond simple calculation, the scope of tool-use extends to an ecosystem of specialized clinical agents that simulate a multidisciplinary medical environment. In this advanced configuration, tools are conceptualized as distinct expert modules. For example, during a simulated consultation, the central LLM acts as a physician who proactively requests findings from a dedicated measurement agent capable of interpreting ECGs or medical imaging, thereby mirroring the referral logic of real diagnostic processes \cite{MEDICALAGENTSIMU}. This interactive capability is further augmented by integrating clinical experience models, which optimize information-gathering strategies to enhance diagnostic precision during the initial phases of care \cite{IMPROVINTEACT}. Furthermore, to address the challenge of diagnostic bias, advanced VWA systems orchestrate tool agents to engage in collaborative reasoning, utilizing specific "critic" agents to challenge premature consensus. This engineered disruption of silent agreement fosters deeper critical thinking and improves robustness in complex cases \cite{SILENCEISNOCONS}. Additionally, tool invocation spans training domains, where agents drive virtual reality simulations for customizable clinical communication practice \cite{DESSIMUCLIN}. 

\subsubsection{Collaboration}\label{subsubsec:VWACL}

In the VWA paradigm, collaboration serves as a computational mirror of the rigid hierarchical interactions found in hospital operations. Unlike the emergent negotiation seen in EP, VWA collaboration is strictly governed by predefined clinical protocols, ensuring that the progression of multi-step tasks—from triage to treatment planning—adheres to the standard of care. The LLM acts not merely as a participant but as the protocol enforcer, facilitating cooperation that prioritizes reliability and patient safety over creative consensus.

\noindent\textbf{Single-agent system}. Conceptually, this architecture functions as a digital general practitioner executing a sequential checklist. Collaboration here is vertical and unidirectional: the agent interacts with a series of external knowledge modules rather than with peer agents. For instance, the workflow might dictate calling a guideline database to retrieve sepsis protocols, then querying a search engine for drug interactions, and finally executing a tool action to update the EHR. The agent serves as the sole active execution unit, with its collaborative scope strictly confined to the pre-configured logic of the clinical toolchain.

\noindent\textbf{Multi-agent system}. To handle the complexity of specialized medicine, the multi-agent architecture simulates the division of labor inherent in a hospital environment. This is predominantly realized through a dominant topology, analogous to an "attending physician" model. Here, a central router agent governs the patient's journey, maintaining strict supervision over the entire workflow to ensure auditability. In the MAKAR framework, this router assesses clinical criteria complexity before dispatching sub-tasks to specialized knowledge-enhancement modules, effectively simulating a referral to a specialist \cite{ENHANCECLITRIAL}. Similarly, TeamMedAgents employs a recruiter agent to dynamically assemble an MDT based on the case profile \cite{TEAMMEDAGENTSE}, while ClinicalAgent utilizes a patient navigator to manage the end-to-end process from triage to consultation \cite{CLINICALLAB}. Crucially, this centralized control mitigates the optimization paradox where individually high-performing specialists fail to produce a cohesive treatment plan due to poor coordination \cite{HILOWAGENT}.

Alternatively, the distributed topology models the linear progression of a clinical care pathway. In this structure, there is no central bottleneck; instead, agents engage in peer-to-peer handoffs based on shared protocols, simulating the transfer of a patient between departments (e.g., from the ED to Radiology). The TAMA framework illustrates this with a strict pipeline model where an agent, upon completing a task like symptom extraction, directly passes the structured output to a reasoning agent according to preset rules, without central mediation \cite{TAMACOLL}. This topology can also incorporate distributed self-correction mechanisms, as seen in TeamMedAgents, where agents monitor peer performance and trigger error corrections based on medical logic \cite{TEAMMEDAGENTSE}. The key difference distinguishing VWA collaboration is its nature of protocol-driven orchestration: whereas EP agents collaborate through negotiation to resolve ambiguity, VWA agents collaborate through rigid handoffs to ensure the clinical workflow is executed without deviation.

\subsubsection{Evolution}\label{subsubsec:VWAEV}

In the VWA paradigm, self-evolution transcends the unconstrained internal knowledge updates typical of implicit navigation models, manifesting instead as a highly structured process of meta-level strategic learning. This mechanism parallels the concept of clinical quality improvement in healthcare, where the objective is not to alter the physician's foundational medical knowledge (the model parameters) but to refine the operational protocols, tool-use strategies, and environmental interaction patterns used in practice. By adhering to the principle of learning-by-doing, the agent systematically distills successes and failures from past diagnostic executions into durable, reusable logic to guide future strategy formulation \cite{HEALTHFLOW}.

One primary aspect of this evolutionary process is the iterative optimization of verifiable workflow components. As a comprehensive automation framework, VWA employs evolutionary algorithms to refine internal agentic workflows, including the adjustment of clinical prompts and tool configurations \cite{EVOAGENTEVOL}. For instance, through meta-tool learning, the agent records and analyzes its interaction history with medical search engines or EHR APIs, thereby autonomously constructing a "procedural knowledge base" that reduces query latency and improves information recall over time \cite{SELFEVOLVINH}. This optimization is strictly grounded in explicit evaluation principles (e.g., maximizing diagnostic accuracy or minimizing test costs), enabling performance enhancement through low-overhead iterations without requiring substantial manual intervention \cite{ZERAZEROINITINS}. Furthermore, by drawing on principles from ensemble learning, VWA evolves superior prompt engineering strategies, such as employing ensemble voting mechanisms to handle biased or ambiguous clinical questions, thereby enhancing the robustness of classification tasks across diverse patient demographics \cite{HEALTHBRANCHE}.

From a broader perspective, the evolutionary mechanism of VWA represents a distinct branch of research focused on behavioral adaptation \cite{SURVEYSELFEVO}. It emphasizes the adjustment of interaction patterns through reflection on historical case outcomes. This characteristic is vital in complex clinical environments where teamwork is essential. In multi-agent collaborative scenarios, an advanced VWA system evolves not just individual competence but also sophisticated collaborative strategies. By observing the dynamics of collective cognition in human-agent hybrid societies, the system learns to adapt communication styles and, crucially, to introduce structured dissent. This evolved capability allows the agent to challenge premature consensus within a diagnostic team, disrupting silent agreement to stimulate deeper critical reasoning and enhance robustness in complex cases \cite{EVOAGENTEVOL}. 

\subsubsection{Summary}

\textbf{In summary, the VWA paradigm defines a clinically conservative approach that prioritizes patient safety and algorithmic accountability by constraining agent behaviors to predefined, auditable care pathways.} Its primary advantage lies in providing the high trustworthiness and traceability required for high-stakes medical execution \cite{TAMACOLL}, a strict operational boundary that distinguishes it from the generative, open-ended planning of the EP paradigm. Furthermore, unlike the GS framework, which focuses on cognitive information synthesis, the VWA is characterized by its commitment to deterministic workflow adherence, effectively transforming the LLM from a passive knowledge interface into an active engine for executing standard operating procedures. However, this reliance on explicit structures introduces rigidity when facing undocumented clinical anomalies, and the optimization paradox reveals critical gaps in ensuring holistic agent compatibility within medical teams \cite{HILOWAGENT}. Addressing these architectural constraints necessitates rigorous validation standards and specialized engineering resources to ensure that such constrained behaviors are viable in practice. This necessity naturally bridges our discussion to the subsequent systematic review of implementation tools, datasets, and evaluation metrics that underpin the development and assessment of these reliable clinical agents. For clarity, we summarize the \textbf{key differences of the four paradigms} in Table \ref{tab:parasum}.

\begin{table}[!htp]
\centering
\setlength{\abovecaptionskip}{0cm}   
\setlength{\belowcaptionskip}{0cm}   
\caption{Architectural deconstruction of the four agentic paradigms. This table summarizes the distinct technical mechanisms employed by each paradigm across the five core cognitive components.}
\label{tab:parasum}
\resizebox{0.9\textwidth}{!}{%
\begin{tabular}{cccclcclccclccc}
\hline
\multicolumn{2}{c}{\multirow{2}{*}{\textbf{Paradigm}}} & \multicolumn{2}{c}{\textbf{Strategic Planning}} &  & \multicolumn{2}{c}{\textbf{Memory Manangement}} &  & \multicolumn{3}{c}{\textbf{Action Execution}} &  & \multicolumn{2}{c}{\textbf{Cooperation}} & \multirow{2}{*}{Evolution} \\ \cline{3-4} \cline{6-7} \cline{9-11} \cline{13-14}
\multicolumn{2}{c}{} & \textbf{Decomposition} & \textbf{Iteration} &  & \textbf{Parametric} & \textbf{Non-parametric} &  & \textbf{Knowledge-based} & \textbf{Search engine based} & \textbf{Tool-use} &  & \textbf{Dominant} & \textbf{Distributed} &  \\ \hline
\multicolumn{1}{c|}{\multirow{2}{*}{\textbf{LSC}}} & Goal & \begin{tabular}[c]{@{}c@{}}Architect\\ cognition\end{tabular} & \begin{tabular}[c]{@{}c@{}}Ensure \\ consistency\end{tabular} &  & \begin{tabular}[c]{@{}c@{}}Internalized medical \\ curriculum\end{tabular} & \begin{tabular}[c]{@{}c@{}}Context \\ extension\end{tabular} &  & - & - & - &  & \begin{tabular}[c]{@{}c@{}}Simulating \\ consultation\end{tabular} & Peer review & \begin{tabular}[c]{@{}c@{}}Prevent \\ forgetting\end{tabular} \\ \cline{2-15} 
\multicolumn{1}{c|}{} & Pattern & \begin{tabular}[c]{@{}c@{}}Internal reasoning\\  chain\end{tabular} & \begin{tabular}[c]{@{}c@{}}Deliberative\\  cycle\end{tabular} &  & \begin{tabular}[c]{@{}c@{}}Implicit weight \\ encoding\end{tabular} & \begin{tabular}[c]{@{}c@{}}Recursive \\ summary\end{tabular} &  & - & - & - &  & \begin{tabular}[c]{@{}c@{}}Central \\ orchestrator\end{tabular} & \begin{tabular}[c]{@{}c@{}}Multi-round \\ discussion\end{tabular} & \begin{tabular}[c]{@{}c@{}}Continual \\ learning\end{tabular} \\ \hline
\multicolumn{1}{c|}{\multirow{2}{*}{\textbf{EP}}} & Goal & \begin{tabular}[c]{@{}c@{}}Cognitive \\ self-guidance\end{tabular} & \begin{tabular}[c]{@{}c@{}}Dynamic \\ adjustment\end{tabular} &  & \begin{tabular}[c]{@{}c@{}}Strategic \\ intuition\end{tabular} & \begin{tabular}[c]{@{}c@{}}Tactical \\ workspace\end{tabular} &  & - & - & - &  & Meta-planning & \begin{tabular}[c]{@{}c@{}}Conflict \\ resolution\end{tabular} & \begin{tabular}[c]{@{}c@{}}Strategy\\  refinement\end{tabular} \\ \cline{2-15} 
\multicolumn{1}{c|}{} & Pattern & \begin{tabular}[c]{@{}c@{}}Cognitive \\ map\end{tabular} & Self-reflection &  & \begin{tabular}[c]{@{}c@{}}Procedural \\ schema\end{tabular} & Dynamic log &  & - & - & - &  & \begin{tabular}[c]{@{}c@{}}Task\\  delegation\end{tabular} & Negotiation & Self-repair \\ \hline
\multicolumn{1}{c|}{\multirow{2}{*}{\textbf{GS}}} & Goal & \begin{tabular}[c]{@{}c@{}}Logical inquiry\\  path\end{tabular} & Gap-filling &  & \begin{tabular}[c]{@{}c@{}}Semantic \\ translator\end{tabular} & \begin{tabular}[c]{@{}c@{}}Evidential \\ ledger\end{tabular} &  & \begin{tabular}[c]{@{}c@{}}Logical \\ grounding\end{tabular} & \begin{tabular}[c]{@{}c@{}}Contextual\\ evidence\end{tabular} & \begin{tabular}[c]{@{}c@{}}Offload \\ calculation\end{tabular} &  & \begin{tabular}[c]{@{}c@{}}Evidence \\ aggregation\end{tabular} & \begin{tabular}[c]{@{}c@{}}Data \\ pipeline\end{tabular} & \begin{tabular}[c]{@{}c@{}}Optimize \\ inquiry\end{tabular} \\ \cline{2-15} 
\multicolumn{1}{c|}{} & Pattern & \begin{tabular}[c]{@{}c@{}}Tree of \\ queries\end{tabular} & \begin{tabular}[c]{@{}c@{}}Info-query\\  loop\end{tabular} &  & \begin{tabular}[c]{@{}c@{}}Pattern \\ mapping\end{tabular} & \begin{tabular}[c]{@{}c@{}}Auditable\\  trace\end{tabular} &  & \begin{tabular}[c]{@{}c@{}}Structured \\ query\end{tabular} & \begin{tabular}[c]{@{}c@{}}Forage \\ constrain\end{tabular} & \begin{tabular}[c]{@{}c@{}}Deterministic\\  call\end{tabular} &  & Hub-and-spoke & DAG flow & \begin{tabular}[c]{@{}c@{}}Policy\\  refinement\end{tabular} \\ \hline
\multicolumn{1}{c|}{\multirow{2}{*}{\textbf{VWA}}} & Goal & \begin{tabular}[c]{@{}c@{}}Protocol \\ alignment\end{tabular} & \begin{tabular}[c]{@{}c@{}}Adaptive \\ navigation\end{tabular} &  & \begin{tabular}[c]{@{}c@{}}Instruction \\ parser\end{tabular} & \begin{tabular}[c]{@{}c@{}}State\\  register\end{tabular} &  & \begin{tabular}[c]{@{}c@{}}Verifiable \\ inference\end{tabular} & \begin{tabular}[c]{@{}c@{}}Programmatic \\ acquisition\end{tabular} & \begin{tabular}[c]{@{}c@{}}Clinical \\ orchestration\end{tabular} &  & \begin{tabular}[c]{@{}c@{}}Strict\\  supervision\end{tabular} & \begin{tabular}[c]{@{}c@{}}Workflow\\  handoff\end{tabular} & \begin{tabular}[c]{@{}c@{}}Meta-level \\ learning\end{tabular} \\ \cline{2-15} 
\multicolumn{1}{c|}{} & Pattern & \begin{tabular}[c]{@{}c@{}}Node \\ mapping\end{tabular} & \begin{tabular}[c]{@{}c@{}}Pathfinding \\ policy\end{tabular} &  & \begin{tabular}[c]{@{}c@{}}Syntax \\ navigation\end{tabular} & \begin{tabular}[c]{@{}c@{}}Clinical \\ axioms\end{tabular} &  & \begin{tabular}[c]{@{}c@{}}Vector \\ retrieval\end{tabular} & \begin{tabular}[c]{@{}c@{}}Query \\ generation\end{tabular} & \begin{tabular}[c]{@{}c@{}}Algorithmic \\ oracle\end{tabular} &  & \begin{tabular}[c]{@{}c@{}}Attending \\ physician\end{tabular} & \begin{tabular}[c]{@{}c@{}}Linear \\ transfer\end{tabular} & \begin{tabular}[c]{@{}c@{}}Meta-tool \\ tuning\end{tabular} \\ \hline
\end{tabular}%
}
\end{table}

\section{Benchmark Datasets and Evaluations}\label{sec:data_eval}
In this section, we provide a detailed overview of the types of clinical dialogue and popular evaluation methods.
\subsection{Datasets}

Researchers have developed numerous benchmark datasets to train and evaluate clinical dialogue agents across diverse communication scenarios. These datasets not only measure model performance but also reflect key tasks and challenges in the field. As comprehensively listed in Table~\ref{tab:comp}, we classify these existing resources into \textbf{five} distinct categories based on core clinical dialogue functions: \textbf{QA dialogue}, \textbf{task-oriented dialogue}, \textbf{recommendation dialogue}, \textbf{supportive dialogue}, and \textbf{hybrid-function dialogue}. This classification clarifies the capabilities targeted by each dataset and supports the selection of appropriate resources for training and evaluation in specific applications.

\begin{table}[!htp]
\centering
\setlength{\abovecaptionskip}{0cm}   
\setlength{\belowcaptionskip}{0cm}   
\caption{Collection of existing agentic medical dialogue datasets for distinct tasks.}
\label{tab:comp}
\resizebox{0.8\textwidth}{!}{
\begin{tabular}{ccc}
\hline
\textbf{Dialogue Type} & \textbf{Specific Task} & \textbf{Dataset} \\ \hline
\multicolumn{1}{c|}{\multirow{3}{*}{QA Dialgoue}} & \multicolumn{1}{c|}{Medical Examination} & \thead{MedQA\cite{MEDQA20}, MedMCQA\cite{MedMCQA22}, cMedQA2\cite{cMedQA2}, \\
            CMExam\cite{CMExam},Medbullets\cite{MedBullet}, HeadQA\cite{HeadQA}, \\
            CasiMedicos-Arg\cite{CASI24},  Huatuo-26{M}\cite{HUATUO26M} \\
            }  \\ \cline{2-3} 
\multicolumn{1}{c|}{} & \multicolumn{1}{c|}{Literature-based} & \thead{PubMedQA\cite{PUBMEDQA}, CliCR\cite{CliCR}, MEDIQA-2019\cite{MEDIQA19}, \\MEDIQA-Ans\cite{MEDIQA-An}, BioASQ\cite{BioASQ}, MedC-I\cite{PMCLL}, \\
             Medical Meadow\cite{MEDALPACA}} \\ \cline{2-3} 
\multicolumn{1}{c|}{} & \multicolumn{1}{c|}{Comsumer Health \& Domain-specific} & \thead{MASH-QA\cite{MASHQA20}, HealthQA\cite{HealthQA19}, webmedQA \cite{webMed} \\ AfriMed-QA\cite{AFRIMEDQA}, RJUA-MedDQA\cite{RJUAMedDQA}, MedAlign\cite{MedAlign}, \\
            MedCalc-Bench\cite{MedCalcBench}, MedHallu\cite{MedHallu}, MedicationQA\cite{MedicationQA},\\  JAMA\cite{MedBullet}, cMedQA2\cite{cMedQA2}
            } \\ \hline
\multicolumn{1}{c|}{\multirow{3}{*}{Task-oriented Dialogue}} & \multicolumn{1}{c|}{Symptom Diagnosis} &  \thead{
MedDialog\cite{MedDialog20}, DialoAMC\cite{DialoAMC23}, MedDG\cite{MEDDG}, MZ\cite{MZ18},\\
 CMDD\cite{CMDD19}, DX\cite{DX19}, CovidDialog\cite{CovidDialog}, iCliniq\cite{CHATD}, \\ Ext-CovidDialog\cite{ExtCovidDialog}, IMCS-21\cite{DialoAMC23}, HealthCareMagic\cite{CHATD}}\\ \cline{2-3} 
\multicolumn{1}{c|}{} & \multicolumn{1}{c|}{Entity Recognition \& Extraction} & \thead{
           Mie\cite{MIE}, GENIA\cite{GENIA}, ADE\cite{DEVE}, BC5CDR\cite{bbb},\\
            NCBI\cite{ncbi}, PHEE\cite{PHEE}, MultiCochrane\cite{MULTIC}, CADEC\cite{CADEC},\\
             MedDG\cite{MEDDG}, NoteCHAT\cite{NoteCHAT},S2ORC\cite{S2ORC}
}  \\ \cline{2-3} 
\multicolumn{1}{c|}{} & \multicolumn{1}{c|}{Instruction Following} & \thead{
          MEDEC\cite{MEDEC}, MedInstruct\cite{MedInstruct}, BianqueCorpus\cite{Bianque},\\
            GPT-3-ENS\cite{GPT3ENS}, MedNLI\cite{MedNLI}, MedSynth\cite{MEDSYN},\\
            MeQSum\cite{MeQSum}, MedSTS\cite{MedSTS}
} \\ \hline
\multicolumn{2}{c|}{Recommendation Dialogue} & \thead{DialMed\cite{DIALMED22}, ReMeDi\cite{REMEDI}, DrugBank\cite{DRUGAGEN}, CMeKG\cite{HUATUO23},\\ MIMIC-III\cite{MEDRAG25}, TCM\cite{TCM}, PubMed\cite{MEDPALM225},NHANES\cite{NHANES},  \\
            ProKnow-data\cite{ProKnowdata}, MedDG\cite{MEDDG}, KaMed\cite{HUATUO225}} \\ \hline
\multicolumn{1}{c|}{\multirow{2}{*}{Supportive Dialogue}} & \multicolumn{1}{c|}{General Empathetic} & \thead{
            EmpatheticDialogues\cite{EMPATH19}, efaqa, Iemocap\cite{IEMO08}, MELD\cite{MELD},\\ DailyDialog\cite{DAILY},  EmotionLines\cite{LINES}, EmoContext\cite{EMOTEXT}}  \\ \cline{2-3} 
\multicolumn{1}{c|}{} & \multicolumn{1}{c|}{Mental Health \& Specific Scenario} & \thead{SUPPORT\cite{SUPPORT}, MTS-dialogue\cite{MTS}, SoulChat-Corpus\cite{SOULCHAT23},\\
            SMILECHAT\cite{SMILE24}, ESConv\cite{ESCONV}, PsyQA\cite{PSYQA},  \\ EDOS\cite{EDOS}, DeepDialogue\cite{DEEP}} \\ \hline
\multicolumn{2}{c|}{Hybrid Function} & \thead{MidMed\cite{MidMed23}, MeQSum\cite{MeQSum}, CHARD\cite{CHARDART} ChiDrug\cite{CHIDRUGBEL}\\
            MedEval\cite{MEDEVAL25}, PathText\cite{MEDICIMAGING}, MENTAT\cite{MVIONGASJD}, MedTrinity\cite{MEDTRINITYM}, \\MedAlpaca\cite{MEDALPACA}} \\ \hline
\end{tabular}}
\end{table}

\subsubsection{QA Dialogue}

QA dialogue datasets assess the knowledge base and factual accuracy of medical intelligent agents by testing their ability to provide correct answers to user queries. Drawn from medical licensing exams to consumer health questions, these datasets enable comprehensive, multi-dimensional evaluation.

\noindent\textbf{Medical examination benchmarks}. To quantify the mastery of professional medical knowledge in models, a significant body of research has utilized datasets based on national medical licensing examinations. MedQA \cite{MEDQA20} is a prominent example, integrating multiple-choice questions from licensing exams in the United States, mainland China, and Taiwan, which rigorously assesses the breadth and depth of a model's clinical knowledge. Building on this, MedMCQA \cite{MedMCQA22} collects questions from Indian medical entrance exams, further expanding the geographical diversity of evaluation benchmarks. To address the need for multilingual capabilities, HeadQA \cite{HeadQA} provides medical exam questions in Spanish, while CasiMedicos-Arg \cite{CASI24} focuses on the Argentinian medical examination context. Additionally, CMExam \cite{CMExam} and CMtMedQA \cite{ZHONGJ} offer important multiple-choice question benchmarks for medical knowledge assessment in Chinese. Due to their standardized format and definitive answers, these datasets have become standard tools for scientifically measuring the professional proficiency of models.

\noindent\textbf{Literature-based QA}. This category of datasets aims to evaluate a model's ability to read, comprehend, and extract key information from unstructured biomedical literature. PubMedQA \cite{PUBMEDQA} requires the model to answer "yes, no, or maybe" questions after reading PubMed abstracts, directly testing its information verification capabilities. BioASQ \cite{BioASQ} presents a more complex challenge, often requiring the model to synthesize answers from multiple literature sources. As a series of evaluation tasks for medical text understanding and question answering, MEDIQA-2019 \cite{MEDIQA19} and MEDIQA-Ans \cite{MEDIQA-An} have advanced the development of models in natural language inference and answer summarization. CliCR \cite{CliCR} focuses on a more specialized domain, requiring the model to verify medical claims based on clinical trial reports.

\noindent\textbf{Consumer health and domain-specific QA}. In contrast to professional examinations, these datasets are more aligned with the daily health needs of the general public. HealthQA \cite{HealthQA19} contains a large volume of health-related questions from real users. MedQuAD \cite{MEDSQ} extracts numerous question-answer pairs from authoritative websites such as the National Institutes of Health (NIH). Similarly, webmedQA \cite{webMed} and Medical Meadow \cite{MEDALPACA} provide rich corpora for general health QA. In more specific domains, MedicationQA \cite{MedicationQA} concentrates on drug-related questions. To address the issue of model-generated hallucinations, MedHallu \cite{MedHallu} was constructed to detect and evaluate such outputs.

\subsubsection{Task-oriented Dialogue}
Task-oriented dialogue datasets are intended for training and evaluating an agent's ability to complete specific clinical or administrative workflows. These dialogues extend beyond single-turn QA, typically involving multi-turn, stateful interactions that require the agent to possess capabilities in contextual understanding, dialogue flow management, and goal-driven action.

\noindent\textbf{Simulated consultation and symptom diagnosis}. This is the most central scenario within task-oriented dialogue datasets. MedDialog \cite{MedDialog20} provides a massive collection of real-world online conversations between doctors and patients, serving as a foundational resource for training simulated diagnostic models. Likewise, datasets such as CMDD \cite{CMDD19}, DialoAMC \cite{DialoAMC23}, MZ \cite{MZ18}, and DX \cite{DX19} offer rich corpora for the online medical consultation scenario. MedDG \cite{MEDDG} not only provides dialogue data but also includes explicit diagnostic and recommendation labels, making it more suitable for training end-to-end diagnostic systems. In response to public health emergencies, datasets like CovidDialog \cite{CovidDialog} and Ext-CovidDialog \cite{ExtCovidDialog} were rapidly developed for the preliminary screening and consultation of COVID-19. Data from ChatDoctor \cite{CHATD} and iCliniq, also sourced from real online medical platforms, possess a high degree of authenticity.

\noindent\textbf{Medical entity recognition and information extraction}. Accurately identifying medical entities such as symptoms, drugs, and tests from dialogue is a prerequisite for task completion. Many classic biomedical named entity recognition and relation extraction datasets provide foundational training for this purpose. Key resources for training an agent's slot-filling capabilities include the disease corpus NCBI \cite{ncbi}, chemical and disease dataset BC5CDR \cite{bbb}, adverse drug events dataset CADEC \cite{CADEC}, adverse drug effects dataset ADE \cite{DEVE}, phenotypic information dataset PHEE \cite{PHEE}, and medical entity dataset GENIA \cite{GENIA}.

\noindent\textbf{Instruction following and note generation}. The powerful instruction-following capabilities of modern LLMs have spurred the creation of new tasks and datasets. MedInstruct \cite{MedInstruct} is a large-scale medical instruction dataset covering over 50 task types, designed to train models to serve as capable assistants to physicians. Medical SOAP and MedSynth \cite{MEDSYN} focus on a critical step in the clinical workflow: generating progress notes in the SOAP format. MeQSum \cite{MeQSum} provides a task for medical dialogue summarization, while MedSTS \cite{MedSTS} enhances a model's understanding of patient chief complaints through a semantic similarity matching task. Large corpora such as BianqueCorpus \cite{Bianque} and GPT-3-ENS \cite{GPT3ENS} also support the training of more generalized task-execution capabilities.

\subsubsection{Recommendation Dialogue}

Recommendation dialogue datasets focus on training agents to provide personalized suggestions, such as recommending suitable drugs, treatment plans, or medical specialists. The challenge in these tasks lies in accurately matching the personalized, unstructured needs expressed by users in natural language with structured medical knowledge bases or large-scale databases.

\noindent\textbf{Drug and treatment recommendations}. DialMed \cite{DIALMED22} is a representative drug recommendation dialogue dataset that requires the model to recommend appropriate medications based on the user's description of diseases and symptoms. The completion of such tasks often relies on robust external knowledge bases, with DrugBank being the most commonly used source that provides comprehensive drug information. ReMeDi \cite{REMEDI} is another dataset that also focuses on drug recommendation.

\noindent\textbf{EHR and KG-based recommendations}. Complex recommendation scenarios require reference to a patient's complete medical history. Large-scale anonymized EHR databases, such as MIMIC-III, are a goldmine of data for building real-world recommendation tasks, although they are not in dialogue format themselves. Researchers can simulate recommendation scenarios based on the diagnostic, medication, and examination records contained within. Similarly, medical KGs like CMeKG, as well as knowledge bases containing traditional medical knowledge such as TCM \cite{TCM}, provide the structured knowledge necessary for agents to perform reasoning and make recommendations. The MedDG \cite{MEDDG} dataset also explicitly integrates dialogue with KGs and can be used for recommendation tasks.

\noindent\textbf{Multi-source recommendation}. To provide more comprehensive health advice, models also need to integrate a broader range of data sources. The vast collection of literature in PubMed can provide evidence for recommending cutting-edge treatment options, while public health data from the NHANES (National Health and Nutrition Examination Survey) database can support recommendations for lifestyle and nutritional supplements. Data from ProKnow-data \cite{ProKnowdata} and the NBME (National Board of Medical Examiners) can also be used to build learning resource recommendation systems for medical students or young doctors.

\subsubsection{Supportive Dialogue}
Supportive dialogue datasets aim to train agents to exhibit empathy, provide emotional comfort, and offer psychological guidance. In these dialogues, technical precision gives way to humanistic warmth, with the core objectives being effective listening and the provision of appropriate emotional feedback.

\noindent\textbf{General empathetic dialogue}. As professional psychological counseling dialogue data are extremely sensitive and difficult to obtain, research typically begins with large-scale, general empathetic dialogue data. EmpatheticDialogues \cite{EMPATH19} is a cornerstone dataset in this domain, providing open-domain dialogues annotated with 32 fine-grained emotions. DailyDialog \cite{DAILY} and EmotionLines \cite{LINES} are also widely used for pre-training models' emotional perception and generation abilities due to their rich everyday conversations and emotional expressions. Iemocap \cite{IEMO08} and MELD \cite{MELD} are classic datasets for multimodal emotion recognition, enhancing a model's ability to capture emotion by combining speech and text.

\noindent\textbf{Mental health support}. In response to the growing need for mental health support, a series of specialized datasets has emerged. PsyQA \cite{PSYQA} collects a large number of real questions and answers from online mental health communities. SoulChat-Corpus \cite{SOULCHAT23} and SMILECHAT \cite{SMILE24} are dialogue corpora developed for creating mental health support chatbots in Chinese and English, respectively. ESConv \cite{ESCONV} explicitly annotates dialogue strategies to train models on how to provide more effective emotional support. The EDOS \cite{EDOS} dataset focuses on the early detection of suicide risk from social media texts.

\noindent\textbf{Specific clinical scenarios}. The SUPPORT \cite{SUPPORT} dataset is a landmark work that records serious conversations about end-of-life care among critically ill patients, their families, and physicians, making it an invaluable resource for training agents to handle complex, high-stakes emotional interactions. MTS-dialogue \cite{MTS} contains motivational interviewing dialogues aimed at motivating patient behavior change. Additionally, cross-domain supportive dialogue data, such as Twitter customer support and DeepDialogue \cite{DEEP}, provide references for models to learn generic soothing and problem-solving linguistic patterns.

\subsubsection{Hybrid-Function Dialogue}
Hybrid-function dialogue datasets are the most complex category and closely simulate the dynamics of real-world clinical interactions. A complete physician-patient communication often seamlessly integrates emotional support, symptom gathering, answering patient questions, and ultimately providing a diagnosis and recommendations. Such datasets are designed to cultivate and evaluate an agent's comprehensive ability to handle dynamic, multi-objective dialogues.

\noindent\textbf{Full-process simulated consultation}. This type of dataset attempts to fully replicate a consultation session from start to finish. MedTrinity \cite{MEDTRINITYM}, through a well-designed, multi-stage framework, utilizes LLMs to generate high-quality, full-workflow simulated dialogues that cover multiple stages from initial consultation to health education. Similarly, MedAlpaca \cite{MEDALPACA} is a dataset built through instruction fine-tuning, aimed at simulating comprehensive, multi-functional interactions between doctors and patients. MidMed \cite{MidMed23} and MENTAT \cite{MVIONGASJD} are also dedicated to constructing comprehensive and functionally rich medical dialogue data.

\noindent\textbf{Comprehensive understanding and complex reasoning}. In complex, lengthy dialogues, the synthesis of and reasoning over information is crucial. The MeQSum \cite{MeQSum} dataset requires the model to generate a summary for a medical QA session, a task that inherently integrates the capabilities of question understanding and summary generation. Although CHARD \cite{CHARDART} and PathText \cite{MEDICIMAGING} focus more on the comprehension of complex medical records or texts, the capabilities they cultivate are foundational for handling hybrid-function dialogues. ChiDrug \cite{CHIDRUGBEL} may involve querying, recommending, and explaining traditional Chinese medicine, representing a typical hybrid-function scenario.



%

%
\subsection{Evaluation Metrics}

The evaluation of LLM-driven clinical dialogue agents presents a complex and multidimensional challenge, as it requires not only an assessment of their linguistic capabilities but also a measure of their diagnostic reliability, decision-making safety, and interactional efficiency in simulated real-world clinical scenarios \cite{RELIABLEZHOU}. While traditional static benchmarks can gauge a model's foundational knowledge, they fail to capture the dynamic and interactive nature of clinical workflows \cite{MEDBENCHM}. Furthermore, many healthcare applications necessitate the integration of heterogeneous data types, including text, images, and laboratory results, which places higher demands on the evaluation framework. Recent studies have shown that simple accuracy metrics can be misleading, as models may exhibit vulnerability when facing adversarial or counterfactual questions, revealing significant deficiencies in their true diagnostic reliability \cite{EVENWORSERAN}. Therefore, a comprehensive evaluation framework must integrate automated, quantitative standard metrics with custom indicators that focus more on clinical utility, metacognitive abilities, and safety. To facilitate this, we systematically organize and categorize these existing metrics in Table~\ref{tab:eval}.

\begin{table}[!htp]
\centering
\setlength{\abovecaptionskip}{0cm}   
\setlength{\belowcaptionskip}{0cm}   
\caption{Collection of existing agentic medical dialogue metrics and evaluation for distinct tasks.}
\label{tab:eval}
\resizebox{0.8\textwidth}{!}{%
\begin{tabular}{cccc}
\hline
\textbf{Metric Category} & \textbf{Specific Application of Metric} & \textbf{Name of Metric or Evaluation} & \textbf{Relvant Essays} \\ \hline
\multicolumn{1}{c|}{\multirow{10}{*}{Standard Metrics}} & \multicolumn{1}{c|}{\multirow{4}{*}{Exact Match}} & \multicolumn{1}{c|}{Accuracy} & Med-PaLM\cite{MEDPALM225}, MedBench\cite{MEDBENCHM}, HuatuoGPT\cite{HUATUO225},  \\ \cline{3-4} 
\multicolumn{1}{c|}{} & \multicolumn{1}{c|}{} & \multicolumn{1}{c|}{Precision} & MedBench\cite{MEDBENCHM}, HuatuoGPT\cite{HUATUO225}, MedAgent\cite{MEDAGENT24}\\ \cline{3-4} 
\multicolumn{1}{c|}{} & \multicolumn{1}{c|}{} & \multicolumn{1}{c|}{Recall} & MedBench\cite{MEDBENCHM}, HuatuoGPT\cite{HUATUO225}, MedAgent\cite{MEDAGENT24} \\ \cline{3-4} 
\multicolumn{1}{c|}{} & \multicolumn{1}{c|}{} & \multicolumn{1}{c|}{F1-Score} & MedBench\cite{MEDBENCHM}, HuatuoGPT\cite{HUATUO225}, MedAgent\cite{MEDAGENT24} \\ \cline{2-4} 
\multicolumn{1}{c|}{} & \multicolumn{1}{c|}{\multirow{6}{*}{\begin{tabular}[c]{@{}c@{}}Content Overlap and \\ Semantic Similarity\end{tabular}}} & \multicolumn{1}{c|}{BLEU\cite{BLEUMET}} & ChatDoctor\cite{CHATD}, MedicalGLM\cite{QEURYEVAL},SoulChat\cite{SOULCHAT23} \\ \cline{3-4} 
\multicolumn{1}{c|}{} & \multicolumn{1}{c|}{} & \multicolumn{1}{c|}{ROUGE\cite{ROUGEMET}} & ChatDoctor\cite{CHATD}, MedicalGLM\cite{QEURYEVAL},SoulChat\cite{SOULCHAT23} \\ \cline{3-4} 
\multicolumn{1}{c|}{} & \multicolumn{1}{c|}{} & \multicolumn{1}{c|}{METEOR\cite{METEORMET}} & ChatDoctor\cite{CHATD}, MedDialog\cite{MedDialog20}, HuatuoGPT\cite{HUATUO225} \\ \cline{3-4} 
\multicolumn{1}{c|}{} & \multicolumn{1}{c|}{} & \multicolumn{1}{c|}{BERTScore\cite{BERTSCOREMET}} & MedAlpaca\cite{MEDALPACA}, Bianque\cite{Bianque} \\ \cline{3-4} 
\multicolumn{1}{c|}{} & \multicolumn{1}{c|}{} & \multicolumn{1}{c|}{MoverScore\cite{MOVERSCORE}} & Huatuo\cite{HUATUO23}, Med-PaLM\cite{MEDPALM225}, BioBART\cite{BioBART22} \\ \cline{3-4} 
\multicolumn{1}{c|}{} & \multicolumn{1}{c|}{} & \multicolumn{1}{c|}{RadGraph F1\cite{RADGRAPHMET}} & MEPNet\cite{MEPNET}, Medical AI Consensus\cite{MEDICONSENS} \\ \hline
\multicolumn{1}{c|}{\multirow{10}{*}{Language Generation Quality}} & \multicolumn{1}{c|}{\multirow{8}{*}{Human Evaluation}} & \multicolumn{1}{c|}{Relevancy\cite{AUTOMEDEVAL}} & AIME\cite{AIME25},  Med-PaLM\cite{MEDPALM225}, HuatuoGPT\cite{HUATUO225} \\ \cline{3-4} 
\multicolumn{1}{c|}{} & \multicolumn{1}{c|}{} & \multicolumn{1}{c|}{Fluency\cite{AUTOMEDEVAL}} & AIME\cite{AIME25},  Med-PaLM\cite{MEDPALM225}, HuatuoGPT\cite{HUATUO225}  \\ \cline{3-4} 
\multicolumn{1}{c|}{} & \multicolumn{1}{c|}{} & \multicolumn{1}{c|}{Knowledge Correctness\cite{AUTOMEDEVAL}} & AIME\cite{AIME25},  Med-PaLM\cite{MEDPALM225}, HuatuoGPT\cite{HUATUO225}  \\ \cline{3-4} 
\multicolumn{1}{c|}{} & \multicolumn{1}{c|}{} & \multicolumn{1}{c|}{FKGL\cite{MEDREADME}} & MedReadMe\cite{MEDREADME}, Patient-Friendly Reports\cite{WORKFLOWLLM}\\ \cline{3-4} 
\multicolumn{1}{c|}{} & \multicolumn{1}{c|}{} & \multicolumn{1}{c|}{ARI\cite{MEDREADME}} &  MedReadMe\cite{MEDREADME}, Patient-Friendly Reports\cite{WORKFLOWLLM}\\ \cline{3-4} 
\multicolumn{1}{c|}{} & \multicolumn{1}{c|}{} & \multicolumn{1}{c|}{SMOG\cite{MEDREADME}} & MedReadMe\cite{MEDREADME}, Patient-Friendly Reports\cite{WORKFLOWLLM} \\ \cline{3-4} 
\multicolumn{1}{c|}{} & \multicolumn{1}{c|}{} & \multicolumn{1}{c|}{RSRS\cite{MEDREADME}} & MedReadMe\cite{MEDREADME}, Patient-Friendly Reports\cite{WORKFLOWLLM} \\ \cline{3-4} 
\multicolumn{1}{c|}{} & \multicolumn{1}{c|}{} & \multicolumn{1}{c|}{SSI\cite{FOUNDATIONMET}} &  MedSentry\cite{MEDSENTRY}, ConfAgent\cite{COSTEFF}, Safety-Prompt\cite{SAFERAG25}\\ \cline{2-4} 
\multicolumn{1}{c|}{} & \multicolumn{1}{c|}{\multirow{2}{*}{LLM-as-a-Judge}} & \multicolumn{1}{c|}{AutoMedEval\cite{AUTOMEDEVAL}} &  AutoMedEval\cite{AUTOMEDEVAL}\\ \cline{3-4} 
\multicolumn{1}{c|}{} & \multicolumn{1}{c|}{} & \multicolumn{1}{c|}{LLM-Judges\cite{IMPROVINGAUTO}} & LLM-Judges\cite{IMPROVINGAUTO} \\ \hline
\multicolumn{2}{c|}{\multirow{7}{*}{Conversational Efficiency Metrics}} & \multicolumn{1}{c|}{Task Success Rate\cite{RELIABLEZHOU}} & AIME\cite{AIME25}, MedAgent\cite{MEDAGENT24}, MeDI-TODER\cite{MEDTOD25} \\ \cline{3-4} 
\multicolumn{2}{c|}{} & \multicolumn{1}{c|}{Number of Turns\cite{RELIABLEZHOU}} & AIME\cite{AIME25}, MedAgent\cite{MEDAGENT24}, MAM\cite{MAM25}, Bianque\cite{Bianque} \\ \cline{3-4} 
\multicolumn{2}{c|}{} & \multicolumn{1}{c|}{Latency\cite{FOUNDATIONMET}} & PMC-LLAMA\cite{PMCLL}, Efficient QA\cite{KNOWLEDGEARGU} \\ \cline{3-4} 
\multicolumn{2}{c|}{} & \multicolumn{1}{c|}{Memory Efficiency\cite{FOUNDATIONMET}} & MedAgentSim\cite{MEDICALAGENTSIMU}, Recursive Sumarizing\cite{RECUR25} \\ \cline{3-4} 
\multicolumn{2}{c|}{} & \multicolumn{1}{c|}{FLOP\cite{FOUNDATIONMET}} &  BioMedLM\cite{BIOMEDLM24}, Efficient QA\cite{KNOWLEDGEARGU}, FreshLLMs\cite{FRESHLLM}\\ \cline{3-4} 
\multicolumn{2}{c|}{} & \multicolumn{1}{c|}{Token Limit\cite{FOUNDATIONMET}} & PMC-LLAMA\cite{PMCLL}, Clinical ModernBERT\cite{CLIMODBERT25} \\ \cline{3-4} 
\multicolumn{2}{c|}{} & \multicolumn{1}{c|}{Number of Parameters\cite{FOUNDATIONMET}} & BioMedLM\cite{BIOMEDLM24}, MedAlpaca\cite{MEDALPACA}, TinyLlama\cite{HUATUO23} \\ \hline
\multicolumn{1}{c|}{\multirow{13}{*}{Custom Indicators}} & \multicolumn{1}{c|}{\multirow{4}{*}{Clinical Safety and Robustness}} & \multicolumn{1}{c|}{Correct Diagnosis Rate\cite{QUALITYSAFE}} &  AIME\cite{AIME25}, ClinicalLab\cite{CLINICALLAB}, ERNIE Bot\cite{QUALITYSAFE}\\ \cline{3-4} 
\multicolumn{1}{c|}{} & \multicolumn{1}{c|}{} & \multicolumn{1}{c|}{Checklist Completion\cite{QUALITYSAFE} } & ERNIE Bot\cite{QUALITYSAFE}, MediQ\cite{MEDIQQA}, Med-PaLM\cite{MEDPALM225} \\ \cline{3-4} 
\multicolumn{1}{c|}{} & \multicolumn{1}{c|}{} & \multicolumn{1}{c|}{Disparity Amplification\cite{EVENWORSERAN}} & Bias Evaluation\cite{BIASRAGQA}, HealthFlow\cite{SILENCEISNOCONS} \\ \cline{3-4} 
\multicolumn{1}{c|}{} & \multicolumn{1}{c|}{} & \multicolumn{1}{c|}{Adversarial Robustness\cite{EVENWORSERAN}} & Med-PaLM\cite{MEDPALM225}, MedSentry\cite{MEDSENTRY} \\ \cline{2-4} 
\multicolumn{1}{c|}{} & \multicolumn{1}{c|}{\multirow{3}{*}{Metacognition and Confidence Calibration}} & \multicolumn{1}{c|}{Confidence Accuracy\cite{LLMLACK}} & MetaMedQA\cite{LLMLACK}, MedConMA\cite{MEDCONMA}, Uncertainty Estimation\cite{UNCERTAIN} \\ \cline{3-4} 
\multicolumn{1}{c|}{} & \multicolumn{1}{c|}{} & \multicolumn{1}{c|}{Expected Calibration Error\cite{ADVANCINGCHNE}} & MR\cite{ADVANCINGCHNE}, MedConMA\cite{MEDCONMA}, Uncertainty Estimation\cite{UNCERTAIN} \\ \cline{3-4} 
\multicolumn{1}{c|}{} & \multicolumn{1}{c|}{} & \multicolumn{1}{c|}{Negative Log-Likelihood\cite{ADVANCINGCHNE}} &   MR\cite{ADVANCINGCHNE}, MedConMA\cite{MEDCONMA}, Uncertainty Estimation\cite{UNCERTAIN}\\ \cline{2-4} 
\multicolumn{1}{c|}{} & \multicolumn{1}{c|}{\multirow{6}{*}{Attribution and Completeness}} & \multicolumn{1}{c|}{Statement-level Support\cite{ANAUTOFRAME}} & SourceCheckup\cite{ANAUTOFRAME}, SearchRAG\cite{SEARCHRAG}, HALO\cite{HALO241} \\ \cline{3-4} 
\multicolumn{1}{c|}{} & \multicolumn{1}{c|}{} & \multicolumn{1}{c|}{Response level Support\cite{ANAUTOFRAME}} & SourceCheckup\cite{ANAUTOFRAME}, SearchRAG\cite{SEARCHRAG}, HALO\cite{HALO241} \\ \cline{3-4} 
\multicolumn{1}{c|}{} & \multicolumn{1}{c|}{} & \multicolumn{1}{c|}{Citation Recall\cite{DOCLENS}} & DOCLENS\cite{DOCLENS}, MedRAG\cite{MEDRAG25} \\ \cline{3-4} 
\multicolumn{1}{c|}{} & \multicolumn{1}{c|}{} & \multicolumn{1}{c|}{Citation Precision\cite{DOCLENS}} & DOCLENS\cite{DOCLENS}, MedRAG\cite{MEDRAG25} \\ \cline{3-4} 
\multicolumn{1}{c|}{} & \multicolumn{1}{c|}{} & \multicolumn{1}{c|}{Claim Recall\cite{DOCLENS}} & DOCLENS\cite{DOCLENS}, Atomic Fact Checking\cite{ATOMRAG} \\ \cline{3-4} 
\multicolumn{1}{c|}{} & \multicolumn{1}{c|}{} & \multicolumn{1}{c|}{Claim Precision\cite{DOCLENS}} & DOCLENS\cite{DOCLENS}, Atomic Fact Checking\cite{ATOMRAG} \\ \hline
\end{tabular}%
}
\end{table}

\subsubsection{Standard Metrics}

Standard metrics, originating from the fields of NLP and information retrieval, provide a baseline for assessing the fundamental linguistic capabilities and information processing accuracy of an agent.

\noindent\textbf{Exact match metrics}. These metrics are applicable to tasks with definitive, clear-cut answers, such as relation extraction from text or multiple-choice questions. They are commonly used to evaluate model performance on standardized examination benchmarks \cite{MEDBENCHM}. Specific metrics include accuracy, precision, recall, and F1-Score. In specialized domains such as biomedicine, however, LLMs may generate responses using synonyms or abbreviations that differ from the gold-standard answers. This causes exact match metrics to fail as they cannot capture semantic equivalence, which is a primary reason why human evaluation remains crucial.

\noindent\textbf{Content overlap and semantic similarity metrics}. These metrics are used to evaluate text generation tasks, such as clinical report generation or diagnostic summarization, by comparing the similarity between the generated text and a gold-standard reference. Metrics based on content overlap include BLEU \cite{BLEUMET}, which assesses precision by counting the number of co-occurring n-grams between the generated and reference texts. ROUGE \cite{ROUGEMET} calculates an F1-score based on n-grams, with a greater emphasis on recall. METEOR \cite{METEORMET} builds upon BLEU by considering synonyms and stemming to calculate an alignment-based F1-score. Metrics based on semantic similarity include BERTScore \cite{BERTSCOREMET}, which evaluates the semantic proximity by computing a cosine similarity matrix between the BERT embeddings of each word in the generated and reference texts. MoverScore \cite{MOVERSCORE} is similar to BERTScore but considers many-to-one word correspondences. In tasks like radiology reporting, RadGraph F1 \cite{RADGRAPHMET} assesses the accuracy of critical clinical information by comparing the F1-score of entities and relations extracted from both the generated and reference texts.

Although widely used, these metrics are limited by the fact that mere textual overlap does not fully reflect clinical correctness. A summary with a high ROUGE score may still omit critical negative findings or fabricate facts. Conversely, two clinically identical statements can have very low BLEU and ROUGE scores. This highlights the inability of these metrics to effectively capture semantic meaning and clinical context.

\subsubsection{Language Generation Quality}
These metrics transcend simple lexical matching to focus on the intrinsic logical properties of the agent's generated content and its adherence to human language conventions. Initially, this was a subjective evaluation dimension dependent on human judgment, but it has since given rise to numerous model-based evaluation methods.

\noindent\textbf{Human evaluation}. Human experts are asked to rate the quality of the generated content across multiple dimensions. Common dimensions include relevancy, which assesses whether the content is on-topic and answers the user's question; fluency, which indicates whether the text is grammatically correct and easy to read; knowledge correctness, which evaluates the accuracy of the medical knowledge contained within the content \cite{AUTOMEDEVAL}; and readability, which assesses the ease with which the text can be understood, ensuring that information is accessible to non-professionals. Researchers have developed specialized medical text readability datasets and measured them using classic formulas like FKGL, ARI, and SMOG, as well as model-based metrics like RSRS \cite{MEDREADME}.

Deeper evaluations include sensibility, specificity, interestingness(SSI) \cite{FOUNDATIONMET}, a comprehensive extrinsic metric used to assess the overall logical flow and coherence of the generated text, ensuring that the response aligns with human behavior and conversational logic. Sensibility determines whether the model's response is logical and understandable. Specificity assesses whether the response is concrete, neither non-committal nor overly general. Interestingness judges whether the response is engaging enough to encourage further interaction. Concise is another key extrinsic metric that evaluates the efficiency and clarity of the model's communication, requiring answers to be both brief and direct while avoiding unnecessary verbosity and repetition. This is particularly crucial in medical scenarios where misunderstandings or contextual confusion can have serious consequences.

\noindent\textbf{LLM-as-a-judge}. In recent years, researchers have begun using advanced language models as proxy evaluators to automate the assessment process. Through carefully designed prompts, a powerful LLM is instructed to evaluate the text quality generated by another LLM \cite{IMPROVINGAUTO}. The AutoMedEval model, for example, is an open-source evaluation model specifically trained for medical question answering \cite{AUTOMEDEVAL}. The validity of such methods is typically confirmed by calculating the Pearson or Spearman correlation coefficient with human expert ratings; a high correlation indicates strong agreement. While significantly more efficient, the accuracy of LLM judges in evaluating complex tasks like biomedical relation extraction can be below 50\%, indicating a need for improved reliability in specialized domains. Various methods aim to ameliorate this; for instance, performance can be significantly improved through structured outputs \cite{IMPROVINGAUTO}.

\subsubsection{Conversational Efficiency Metrics}

These metrics measure the efficiency, resource consumption, and usability of the agent during interaction, which are directly related to user experience and the practical feasibility of deployment. Basic metrics include the task success rate, which measures the proportion of successfully completed clinical tasks and is a core indicator of behavioral effectiveness, and the number of turns, which is used to evaluate the average number of dialog rounds required to complete a task, where fewer turns typically signify higher efficiency \cite{RELIABLEZHOU}. Additionally, there are metrics related to the operational environment, such as latency, memory efficiency, and computational efficiency \cite{FOUNDATIONMET}. Latency measures the round-trip time from the user request to the response, where low latency is crucial for effective communication and user experience. Memory efficiency quantifies the memory usage of the chatbot, which is vital for deployment on resource-constrained devices, such as mobile phones or embedded systems. Floating-point operations (FLOP) quantify the floating-point operations required for a single dialogue instance, serving as a key indicator of the model's computational efficiency and latency. Furthermore, the model's scale also impacts its performance, including the Token Limit, which evaluates the number of tokens the model can process in a multi-turn interaction and directly affects its contextual understanding and resource consumption \cite{FOUNDATIONMET}; and the number of parameters, which signifies the model's size and complexity, influencing its data processing and learning capacity, as well as its memory and computational requirements \cite{FOUNDATIONMET}.

\subsubsection{Custom Indicators}

These indicators are closely tailored to specific application scenarios and requirements, relating to the reliability, safety, ethics, and user acceptance of clinical applications. Given the high-risk and specialized nature of the medical field, it is imperative to design custom evaluation indicators that go beyond the scope of traditional NLP and are directly linked to clinical value.

\noindent\textbf{Clinical safety and robustness}. These metrics directly measure the model's clinical performance, ensuring safety and stability when handling clinical tasks. Safety indicators quantify the frequency of unsafe AI behaviors and include metrics such as correct diagnosis rate and checklist completion. The former evaluates whether the AI's diagnosis and prescription recommendations adhere to clinical guidelines \cite{QUALITYSAFE}, while the latter assesses the comprehensiveness of the AI's inquiry by measuring the coverage of key questions in a simulated consultation \cite{QUALITYSAFE}. Robustness metrics measure the agent's adaptability in complex clinical dialogues and are key to eliminating bias. Disparity amplification evaluates whether the AI treats simulated patients of different ages or economic statuses differently. For instance, by ordering more tests for wealthier patients. Adversarial robustness tests the model's ability to distinguish fact from falsehood rather than relying on superficial cues by using adversarial question pairs that include one stating a fact and the other a slightly altered fallacy. Even advanced models may perform below random chance on such probing evaluations \cite{EVENWORSERAN}.

\noindent\textbf{Metacognition and confidence calibration}. Metacognitive abilities refer to the model's capacity to "know what it doesn't know," which is critical for high-stakes medical decision-making. Benchmarks like MetaMedQA are specifically designed to test this by introducing unanswerable or fictitious questions \cite{LLMLACK}. Confidence calibration is an important measure of a model's reliability, with specific metrics including expected calibration error and negative log-likelihood, which measure the consistency between a model's confidence and its actual accuracy \cite{ADVANCINGCHNE}.

\noindent\textbf{Attribution and completeness}. Attribution metrics evaluate whether each statement generated by the model can be substantiated by its cited literature. Frameworks like SourceCheckup \cite{ANAUTOFRAME} and DOCLENS \cite{DOCLENS} have proposed fine-grained metrics for this purpose. Statement-level Support reflects the proportion of statements supported by at least one cited source \cite{ANAUTOFRAME}. Response level support indicates the proportion of complete responses in which all statements are supported by citations \cite{ANAUTOFRAME}. Citation recall \cite{DOCLENS} measures whether each statement is supported by its corresponding citations, while citation precision \cite{DOCLENS} assesses whether each given citation is necessary and relevant for supporting its corresponding statement. Completeness metrics are used for the fine-grained evaluation of dialogue generation quality by atomically decomposing the answer for comparison. These include claim recall and claim precision, which measure the proportion of key information points from the reference answer that are covered by the generated content, and the proportion of information points in the generated content that are accurate and from the reference, respectively.


%


\section{Real-World Applications}\label{sec:appl} 
The practical application of LLM-enabled agents in healthcare is rapidly evolving, moving from theoretical potential to tangible implementation. To systematically map this burgeoning landscape, we situate current real-world applications within the agentic paradigm framework defined by the orthogonal axes of knowledge and agency. This exercise not only validates the proposed taxonomy by demonstrating its ability to categorize and differentiate existing systems but also illuminates the epicenters of current research and development, while simultaneously revealing nascent frontiers that are yet to be fully explored. By analyzing how current systems populate this matrix, we can discern clear trends in the trade-offs between reliability and creativity, and between safety and autonomy.

\begin{itemize}
    \item \textbf{Application of LSC.} The LSC quadrant represents the most accessible and widely explored application of LLMs in medicine to date \cite{HSCR25}. These agents primarily leverage the vast, unstructured knowledge embedded within the LLM's parameters during pre-training to function as sophisticated medical consultants. A prime example is the use of general-purpose models like ChatGPT to answer board-style examination questions across various specialties, such as pathology \cite{INTERSECAI}. In these scenarios, the agent's role is to comprehend a clinical query and furnish an accurate, explanatory response, thereby demonstrating its grasp of the clinical context. Its strength lies in its broad, generalist knowledge, making it a powerful tool for medical education, preliminary clinical inquiry, and the generation of patient-friendly summaries of complex medical reports. However, because their reasoning is not anchored to verifiable external sources, these agents carry an inherent risk of hallucination, posing a significant barrier to their direct use in high-stakes clinical decision-making.

    \item \textbf{Application of GS.} In stark contrast, the GS quadrant prioritizes reliability and verifiability by anchoring the agent's function in explicit, curated knowledge sources \cite{MEDIATOR}. These agents act as intelligent interfaces, translating natural language into structured queries to provide users with reliable, context-aware syntheses of information. The DrBioRight 2.0 \cite{DrBioRight} system serves as a quintessential example, functioning as an LLM-powered bioinformatics chatbot that enables researchers to interact with a large-scale, proprietary cancer proteomics database through intuitive dialogue. Instead of relying on its internal knowledge, the agent grounds its entire analytical and visualization process on this external, verifiable database, thus mitigating the risk of factual inaccuracies. This paradigm is pivotal for applications where data integrity is paramount, such as interacting with EHRs, querying clinical guidelines, or analyzing administrative data to optimize hospital resource allocation. By acting as a reliable cognitive tool, the GS enhances the accessibility and utility of complex, structured data without sacrificing trustworthiness.

    \item \textbf{Application of VWA.} The VWA quadrant marks a significant leap in agentic capability, coupling explicit knowledge grounding with proactive goal execution \cite{RULERELIABLEMODL}. Agents in this category move beyond mere cognition to actively participate in and automate clinical workflows. The daGOAT system, an autonomous AI agent designed to prevent severe graft-versus-host disease (GvHD), stands as a pioneering example in this domain \cite{AUTOAIINTEL}. This system autonomously monitors real-time, multimodal patient data from the hospital's information system, predicts the risk of GvHD, and independently initiates a preventive drug prescription, which becomes the default action unless vetoed by a clinician. Here, the agent is not a consultant but an active collaborator, executing a critical, multi-step clinical task. Its decisions are both verifiable, as they are based on explicit patient data, and autonomous, as they directly propel the clinical workflow forward. Such systems, which require sophisticated architectures for strategic planning, memory management, and action execution, represent a tangible shift towards the autonomous scientific partners envisioned in next-generation healthcare, automating not just data analysis but clinical intervention itself.

    \item \textbf{Application of EP.} Finally, the EP quadrant remains the most speculative and sparsely populated area in real-world medical applications \cite{MOTOR24}. An agent in this category would leverage the creative, generative power of its implicit knowledge to autonomously devise and execute novel, multi-step clinical plans, potentially without being constrained by existing protocols. For instance, such an agent might devise a novel therapeutic strategy for a rare disease by synthesizing disparate insights from across the medical literature. While this holds immense transformative potential, it also carries substantial risks, as ungrounded, autonomous actions could lead to clinically unsafe or unpredictable outcomes \cite{EVALLLMMED}. The technical and ethical hurdles to developing and deploying such agents are formidable, requiring breakthroughs in managing hallucinations, ensuring goal alignment, and creating robust validation frameworks. Currently, this quadrant serves less as a category of existing applications and more as a critical beacon for future research, highlighting the ultimate challenge and promise of agentic AI in truly reinventing clinical dialogue and practice.
\end{itemize}

\section{Open Challenges \& Future Opportunities}\label{sec:futurecha}
\begin{figure}[!htp]
    \centering
\setlength{\abovecaptionskip}{0cm}   
\setlength{\belowcaptionskip}{0cm}   
    \includegraphics[width=0.3\linewidth,height=0.3\linewidth]{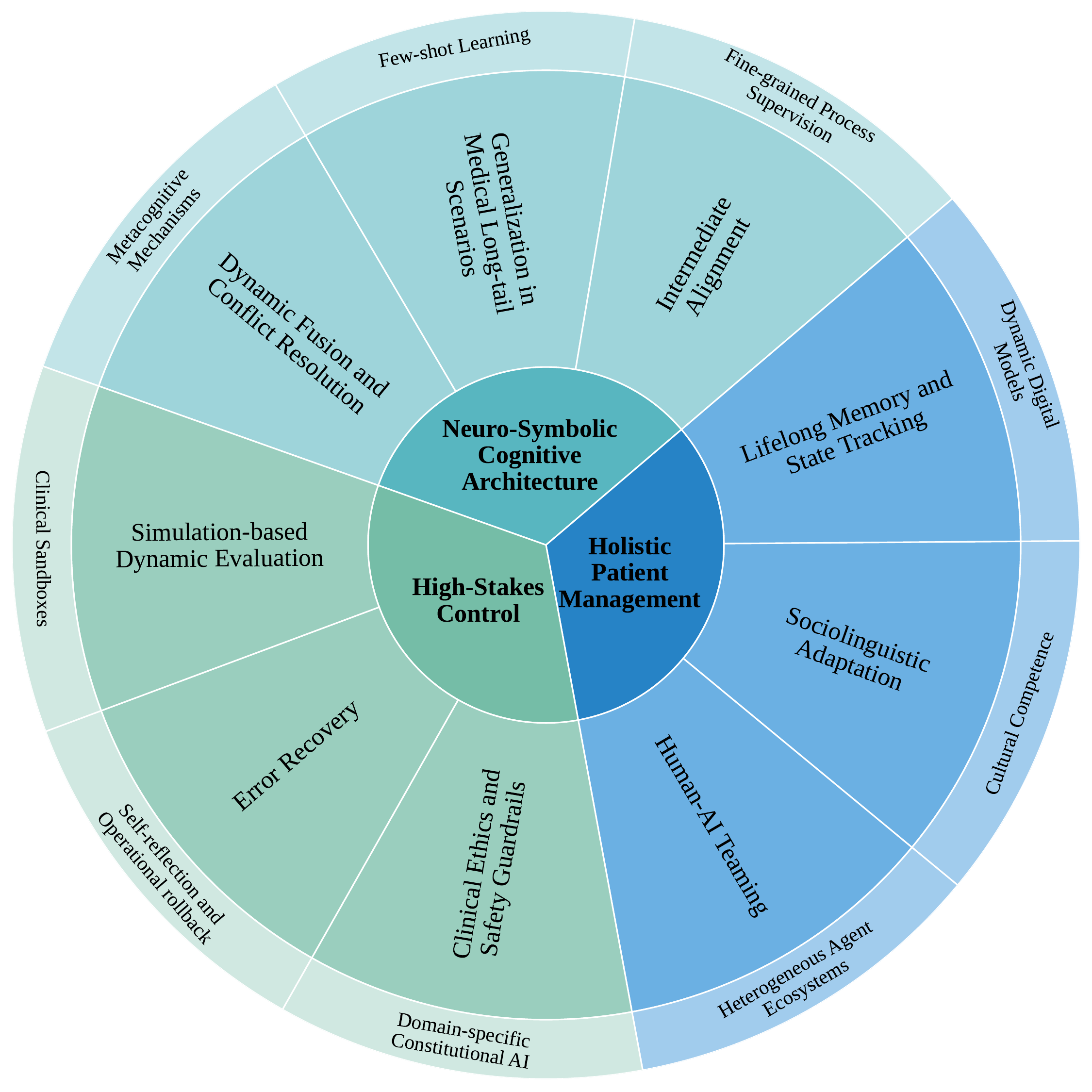}
    \caption{The challenges of agentic clinical dialogue of three perspectives.}
    \label{fig:chall}
\end{figure}

While LLM-based agents have demonstrated transformative potential in shifting healthcare communication from passive information retrieval to active clinical intervention, realizing truly reliable, autonomous, and ethically aligned digital doctors presents formidable challenges. As represented in Fig.~\ref{fig:chall}, this section delineates three critical research trajectories and their associated hurdles.

\begin{itemize}
    \item \textbf{Neuro-symbolic cognitive architectures}. The current dichotomy between creative but hallucination-prone agents and reliable yet rigid systems necessitates a hybrid approach. This hybrid approach aims to reconcile the intuitive reasoning of LLMs with the logical rigor of symbolic systems, primarily through dynamic fusion and conflict resolution. A critical challenge lies in engineering metacognitive mechanisms capable of adaptively arbitrating between parametric memory and non-parametric external knowledge, particularly when model intuition contradicts clinical guidelines \cite{SELFEVOLVINH}. This demands agents possess intrinsic self-doubt and cognitive boundary awareness beyond simple retrieval augmentation. Furthermore, addressing generalization in medical long-tail scenarios, such as rare diseases and complex comorbidities, requires agents to transcend predefined workflows \cite{RAREDIS, TOGENORRET}. By leveraging few-shot learning, medical ontologies, and causal reasoning, agents must autonomously construct logical diagnostic pathways in unstructured environments. Finally, ensuring intermediate alignment remains paramount \cite{TRANSMEDDATA}. Future research must anchor every reasoning step to verifiable medical facts. Developing fine-grained process supervision is essential to ensure that intermediate cognitive actions, from symptom extraction to differential diagnosis, are not only outcome-correct but also logically sound and auditable.
    \item \textbf{Holistic patient management.} Current session-based systems fail to integrate the longitudinal data necessary for comprehensive care. Establishing a genuine therapeutic alliance requires agents to master lifelong memory and state tracking, processing care trajectories that span months or years \cite{KPLTRAININGFREE}. The engineering challenge involves designing efficient memory compression and retrieval mechanisms to extract evolving clinical features, such as disease progression or medication tolerance, from massive EHRs, thereby mitigating catastrophic forgetting and advancing toward dynamic models. Moreover, effective healthcare demands sociolinguistic adaptation, recognizing that clinical dialogue is an interplay of information, emotion, and culture \cite{MEDREADME}. Agents must be endowed with cultural competence to adjust communication strategies based on patient dialects and socioeconomic backgrounds, thereby preventing stereotyping and ensuring equity. Ultimately, the field must evolve toward human-AI teaming, reflecting the multi-disciplinary nature of complex medical decision-making \cite{HUMANAI}. 
    Future directions involve constructing heterogeneous agent ecosystems that seamlessly integrate human expertise, requiring robust communication protocols to prevent inter-agent consensus hallucination and clearly delineated boundaries of accountability.
    \item \textbf{High-stakes control}. As agents transition from passive advisors to active executors, traditional evaluation metrics become obsolete, and safety and ethical alignment must be prioritized. Establishing clinical ethics and safety guardrails is critical, as general reinforcement learning is insufficient for complex medical ethics cite{EQUITY25}. The challenge lies in constructing domain-specific constitutional AI frameworks that encode medical ethics as fundamental constraints, enabling agents to navigate ethical dilemmas and resist adversarial inputs. Simultaneously, error recovery mechanisms are essential for managing the uncontrollable risks associated with external tool use \cite{TOOLARGU}. Research must prioritize robustness, equipping agents with the capacity for self-reflection, operational rollback, and soliciting human intervention to prevent cascading failures in automated workflows. Consequently, evaluation paradigms must shift toward simulation-based dynamic evaluation \cite{MEDICALAGENTSIMU}. Moving beyond static benchmarks, the field requires high-fidelity clinical sandboxes populated by realistic patient simulators. These zero-risk environments are essential for rigorously assessing an agent's interactive strategies, long-term planning, and adaptability to disease progression and emergent conditions.

\end{itemize}

\section{Conclusion}\label{sec:conclu} 
This survey provides a comprehensive review of the ongoing paradigm shift in automated healthcare communication, charting the evolution from traditional pipeline-based systems to the era of autonomous, LLM-driven clinical agents. The implications of this transition are profound, demanding a principled understanding of the new design space and its inherent trade-offs. To this end, we introduced a novel taxonomy structured along two orthogonal axes, which yields four distinct agentic paradigms. For each paradigm, we conduct a granular analysis of its core technical components, including strategic planning, memory management, action execution, collaboration, and evolution, systematically evaluating their respective architectural patterns, technical implementations, advantages, and limitations. By providing this structured framework, this survey offers more than a mere catalog of existing work. It furnishes a conceptual lens to analyze current systems and a roadmap to guide the future development of clinical dialogue agents that are not only powerful and autonomous but also reliable, safe, and aligned with the core tenets of patient care.
\bibliographystyle{ACM-Reference-Format}
\bibliography{main}

\end{document}